\documentclass[letter,dvips]{article}
\usepackage{cite}
\usepackage{graphicx}
\usepackage{subfigure}
\usepackage{xspace}
\usepackage{mathptmx}
\usepackage{helvet}
\usepackage{color}
\usepackage{amsmath}
\usepackage{amsthm}

\newcommand{\bh}{{\sc BumpHunter}\xspace}
\newcommand{\tailHunter}{{\sc TailHunter}\xspace}
\newcommand{\pval}{\ensuremath{p{\mbox{-value}}}\xspace}
\newcommand{\pvals}{\ensuremath{p{\mbox{-values}}}\xspace}
\newcommand{\Ho}{\ensuremath{H_0}\xspace}

\title{On hypothesis testing, trials factor, hypertests and the \bh}
\date{\today}

\author{
Georgios Choudalakis \thanks{gchouda@alum.mit.edu} \\ \emph{University of Chicago, Enrico Fermi Institute}
}

\begin{document}
\maketitle
\begin{abstract}
A detailed presentation of hypothesis testing is given.  The ``look elsewhere'' effect is illustrated, and a treatment of the trials factor is proposed with the introduction of hypothesis hypertests.  An example of such a hypertest is presented, named \bh, which is used in ATLAS \cite{dijetResonanceSearch}, and in an earlier version also in CDF \cite{vistaPRDRC}, to search for exotic phenomena in high energy physics.
As a demonstration, the \bh is used to address Problem~1 of the Banff Challenge \cite{BanffChallenge}.
\end{abstract}

\tableofcontents

\section{Introduction}

The goal of the \bh is to point out the presence of a local data excess like those caused by resonant production of massive particles in Particle Physics \cite{dijetResonanceSearch}.  Such features are colloquially called ``bumps'', hence the name ``\bh''.  More specifically, the \bh is a test that locates the most significant bump, where the data are most deviant from the Null hypothesis.  Based on this bump, the test returns a \pval, corresponding to its Type-I error probability.  This is done in a way that accounts for the ``trials factor''.

For the reader who may not be familiar with the terminology of hypothesis tests, a thorough discussion follows.  Another account of hypothesis testing can be found in \cite{Whalley:2002tu}.  A similar discussion on trials factor can be found in \cite{2010EPJC.70.525G}.  

In the following paragraphs we spell out issues that are often misunderstood, such as the interpretation of \pvals and the issue of ``trials factor''.  A solution is provided to account for the latter, by introducing the notion of {\em hypothesis hypertest}.  The discussion that follows is not limited only to the \bh; the latter is a practical application.

After presenting the \bh algorithm, a demonstration is made, based on Problem 1 of the Banff Challenge \cite{BanffChallenge}.

\subsection{Hypothesis tests and \pvals}
\label{sec:allTests}

There are several statistical tests to evaluate if some data are consistent with a specific hypothesis.  Two famous examples are Pearson's $\chi^2$, and the Kolmogorov-Smirnov (KS) test.  
The \bh is one more test in this category.   

In all tests of this kind, often called ``hypothesis tests'' or ``goodness of fit tests'', one has some data $D$ and a hypothesis, which typically is the ``Null'', or ``0-signal'', or ``background'' hypothesis, denoted \Ho.   One could test the consistency of $D$ with {\em any} hypothesis, but \Ho is usually chosen, because typically a discovery can be claimed by establishing that the data are inconsistent with the ``standard'' theory, without having to show necessarily that they are consistent with some alternative theory.  Once inconsistency with \Ho is established, several alternative signal hypotheses can be tested to characterize the discovery.  For example, we can assume that the signal follows a specific distribution, and estimate its amount, either by bayesian inference, or by defining frequentist confidence intervals (CIs).
It helps, conceptually, to distinguish hypothesis tests, like $\chi^2$ or the \bh, from bayesian inference and frequentist CI-setting methods\footnote{\label{foot:CI} There is, actually, a connection between hypothesis tests and frequentist CIs, which will be explained in this footnote, hoping to avoid confusion.  One can assume {\em any} kind of signal, and set a {\em lower} limit to the amount of this signal that may exist in $D$, using the classical Neyman construction, where the statistic of some test is used as observable; to be specific, let's say $\chi^2$ is used to construct the Nayman band.  If the resulting CI doesn't contain the value 0 for signal, then \Ho is excluded, in the frequentist sense, namely in the sense that 0-signal is not contained in a CI characterized by some Confidence Level (CL).  The {\em smallest} CL for which the corresponding semi-infinite CI includes the value 0 for signal, is equal to $(1-\pval)$ of the hypothesis test (of the $\chi^2$ test in this case) which compares $D$ to \Ho.  This is the case for any assumed signal shape. }.   One can use as observable the value of $\chi^2$ or of the \bh statistic (defined below) to make a bayesian inference or to set a frequentist CI on the amount of a specific signal that may exist in the data, but the \bh is designed to address a different question, for which only $D$ and \Ho are required, and no specific signal is assumed, hence its model-independence.

All hypothesis tests, including the \bh, work as follows:
\begin{enumerate}
\item {$D$ is compared to \Ho, and their difference is quantified by a single number.  This number is called ``the statistic'' of the test, or ``test statistic'', and in this document it is denoted by $t$.  For example, in the $\chi^2$ test, the statistic is 
\begin{equation}
\label{eq:chi2stat}
t = \sum_{i} \left(\frac{d_i - b_i}{\sqrt{b_i}}\right)^2,
\end{equation}
where $d_i$ denotes the observed events in bin $i$, and $b_i$ the events expected by \Ho in the same bin.  The statistic in the KS test is the biggest difference between the cumulative distribution of the data and the cumulative distribution expected by \Ho.  We will present later the exact definition of the \bh statistic, but it follows the same logic: the bigger the difference between data and \Ho, the bigger the test statistic. }
\item {Pseudo-data are generated, following the expectation of \Ho.  In each pseudo-data spectrum, the same test statistic is computed, comparing the pseudo-data to \Ho.  The distribution of test statistics from pseudo-experiments is made.  The achievement of Pearson, Kolmogorov and Smirnov, was that they calculated analytically the distribution of the statistic of their tests under \Ho.  For example, Pearson showed that, under some assumptions of gaussianity, his $\chi^2$ statistic follows a $\chi^2$-distribution.    Nowadays, computers make it possible to estimate numerically the distribution of any test statistic.}
\item {Calculate the \pval of the test.  The \pval is the probability that, when \Ho is assumed, the test statistic will be equal to, or greater than\footnote{The convention used is that the test statistic becomes greater as the discrepancy increases; otherwise the \pval would be defined as $P(t \le t_o | H_0)$.} the test statistic obtained by comparing the actual data ($D$) to \Ho:
\begin{equation}
\label{eq:pvalDef}
\pval \equiv P(t \ge t_o | H_0),
\end{equation}
where the test statistic $t$ is a random variable since it depends on how pseudo-data fluctuate around \Ho, and $t_o$ is the observed statistic from comparing $D$ to \Ho.
If the exact probability density function (PDF) of $t$ under \Ho is known ($\rho(t|H_0)$), then the \pval is exactly computed as $\int_{t_o}^\infty \rho(t|H_0) dt$.  When $\rho(t|H_0)$ is estimated using pseudo-experiments, as the case is for the \bh, then the \pval is estimated as a binomial success probability.  Using Bayes' theorem, if $N$ pseudo-experiments are produced, of which $S$ had $t \ge t_o$, we infer 
\begin{equation}
p(\pval | N,S) = \binom{N}{S} {\pval}^S(1-\pval)^{N-S} \frac{\pi(\pval)}{\mathcal N},
\end{equation}
where $\mathcal N$ is a normalization constant, and $\pi(\pval)$ is the prior assumed.  If we assume $\pi(\pval) = 1$, which is a reasonable choice, the result becomes
\begin{equation}
p(\pval | N,S) = \binom{N}{S} {\pval}^S(1-\pval)^{N-S} (1+N).
\end{equation}
According to this posterior distribution, the most likely \pval is $\frac{S}{N}$.
}
\end{enumerate}

So, the final product of a hypothesis test of this kind is a \pval.  Ideally, the \pval would be precisely computed, but in practice is has to be estimated from a finite set of pseudo-experiments.  We will explain next how the \pval can be interpreted, and why it is so useful.

\subsection{What does the \pval mean?}
It will be shown that the \pval is interpretable as a false-discovery probability.  To reach systematically to that interpretation, and to clarify what that means, we will first prove a simple theorem.

\subsubsection{A simple theorem about \pvals}
\label{sec:theorem}

Assume a decision algorithm which declares discovery (i.e.\ it rules out \Ho) if  $\pval \le \alpha$, where $\alpha \in [0,1]$ is an arbitrary parameter of the algorithm.  It will be shown that the probability of this algorithm to {\em wrongly} rule out \Ho is $\alpha$, no matter what hypothesis test the \pval is coming from, under one condition; that there be a solution $\zeta$ for which $\int_\zeta^\infty \rho_t(x) dx = \alpha$, where $\rho_t$ is the PDF followed by the test statistic ($t$) under \Ho.

The probability to wrongly rule out \Ho, which is named ``Type-I error'', is the probability to find $\pval \le \alpha$ while \Ho holds, namely
\begin{equation}
P(\mbox{Type-I}) = P(\pval \le \alpha | H_0),
\end{equation}
which can be spelled out more clearly, using the definition of \pval:
\begin{equation}
\label{eq:deriv1}
P(\mbox{Type-I}) = P( P(t \ge t_o|H_0) \le \alpha | H_0).
\end{equation}
In eq.~\ref{eq:pvalDef}, $t$ was a random variable and $t_o$ was a fixed number, which depended on the real data $D$ and on \Ho. 
In eq.~\ref{eq:deriv1}, both $t$ and $t_o$ are random variables, because we don't have a fixed observed dataset $D$; we are instead trying to calculate the probability that $D$ will be such that $t_o$ will satisfy $P(t \ge t_o | H_0) \le \alpha$.
In other words, eq.~\ref{eq:deriv1} is the probability of drawing a random variable $t_o$, such that the random variable $t$ will have probability less than $\alpha$ to be greater than $t_o$.
That happens if $t_o \ge \zeta$, where $\int_\zeta^\infty \rho_t(x) dx = \alpha$, where $\rho_t$ is the PDF followed by $t$.\footnote{\label{foot:theorem} The equation $\int_\zeta^\infty \rho_t(x) dx = \alpha$ needs to have a solution; if $\zeta$ doesn't exist, the rest of the proof fails.  For example, consider $\rho_t(x) = \frac{1}{2}+\frac{1}{2}\delta(x-0.5)$, where $x\in[0,1]$ and $\delta(\cdot)$ is the Kronecker $\delta$ function.  In this case, there is no $\zeta$ that satisfies $\int_\zeta^\infty \rho_t(x) dx = \alpha$ for $\alpha \in [0.25,0.75)$, because if $\zeta > 0.5$ then $\int_\zeta^\infty \rho_t(x) dx < 0.25$, and if $\zeta \le 0.5$ then $\int_\zeta^\infty \rho_t(x) dx \ge 0.75$.  If $\rho_t(x)$ is continuous, then a $\zeta$ exists $\forall \alpha$.  Most test statistics based on event counts don't follow a continuous PDF, due to event counts being discrete.  Another possible reason to not follow continuous PDF is the imposition of conditions as we will see paragraph \ref{sec:bhAlgo}.  So, there are specific values of $\alpha$ for which this theorem is exactly true; in other cases the probability to wrongly exclude \Ho is not exactly $\alpha$.  However, we will explain in paragraph~\ref{sec:singleTest} that this theorem's condition is met if we set $\alpha$ equal to an {\em observed} \pval, which allows any observed \pvals to be exactly interpreted as a Type-I error probabilities.}
The probability for $t_o$ to be greater than $\zeta$ is $\int_{\zeta}^\infty \rho_{t_o}(x) dx$, where $\rho_{t_o}$ is the PDF followed by $t_o$.
So, eq.~\ref{eq:deriv1} can be written
\begin{equation}
P(\mbox{Type-I}) = \int_{\zeta}^\infty \rho_{t_o}(x) dx, \mbox{\ where\ } \int_\zeta^\infty \rho_t(x) dx = \alpha
\label{eq:deriv2}
\end{equation}
By looking back at eq.~\ref{eq:deriv1}, we see that $t$ fluctuates according to how pseudo-data fluctuate around \Ho, as implied by the conditional in $P(t \ge t_o|H_0)$.
At the same time, $t_o$ fluctuates according to how pseudo-data fluctuate around \Ho, as implied by the rightmost conditional in eq.~\ref{eq:deriv1}.
So, both $t$ and $t_o$ are drawn from the same distribution, namely  $\rho_{t_o}(x) = \rho_t(x)$.
Therefore, eq.~\ref{eq:deriv2} becomes
\begin{eqnarray}
\nonumber P(\mbox{Type-I}) &=& \int_{\zeta}^\infty \rho_{t}(x) dx, \mbox{\ where\ } \int_\zeta^\infty \rho_t(x) dx = \alpha \\
\Rightarrow\ P(\mbox{Type-I}) &=& \alpha
\end{eqnarray}

This is an important result, and it is what makes \pvals useful.  We showed that, no matter how we define the test statistic $t$, if we use the resulting \pval in a discovery algorithm that declares discovery when $\pval \le \alpha$,  the Type-I error probability of that algorithm will be equal to $\alpha$.   The only requirement is for a $\zeta$ to exist that satisfies $\int_{\zeta}^\infty \rho_t(x) dx = \alpha$.  

\paragraph{Corollary:} If a test statistic $t$ follows a continuous PDF $\rho_t$ under \Ho, then the condition of the above theorem is satisfied for any value $\alpha \in [0,1]$, therefore  $P(\pval \le \alpha | H_0) = \alpha$ $\forall \alpha$, therefore the \pval of any such hypothesis test is a random variable that follows a uniform distribution between 0 and 1, when \Ho is true.  

Note that if $\rho_t$ is discontinuous, then this corollary does not follow, i.e.\ the \pval does not follow a uniform distribution between 0 and 1, but the previous theorem is still valid for $\alpha$ values for which $\int_\zeta^\infty \rho_t(x) dx = \alpha$ has a solution.  This is important, because it is often wrongly thought that if a \pval doesn't follow a uniform distribution under \Ho, then it can not be correctly interpreted as a Type-I error probability.  That is not true.  In paragraph~\ref{sec:singleTest} we will see why.

\subsubsection{Interpretation of the \pval of a test}
\label{sec:singleTest}

So, how should we interpret the \pval of a hypothesis test checking the consistency of a dataset $D$ with a hypothesis \Ho?

Exploiting the theorem of paragraph \ref{sec:theorem}, if we observe $\pval = \gamma$ we know that there is a discovery algorithm which would have ruled out \Ho based on this \pval with Type-I error probability equal to $\gamma$.  That algorithm is the one with parameter $\alpha = \gamma$.   If we set $\alpha < \gamma$ then the algorithm wouldn't declare discovery for the observed \pval.  If we set $\alpha > \gamma$ a discovery would still be declared, but such an algorithm would have a larger Type-I error probability, so it would be less reliable.  Therefore, if we observe $\pval = \gamma$, then the discovery algorithm with the smallest Type-I error probability that would still declare discovery would do so with probability $\gamma$ of being wrong.  In this sense, the observed \pval\ {\em is} a false-discovery probability.  It is the smallest false-discovery probability we can have, if we declare \Ho to be false.   

What if the hypothesis test $t$ follows a discontinuous PDF $\rho(t)$?  We saw in \ref{sec:theorem} that in that case there can be some values of $\alpha$ for which the proof can not proceed, because there is no $\zeta$ satisfying $\int_\zeta^\infty \rho(t) dt = \alpha$.  That, however, does not interfere with the interpretation of an observed $\pval = \gamma$ as a Type-I error probability.  The reason is that $\gamma$ will always be such that $\int_\zeta^\infty \rho(t) dt = \gamma$ will have a solution, so, the theorem of paragraph \ref{sec:theorem} will always hold, if we set $\alpha = \gamma$.  How do we know that any observed $\pval = \gamma$ will always be such that $\int_\zeta^\infty \rho(t) dt = \gamma$ will have a solution?  We know, because otherwise $\gamma$ couldn't have been observed.  Let's take, for example, the discontinuous PDF that was mentioned in paragraph \ref{sec:theorem}: $\rho(t) = \frac{1}{2} + \frac{1}{2}\delta(t-0.5),\ t\in[0,1]$.  For this $\rho(t)$, as mentioned earlier, the equation $\int_\zeta^\infty \rho(t) dt = \alpha$ has no solution for $\alpha \in [0.25,0.75)$, but this is precisely the range where $\gamma$ couldn't be in any circumstance.  If $t_o > 0.5$, then the \pval will be $< 0.25$.  If $t_o \le 0.5$, then $\pval \ge 0.75$.  

We showed, therefore, that any observed \pval will always be interpretable, thanks to the theorem of paragraph \ref{sec:theorem}, as the smallest possible Type-I error probability of a discovery algorithm which would have declared discovery on the basis of the observed \pval.  This interpretation will be correct even if the conditions are not met for the corollary of \ref{sec:theorem} to be true, i.e.\ even if the \pval is not distributed uniformly in $[0,1]$ under \Ho.

To prevent a common misinterpretation,  if we find a \pval = 0.7, it doesn't mean that \Ho is right with probability 70\%.  In strictly frequentist terms, the \pval is not a statement about \Ho itself, but about the Type-I error probability of an algorithm that would exclude \Ho, as explained above\footnote{Equivalently, the \pval corresponds to the CL of a specific CI.  See footnote \ref{foot:CI}.}.

\subsection{Interpretation of multiple tests}
\label{sec:manyTests}


If we run the KS test and find \pval = 0.7, we know that even the most reliable decision which would rule out \Ho\ {\em on the grounds of the KS test} would still have 70\% probability to be wrong.    With so high odds of being wrong, we couldn't support a discovery claim.  But the fact KS doesn't identify a big discrepancy doesn't mean no other test will.  For example,  the data $D$ may follow the PDF predicted by \Ho, but have different population.  Since the KS test compares cumulative distributions, it is insensitive to an overall normalization difference, while the $\chi^2$ test would notice it.  So, if the $\chi^2$ test returns \pval = $10^{-6}$, we can say that the most reliable decision which would rule out \Ho\ {\em on the grounds of the $\chi^2$ test} would have probability $10^{-6}$ to be wrong.  With such high confidence, a discovery claim could be supported.  This statement from $\chi^2$ does not contradict the one from KS.  Both are correct, simultaneously.  One says that the $D$ distribution shape agrees with \Ho; the other says that the normalization doesn't.  

The above scenario illustrates why one can benefit from more than one statistical test.  Each test is sensitive to different features, and we may not know a-priori how $D$ may differ from \Ho.  Unless one is willing to limit the scope of his search to only one kind of discrepancy (e.g.\ shape discrepancy or normalization discrepancy), he needs to compare $D$ to \Ho in more than one way.  To do so correctly, he must carefully take into account the ``trials factor'', which is the subject of the next paragraph.

\subsubsection{Ad-hoc tests, and trials factor}

Reading paragraph \ref{sec:manyTests}, one may be tempted to ``engineer'' more hypothesis tests, until one of them gives a small \pval that would allow him to rule out \Ho with great confidence.  For example, imagine that the data $D$ are binned in $10^4$ small bins.  In so many bins, it is only natural for one bin to fluctuate significantly from the \Ho prediction, even if \Ho is true.  If a hypothesis test is engineered to look just at that bin, then the observed statistic ($t_o$) will be very large, and the \pval will be very small, because pseudo-experiments will very rarely have as big a discrepancy in the same bin.  

Even for such an ad-hoc test, everything we proved still holds.  It would be technically correct that {\em based on this a-posteriori decided test} we could rule out \Ho with a tiny chance of being wrong.  And yet, any minimally skeptical scientist should refuse to rule out \Ho based on this result.   All it says, in essence, is that there is one out of the $10^4$ bins that is very discrepant.  If we had stated it like that, it wouldn't have sounded so dramatic, but that's really what it means, and the reason is that the bin had not been chosen a-priori, but after seeing $D$.  If a different bin had fluctuated far from \Ho, then another a-posteriori test would have been quoted, which would again rule out \Ho with high confidence, even if \Ho were true.  This is what physicists refer to as ``the look elsewhere effect'', or ``the trials factor'', implying that each bin counts as a trial with its own chance of triggering a discovery, and the fact there are many such trials has to be taken into account somehow.  It will become clear later that the ``trials'' actually are not due to the many bins, but due to the many possible hypothesis tests one would be interested in considering simultaneously.  In other words, the ``look elsewhere effect'' may better be called ``look in different ways effect''.

\subsubsection{How to account for the trials factor -- Hypertests}
\label{sec:howToAccount}

Continuing the example of the previous paragraph, to see if there is a single-bin fluctuation that is too unlikely under \Ho, without having any prior preference to some bin, we will come up with a statistical test that considers all possible bins on an equal footing.   It will have, like every hypothesis test, a statistic $t$ and a \pval corresponding to the observed statistic $t_o$.  Its \pval will follow the theorem of paragraph \ref{sec:theorem}, it will therefore be interpreted as the Type-I error probability based on this test.

The hypothesis test that looks at all bins can be viewed as a hypertest, which combines all the specialized tests which focus on individual bins.  
These many tests are the many ways in which a discovery could be claimed.
These many tests are the ``trials''.  We will see how to construct such a hypertest.

In our example, where the data $D$ are partitioned in $N=10^4$ bins, one could define $N$ hypothesis tests, each using one bin to define its test statistic.  Of these $N$ hypothesis tests, each can use any test statistic; they don't even have to be the same.  For example, for hypothesis tests that examine odd bins we could define the test statistic
\begin{equation}
t_{i \in {\rm odds}} = (d_i-b_i)^2,
\end{equation}
where $d_i$ and $b_i$ are the observed data and the expectation of \Ho in the bin $i$ where each hypothesis test focuses.  For hypothesis tests that examine even bins, we could define the test statistic
\begin{equation}
t_{i\in {\rm evens}} = (d_i-b_i)^{100}.
\end{equation}
No matter how we define these hypothesis tests, regardless how numerically different their statistics may be, for each one of these $N$ tests there is an observed statistic $t_{io}$, and the corresponding $\pval_i$ in the interval $[0,1]$.  For each one of these $N$ \pvals, the theorem of paragraph \ref{sec:theorem} holds: If \Ho is true, then each one of these tests has probability $\alpha$ to return $\pval_i \le \alpha$.   

In this example the $N$ tests are independent, meaning that $$P(\pval_i \le \alpha | \pval_j \le \alpha) = P(\pval_i \le \alpha) \ \ \forall \{i,j\},$$ so the probability of at least one such hypothesis test giving a $\pval_i \le \alpha$ is 
\begin{eqnarray}
\nonumber P(\mbox{at least one\ test\ \pval} \le \alpha)  &=& 1 - \prod_{i=1}^{N}P(\pval_i > \alpha) \\
 &=& 1 - (1-\alpha)^N 
\end{eqnarray}
In this case we may use the phrase ``the trials factor is $N$'', meaning that this set of hypothesis tests consists of $N$ statistically independent tests.  
If, on the contrary, all $N$ tests were totally correlated, meaning that $$P(\pval_i \le \alpha | \pval_j \le \alpha) = 1 \ \ \forall \{i,j\},$$   then we would have 
\begin{eqnarray}
\nonumber P(\mbox{at least one\ test\ \pval} \le \alpha)  &=& P(\pval_i \le \alpha)\ \ \ \forall i \\
 &=& \alpha = 1 - (1-\alpha)^1 
\end{eqnarray}
In this case, we may say ``the trials factor is 1'', meaning that, although there are many ($N$) hypothesis tests in the set we are considering, they count as 1 because they behave identically.
In any intermediate case of partial independence, we can define a real number $\tilde{N}$ such that 
\begin{eqnarray}
\label{eq:effTrialsFactor}
  \nonumber P(\mbox{at least one\ test\ \pval} \le \alpha)  \equiv 1 - (1-\alpha)^{\tilde{N}}\ \ \Rightarrow \\
  \tilde{N} \equiv \log_{(1-\alpha)}(1-P(\mbox{at least one\ test\ \pval} \le \alpha)).
\end{eqnarray}
We can refer to $\tilde{N}$ as the {\em effective trials factor}, which can take values between 1 and $N$.  The value of $\tilde{N}$ depends on $N$, on the way the $N$ hypothesis tests are correlated, and on $\alpha$.  It should be clear at this point that the trials factor has little to do with how many bins there are in the data, or how many final states we consider in a search for new physics\footnote{More bins and more final states allow one to devise more hypothesis tests, but one doesn't have to.}.  It is really a function of the number of hypothesis tests that we employ, and of how their answers correlate.  

We just showed that a discovery algorithm that says ``declare discovery if any of the $N$ tests gives $\pval \le \alpha$'' does {\em not} have Type-I error probability equal to $\alpha$, but equal to $1-(1-\alpha)^{\tilde{N}}$ which is $\ge \alpha$.  That is why we cannot look at a set of hypothesis tests (e.g.\ $N$ tests, each looking at a different bin), pick the smallest \pval, and interpret that as a Type-I error probability.  

There is a way to account for the trails factor, by defining a new hypothesis test that is sensitive to the union of the features that each of the $N$ tests is sensitive to, and has a \pval which can be interpreted as a Type-I error probability.  This new test will be combining $N$ hypothesis tests, and use as statistic the following:
\begin{equation}
\label{eq:superStatistic}
t = -\log(\min_i \{\pval_i\}).
\end{equation}
In words, this new hypothesis test uses as statistic the smallest $\pval_i$.   The negative $\log$ function is used to make $t$ increase monotonically as $\min\{\pval_i\}$ decreases, following the convention that wants $t$ to increase with increasing discrepancy.   Obviously the $\log$ function could be replaced by any other monotonically increasing function. 

We refer to the new test as a {\em hypertest}, i.e.\ a union of many tests, because its statistic is a \pval of some other hypothesis test from a pre-determined set of hypothesis tests.    

Every hypertest has an observed statistic $t_o$ and a corresponding \pval, found as described in paragraph \ref{sec:allTests}.     This \pval quantifies how often such a small (or smaller) \pval would be returned by at least one of the $N$ hypothesis tests included in the set, under \Ho.  The \pval of this hypertest, like any \pval, obeys the theorem of paragraph \ref{sec:theorem}.  The \pval of this hypertest can be interpreted as described in paragraph \ref{sec:singleTest}.


\subsubsection{Final remarks on the definition of hypertests}
\label{sec:theSet}

In paragraph \ref{sec:howToAccount} we gave a prescription to correctly consider simultaneously a set of hypothesis tests, by defining a hypertest that takes into account the trials factor, and returns a \pval that can be correctly interpreted as a Type-I error probability.  The obvious question is which hypothesis tests to include in the set used to define the hypertest.

There is no unique answer.
By including more (independent) hypothesis tests to the set, the hypertest gains sensitivity to more features.  That can be desirable, especially when we have no prior expectation of how $D$ may differ from \Ho.  The price one pays is that the effective trials factor ($\tilde{N}$) increases, so, the power of the test decreases, namely it would take more signal to obtain the same \pval from the hypertest.

If we knew somehow that $D$ would differ from \Ho in a specific bin, there would be no need to get distracted by looking in any other bin.  

A reasonable strategy, which is adopted also by the \bh, is to specify a set of hypothesis tests which cover a large family of similar features.  For example, the \bh, as we will see, is a hypertest based on the set of hypothesis tests that look for bumps of various widths in various locations of the spectrum.  The interpretation of such a test is rather simple.  If the \pval is not small enough, we conclude that there is no significant bump of any width, at any location.

One final remark is that a hypertest A may be included in the set of hypothesis tests used by a hypertest B.  That doesn't make B a hyper-hypertest or something.  Both B and A are hypertests, because their \pvals are the result of considering simultaneously the \pvals of a set of hypothesis tests (or hypertests).  
It is also trivial to show that if a hypertest A contains in its set just one hypothesis test (or hypertest) B, the \pval of A is identical to the \pval of B, so the distinction between simple hypothesis test and hypertest gets lost in the trivial case.

\section{The \bh}
\label{sec:bhAlgo}

The \bh scans the data ($D$) using a window of varying width, and keeps the window with biggest excess of data compared to the background (\Ho).  This test is designed to be sensitive to local excesses of data.  The same treatment is given to pseudo-data sampled from \Ho, and the \pval is estimated as described in paragraph \ref{sec:allTests}.  

In the language of paragraph \ref{sec:howToAccount} and \ref{sec:theSet},  the \bh is a hypertest that combines hypothesis tests which focus on bumps of various widths at various positions of the spectrum, taking the trials factor into account.

It will become clear that some choices have been made in this implementation of the \bh which could be different.  For example, one may use different sideband definitions, or may search for bumps within some width range.  As explained in paragraph \ref{sec:theSet}, such choices are essentially arbitrary.  They are made based on what we wish the interpretation of the result to be.  

This version of the \bh operates on data that are binned in some a-priori fixed set of bins.  In the limit of infinitesimally narrow bins, the arbitrariness of the binning choice is removed.  If the bins are not infinitesimally small, then their size limits the narrowest bump that one may be sensitive to.  In most applications there is a natural limit to how narrow a bump can be.  For example, in \cite{dijetResonanceSearch} the limit reflects the finite detector resolution.  Practically, one can have very good performance using bins of finite width.  In the case of the Banff Challenge, the information is given that the signal follows a Gaussian distribution with $\sigma = 0.03$, so, we define 40 equal bins between 0 and 1, resulting in bin size 0.025.

Given some data $D$ and some background hypothesis \Ho, the following steps are followed to obtain the test statistic ($t$) of the \bh:

\begin{enumerate}
\item \label{item:setWC} Set the width of the central\footnote{``Central window'' is the window where excess of data is checked for.  The word  ``central'' is used to distinguish that window from its left and right sideband.}  window $W_C$.   In this implementation, where the data are binned, $W_C$ is an integer which specifies how many consecutive bins to include in the central window.  This width is allowed to vary between some values.  In \cite{dijetResonanceSearch}, where the potential signal is of unknown width, $W_C$ is allowed to range from 1 to $\lfloor \frac{N}{2} \rfloor$, where $N$ is the total number of bins from the lowest observed mass to the highest.  To address the Banff Challenge \cite{BanffChallenge}, where the signal is a Gaussian of known $\sigma=0.03$, we constrain $W_C$ between 3 and 5 bins, which fit roughly 68\% to 95\% of this Gaussian signal.

\item \label{item:sb} Set the width of each sideband.  Sidebands are used, optionally, if one wishes to impose quality criteria ensuring that the \bh will focus on excesses surrounded by non-discrepant regions.  In \cite{dijetResonanceSearch} such sidebands were used, and their size (in number of bins) was set to $\max\{1,\lfloor \frac{W_C}{2}\rfloor\}$.  To address the Banff Challenge, we do not use any sidebands, in the interest of speed, and because there is some risk associated with using sidebands when $W_C$ is constrained to small values; this risk is illustrated in paragraph~\ref{sec:powers}.  In the following steps we will describe how sidebands are used, because they constitute part of the \bh algorithm, even though in the Banff Challenge they are not used.

\item Set the position of the central window, which will range from the lowest to the highest observed value\footnote{Dijet mass in the case of \cite{dijetResonanceSearch}, or $x$ in the case of the Banff Challenge.}.

\item Count the data ($d_C$) and background ($b_C$) in the central window.  Obviously $d_C$ is an integer and $b_C$ is a real number, representing the expectation value, according to \Ho, in the central window.  Similarly, count the data ($d_L$, $d_R$) and the background ($b_L$, $b_R$) in the left and right sideband (subscript ``$L$'' and ``$R$'' respectively).

\item \label{item:ps} In this step, which is at the heart of the \bh,  we will make a connection to what was said in paragraph \ref{sec:howToAccount}.   We will define the test statistic $t$ of each one of the hypothesis tests that are combined in the \bh hypertest.  Each local hypothesis test examines the presence of a bump at the location where we are currently placing the central window as we scan the spectrum.  Each such hypothesis test has its statistic $t$, which has an observed value $t_o$ coming from comparing the data $D$ to \Ho, resulting in a \pval.  The smallest of these \pvals will be used in step \ref{item:lastStep} to define the \bh test statistic, according to paragraph \ref{sec:howToAccount}.

  Given the six numbers $d_{\{L,C,R\}}$ and $b_{\{L,C,R\}}$, we define the following test statistic $t$ for the hypothesis test which focuses on the current window and sidebands:
  \begin{equation}
    \label{eq:correctT}
    t = 
    \begin{cases}
      0 & \text{if $d_C \le b_C$ or $\mathcal P(d_L,b_L) \le 10^{-3}$ or $\mathcal P(d_R,b_R) \le 10^{-3}$,} \\
      f(d_C - b_C) & \text{otherwise.} 
    \end{cases}
  \end{equation}
In this definition, $f$ can be any positive, monotonically increasing function, such as $(d_C-b_C)^2$ or $(d_C-b_C)^{100}$.   
Also,
\begin{equation}
  \label{eq:PoissonPval}
\mathcal P(d,b)= 
\begin{cases} \sum_{n=d}^\infty \frac{b^n}{n!}e^{-b} & \text{if $d \geq b$,}
\\ \\
\sum_{n=0}^{d} \frac{b^n}{n!}e^{-b} & \text{if $d < b$.}
\end{cases}
\end{equation}

Ignoring the sidebands is equivalent to using 0 instead of $10^{-3}$ in eq.~\ref{eq:correctT}.  The definition\footnote{As an aside, in footnote \ref{foot:theorem} it was mentioned that paragraph \ref{sec:bhAlgo} would illustrate an example of a test statistic which doesn't follow a continuous distribution $\rho_t$.  Indeed, the test statistic $t$ of eq.~\ref{eq:correctT} is discontinuous at 0.  Due to the condition which may set $t$ to 0 in some cases, the PDF of $t$ contains a peak at 0 which could be formulated as a Kronecker $\delta(t-0)$ multiplied by the probability for $t$ to be 0.} of eq.~\ref{eq:correctT} was carefully designed to have the following characteristics, which make it meaningful and practical:
\begin{itemize}
\item $t \ge 0$.
\item $t=0$, i.e.\ the discrepancy is characterized maximally uninteresting when the data, where the particular hypothesis test focuses and $t$ is computed, do {\em not} meet the following criteria which a bump would be expected to meet:  (a) Have an excess of data in the central window, namely $d_C > b_C$.  And (b), have both sidebands consistent with the background.  That is where the two $\mathcal P(d_{X},b_{X})$ with $X=\{L,R\}$ are employed.  Each one of these is the \pval of a hypothesis test that focusses on just the left or right sideband, and uses as test statistic $t = |d_X-b_X|$, or something similar that increases monotonically with the difference between data and background in each sideband.  By requiring $\mathcal P(d_X,b_X)$ to be greater than $10^{-3}$, we require that \Ho can not be excluded, based on event counts in the sideband, with less than $10^{-3}$ probability of being wrong. The value $10^{-3}$ is arbitrary, and can be set higher or lower to tighten or relax, respectively, the good sidebands requirement.
\item The \pval of this hypothesis test is analytically calculable directly from $d_{\{L,C,R\}}$ and $b_{\{L,C,R\}}$, without even having to calculate $t$ or $t_o$!  We will soon explain how.  This remarkable property allows the \bh statistic to be computed quickly, without needing pseudo-experiments to estimate the \pval of each local hypothesis test that it incorporates.
\end{itemize}

The \pval is computed as follows.  We have the observed events $d_{Co}$, $d_{Lo}$ and $d_{Ro}$.  If $d_{Co} \le b_C$, we don't have an excess, so we know that the observed statistic $t_o$ is 0 according to eq.~\ref{eq:correctT}, therefore any other pseudo-experiment would have $t\ge t_o$, therefore $\pval = 1$.  The same is true, for the same reason, if $\mathcal P(d_{Lo},b_L) \le 10^{-3}$ or $\mathcal P(d_{Ro},b_R) \le 10^{-3}$.  When none of the above happens, $t$ is defined to increase as $d_C$ increases, since $f(d_C-b_C)$, in eq.~\ref{eq:correctT}, is monotonically increasing and $b_C$ is fixed.  So, when $t_o > 0$, we know that the only way $t$ would be $\ge t_o$ is by having $d_C \ge d_{Co}$, while $d_{L}$ and $d_R$ remain consistent with $b_L$ and $b_R$.  To find the \pval, which by definition is $P(t \ge t_o|\Ho)$, we have to compute the probability of these three things happen simultaneously.  The conditions on $d_L$ and $d_R$ were designed to be independent from each other and from $d_C$.  This allows us to express the \pval as the product of 3 probabilities:  $P(d_C \ge d_{Co}|\Ho)$,  $P(\mathcal P(d_L,b_L)>10^{-3}|\Ho)$, and $P(\mathcal P(d_R,b_R)>10^{-3}|\Ho)$.  The first probability is, by definition, $\mathcal P(d_{Co},b_C)$.  The second and third probabilities are equal\footnote{This equality is only approximate, due to $d_X$ being integer.  It is, however, a very good approximation.  Due to $d_X$ taking discrete values, so does $\mathcal P(d_X,b_X)$. For example, if $b_X = 1.5$, then to have $\mathcal P(d_X,1.5) > 10^{-3}$, $d_X$ has to be $< 7$, and that has probability $\sum_{n=0}^{6} \frac{1.5^n}{n!}e^{-1.5} = 0.99074$ instead of 0.999.  If $b_X=0.001$, then the same probability is  $\sum_{n=0}^{0} \frac{0.001^n}{n!}e^{-0.001} = 0.9990005$, and for large values of $b_X$ the approximation becomes better because the discreteness of $d_X$ becomes negligible.} to $(1-10^{-3})$, because of the theorem of paragraph \ref{sec:theorem}, and because $\mathcal P(d_L,b_L)$ and $\mathcal P(d_R,b_R)$ are \pvals.  Putting it all together, we have:
\begin{eqnarray}
\label{eq:correctPval}
 \pval &=& 
\begin{cases}
  1 \ \ \  \text{if $d_{Co} \le b_C$ or $\mathcal P(d_{Lo},b_L) < 10^{-3}$ or $\mathcal P(d_{Ro},b_R) < 10^{-3}$} \\
  \mathcal P(d_{Co},b_C)
  (1-10^{-3})^2 \ \ \  \text{otherwise.}
\end{cases} 
\end{eqnarray}

The term $(1-10^{-3})$ is very close to $1$, but even if it wasn't, it could be ignored because it is constant of all local hypothesis tests, therefore it affects neither which \pval will be the smallest (see step \ref{item:lastStep}), nor the \bh \pval.   

After all, we have shown that the \pval of eq.~\ref{eq:correctPval} depends on three $\mathcal P(d,b)$ values, which are analytically calculable quantities, using the well-known function $\Gamma(d) = \int_0^\infty t^{d-1} e^{-t} dt$ and its normalized lower incomplete version, which is also tabulated in standard computational packages code libraries, like the {\tt ROOT TMath} class \cite{Brun:1997pa}.  The useful relationship that allows this computation is:
\begin{equation}
\sum_{n=d}^{\infty} \frac{b^n}{d!}e^{-b} = \frac{1}{\Gamma(d)}\int_0^b t^{d-1}e^{-t} dt = \Gamma(d,b),
\end{equation}
from which it follows that:
\begin{equation}
  \label{eq:PoissonPval2}
\mathcal P(d,b)= 
\begin{cases} \Gamma(d,b) & \text{if $d \geq b$,}
\\ \\
1-\Gamma(d+1,b) & \text{if $d < b$.}
\end{cases}
\end{equation}

\item Shift the central window, and its sidebands, by a number of bins, and repeat step \ref{item:ps}, namely compute the \pval of the local hypothesis test that focuses on that new location.  In principle, the bins could be infinitesimally narrow, and the translation could be in infinitesimally small steps, to include in the \bh every possible bump candidate (or, equivalently, every possible hypothesis test focusing on a local mass range).  However, in practice there are computational limitations.  Hypothesis tests which focus on roughly the same mass range are highly correlated.  By adding more highly correlated tests not much new information is gained, the effective trials factor $\tilde{N}$ doesn't increase much (see eq.~\ref{eq:effTrialsFactor}), but it takes time to compute the \pvals all these tests.  For this reason, in the implementation of the \bh used in \cite{dijetResonanceSearch} and in the Banff Challenge we use $$\text{step size} = \max\{1,\lfloor \frac{W_C}{2} \rfloor\}.$$  In this way we still consider bump candidates which overlap significantly, but we avoid spending time to consider almost identical bump candidates.
  
\item Repeat the above steps for all desired values of $W_C$, as they were described in step \ref{item:setWC}.  For every choice of $W_C$ and every location of the central window, compute the corresponding \pval as in eq.~\ref{eq:correctPval}.
  
\item \label{item:lastStep} In this last step, the \bh test statistic $t$ is calculated, according to eq.~\ref{eq:superStatistic}:
\begin{equation}
t = -\log \pval^{\min},
\end{equation}
where $\pval^{\min}$ is the smallest of all \pvals found in the previous steps.
\end{enumerate}

\subsection{The background and pseudo-data}

Like in all hypothesis tests (e.g. $\chi^2$, KS etc.), in the \bh the \Ho is an input.  The \bh uses the \Ho, its \pval depends on it, but it doesn't {\em define} \Ho.  Depending on how the analyst defines \Ho, the interpretation of the \bh, or any other hypothesis test, will have different interpretations.

In particle physics, \Ho may come from Monte Carlo (MC) simulation, representing typically the Standard Model prediction.  Then, everything we have discussed so far applies.  The MC-based background is used 
\begin{enumerate}
\item to compare $D$ to \Ho, thus obtaining the observed \bh statistic $t_o$,
\item to generate pseudo-data according to \Ho multiple times, 
\item to obtain the \bh statistic $t$ by comparing each pseudo-data spectrum to \Ho.
\end{enumerate}
Then the \bh \pval is estimated, according to paragraph~\ref{sec:allTests}.

In some cases, it is well-motivated to formulate \Ho as a function of $D$, instead of using MC.  Specifically, in \cite{dijetResonanceSearch} and in the Banff Challenge, the background is not independent of $D$.  It is obtained by fitting a function to $D$.  In the case of Banff Challenge we have the information that the background should follow an exponential spectrum
\begin{equation}
\label{eq:BanffBkg}
B(x) = A e^{-Cx}.
\end{equation}
In the case of \cite{dijetResonanceSearch}, studies showed that there is a more complicated functional form which can fit the Standard Model prediction, but couldn't fit a spectrum with a resonance.  One can define as \Ho the result of fitting this functional form to the data $D$.  This definition of the null hypothesis may be called ``smooth background hypothesis''.

When \Ho depends on $D$, it is necessary to compute \Ho (i.e.\ by re-fitting) not only for the actual data $D$, but also for every pseudo-experiment that will be used to estimate the \pval.  Otherwise \Ho is not consistently defined, which means that in theorem~\ref{sec:theorem} $\rho_t$ and $\rho_{t_o}$ are not identical, thus the \pval is not interpretable as a Type-I error probability.

\subsubsection{Fitting by omitting anomalies}
\label{sec:omitAnomaly}

\begin{figure}[p]
\centering
\includegraphics[width=0.6\textwidth]{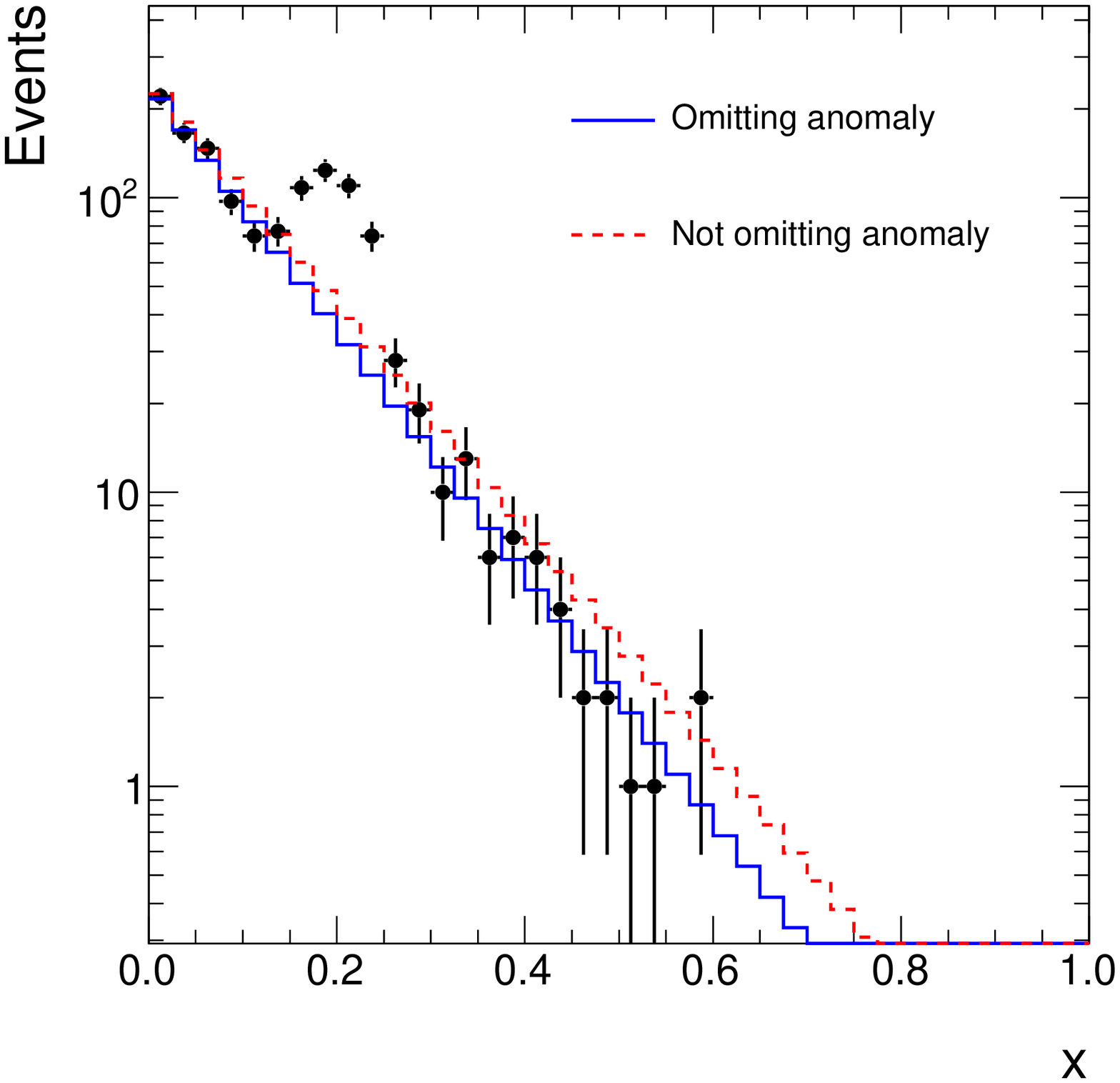}
\caption{\label{fig:fitThroughAnomaly} Fitting an exponential spectrum with a Gaussian signal, like in \cite{BanffChallenge}.  In one case the whole spectrum is fitted, and in the other the algorithm described in paragraph~\ref{sec:omitAnomaly} locates the anomalous region and fits the rest of the spectrum.}
\end{figure}

When \Ho is computed by fitting $D$ there is the concern that, if a bump actually exists, it will influence the fit.  Naturally, the fitted background will try to accommodate part of the signal, even if it doesn't have the flexibility to fully do so.  That can obscure the signal, and cause the fit to not describe the data even where they don't contain signal.  Fig.~\ref{fig:fitThroughAnomaly} shows such an example.

An alternative is to define \Ho as the spectrum obtained by fitting the data, after omitting the window which improves the fit in a pre-determined, algorithmic way.  The algorithm used in the Banff Challenge is to try the fit after omitting various windows, similar to the way the \bh scans the spectrum (paragraph~\ref{sec:bhAlgo}).  The windows that are omitted have size between 3 and 5 bins, corresponding to width of potential signal, and they are considered for exclusion only if they contain an excess of data.  If after the omission of some window the $\chi^2$ test \pval becomes greater than 0.1, then we consider the fit good enough and we stop looking for other windows to possibly omit from the fit.  If the fit is not made better than that after the omission of any window, then we keep the fit which gave the greatest $\chi^2$ \pval, even if it was less than 0.1.  An example of this algorithm in action is shown in Fig.~\ref{fig:fitThroughAnomaly}, where the window with the bump is automatically excluded, resulting in a much better fit of the rest of the spectrum.
The same algorithm, obviously, is used each time we fit pseudo-data.

The advantage of omitting the most discrepant region is that it pronounces the bump, as one sees in Fig.~\ref{fig:fitThroughAnomaly}.  Also, if the goal of the fit is to estimate the background parameters, e.g. the value of $A$ in eq.~\ref{eq:BanffBkg}, then this allows for the fit to find the right value of $A$ without bias caused by the signal.\footnote{However, in the specific case of the Banff Challenge this is not how we estimate $A$, because we have the information that the signal follows a Gaussian of known width, so, it is better to fit the background of eq.~\ref{eq:BanffBkg} simultaneously with a Gaussian.  The primary goal of the \bh is not to estimate parameters, but to test \Ho.}

Besides these advantages, nothing would be wrong about the results of the \bh even if one didn't follow this fit procedure.  If we define \Ho as the result of fitting the whole spectrum, then the \bh (and any other test) returns the right \pval that reflects this definition.  If the \pval indicates a significant discrepancy between $D$ and \Ho, it is clear what \Ho means and what the interpretation is.  In other words, the \bh (like any test) operates with the input $D$ and \Ho, not caring how well-motivated \Ho is; that is up to the analyst.




\section{The Banff Challenge, problem 1}

The Banff Challenge \cite{BanffChallenge}, Problem 1, offers an opportunity to demonstrate \bh's performance.

\Ho is defined as the spectrum obtained by fitting the data with eq.~\ref{eq:BanffBkg}, following the algorithm of paragraph \ref{sec:omitAnomaly}.
The \bh \pval is estimated using the procedure of sec.~\ref{sec:allTests}, generating pseudo-experiments until we are sure (in the bayesian sense described in \ref{sec:allTests}) that the \pval is smaller or greater than 0.01 with probability $\ge 0.999$.  If the \pval is estimated to be  $< 0.01$ (with probability $\ge$ 0.999), we declare discovery; if the \pval is estimated to be $\ge 0.01$ (with probability $\ge$ 0.999), then we don't.

Then comes the challenge of estimating the parameter $A$ of the background and the position $E$ of the signal (if discovery was declared).  We go one step further, and estimate also the amount of signal ($D$).  We do all that by fitting to the data the function
\begin{equation}
\label{eq:bkgAndSig}
f(x) = A e^{-Cx} + D \frac{1}{\sqrt{2 \pi}\ 0.03} e^{-\frac{(x-E)^2}{2\cdot 0.03^2}}.
\end{equation}
This fit has free parameters $\{A,C,D,E\}$. We use the result of the \bh to aid it;  the initial value of $E$ is set to the position where the \bh located the most significant bump.

All data are studied after binning them in 40 equal bins of $x$ between 0 and 1.  (Bin size = 0.025.)  If the actual $A$ is $10^4$ the fit will return roughly $10^4$/40=250.\footnote{This is the result of not using the option {\tt `I'} when fitting in ROOT \cite{Brun:1997pa}.}

We executed the \bh and the subsequent 4-parameter fit to all 20000 distributions handed out with the Challenge.  
The results are tabulated in a separate, long text file, with the columns:
\begin{itemize}
\item Dataset number (from 0 to 19999)
\item Decision : 0 means ``most likely estimated \pval $> 0.01$, thus no discovery claim.''  1 means ``most likely estimated \pval $\le 0.01$, thus discovery is claimed.''
\item \pval estimate.  For example, the string 
\begin{center}
{\tt 0.0666667 = 6/90	P(pval>0.01)= 0.999961}
\end{center}
condenses the following information: 90 pseudo-experiments were generated.  6 of them had a \bh statistic greater than the \bh statistic observed in the actual data.  That means that the most likely value for the \pval is 6/90 = 0.067.  According to the bayesian posterior described in paragraph~\ref{sec:allTests}, the \pval is greater than 0.01 with probability 0.999961\footnote{The actual accuracy of this probability does not extend beyond the third or fourth significant digit.}.  So, it is safely above 0.01, and in this case we don't declare discovery.  Let's see another example:
\begin{center}
{\tt 0 = 0/690	P(pval<0.01)= 0.99904}.
\end{center}
This string means that 690 pseudo-experiments were generated, none of them was more discrepant than the actual data, which means that the most likely \pval is 0, and the bayesian posterior ensures that the \pval is less than 0.01 with probability 0.99904.  In this case we claim discovery.

\item The next three numbers: the most likely signal position $E$, its lower 68\% CI limit and its higher 68\% CI limit.  This interval is obtained from MINUIT, by fitting eq.~\ref{eq:bkgAndSig}, and taking the error of the parameter with {\tt TF1::GetParError} \cite{Brun:1997pa}.

\item The next three numbers: same as the previous three numbers, but for parameter $D$, after fitting eq.~\ref{eq:bkgAndSig}.

\item The last three numbers: same as the previous three numbers, but for parameter $A$, after fitting eq.~\ref{eq:bkgAndSig}.
\end{itemize}

Appendix \ref{sec:log} includes the first 100 lines of the aforementioned text file.

\subsection{A discovery example}

\begin{figure}[p]
\centering
\subfigure[]{
\includegraphics[width=0.48\textwidth]{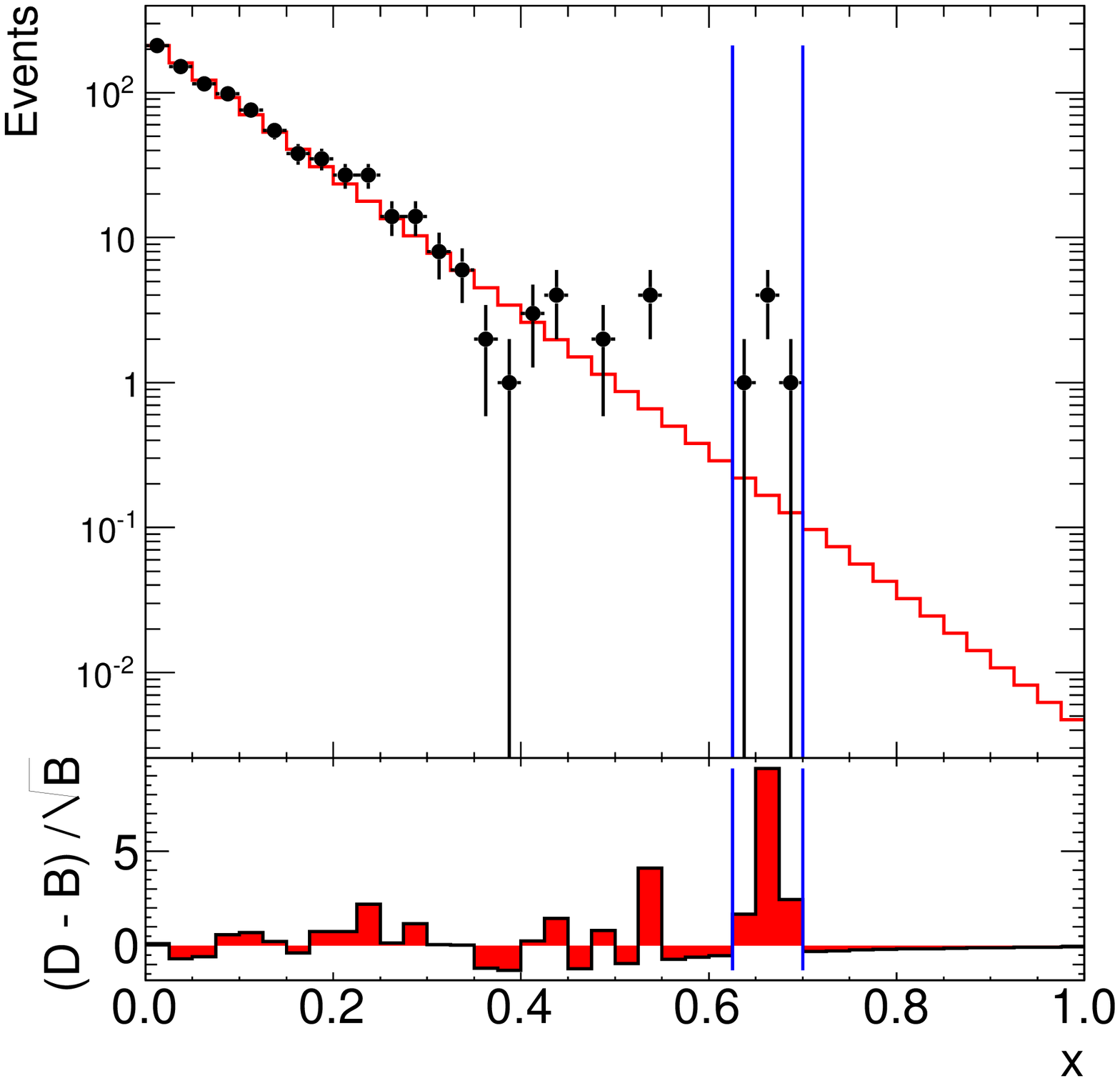}
\label{fig:figures/dataset_10/dataAndFitAndDifference.eps}
}
\subfigure[]{
\includegraphics[width=0.48\textwidth]{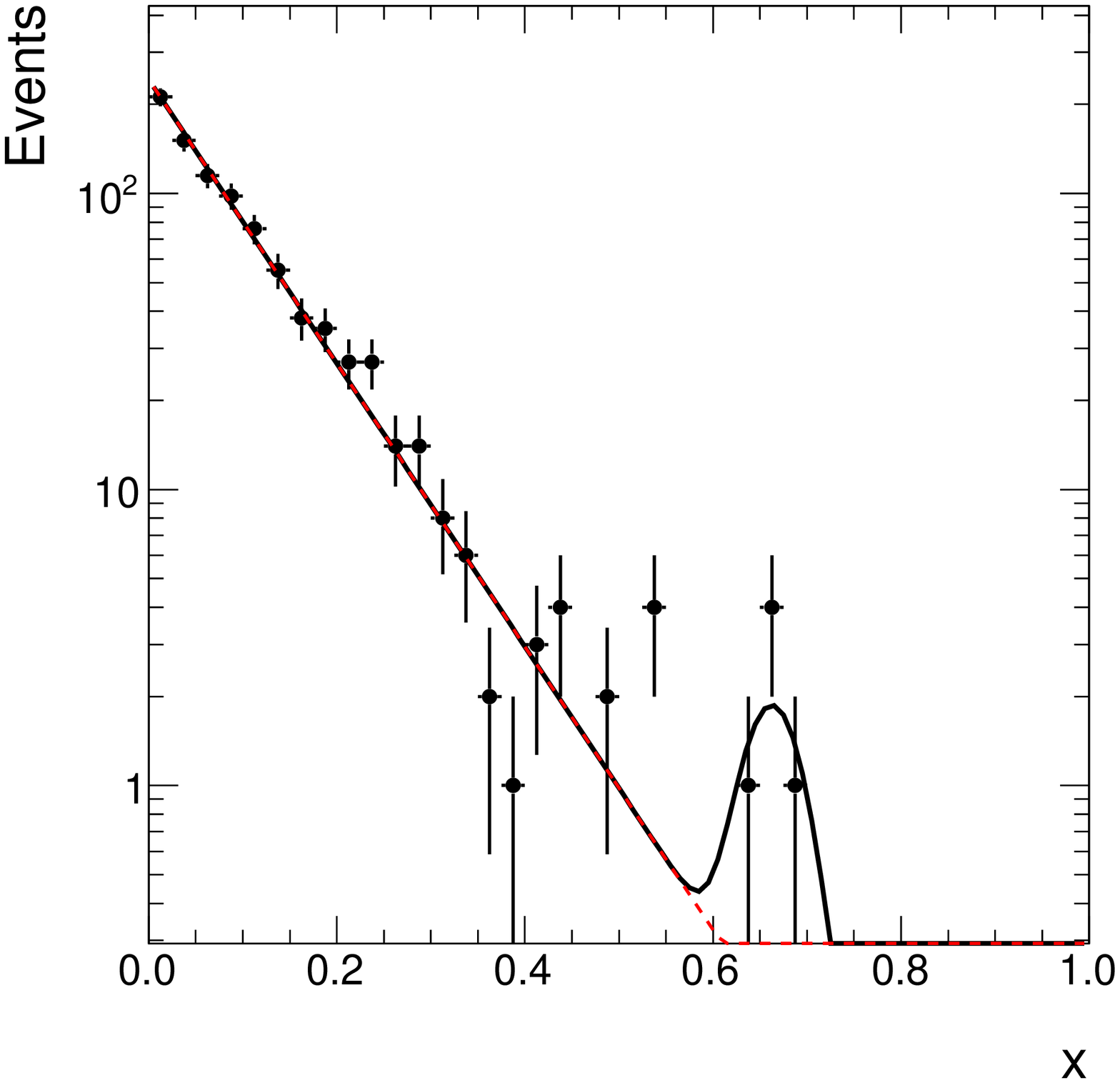}
\label{fig:figures/dataset_10/fittedWithGaussian.eps}
}
\\
\subfigure[]{
\includegraphics[width=0.3\textwidth]{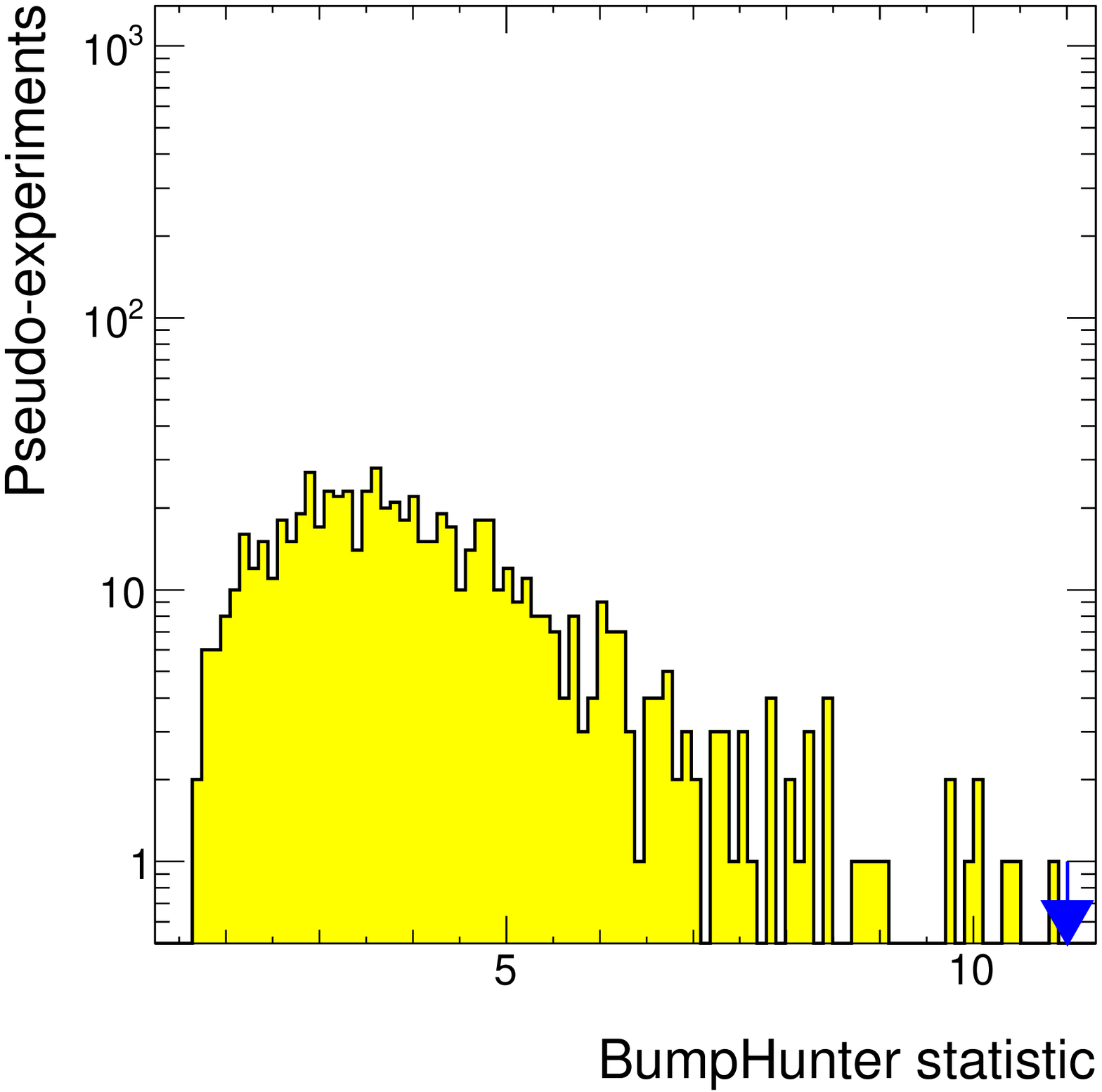}
\label{fig:figures/dataset_10/nullStatistic.eps}
}
\subfigure[]{
\includegraphics[width=0.3\textwidth]{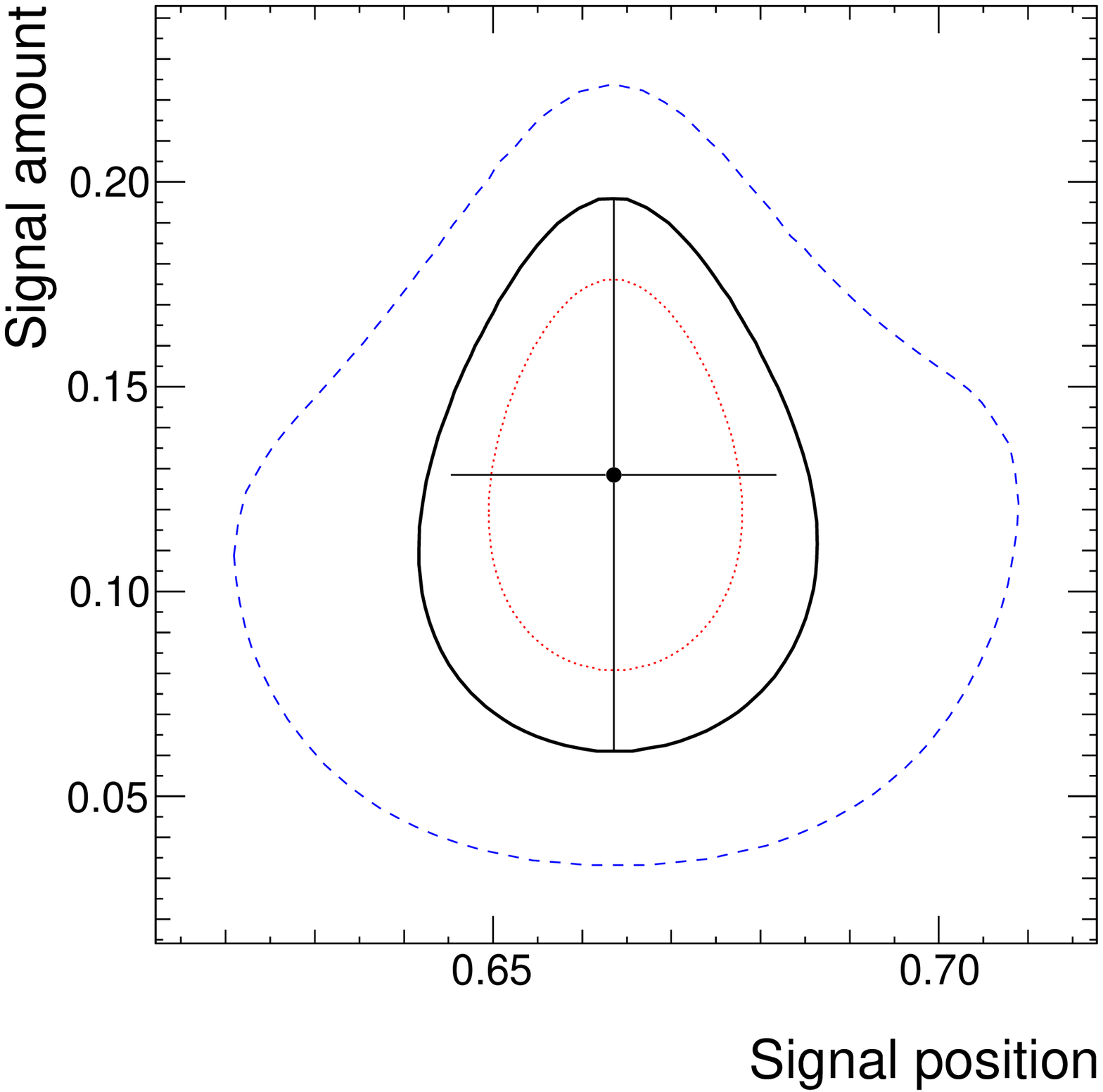}
\label{fig:figures/dataset_10/contourDE.eps}
}
\subfigure[]{
\includegraphics[width=0.3\textwidth]{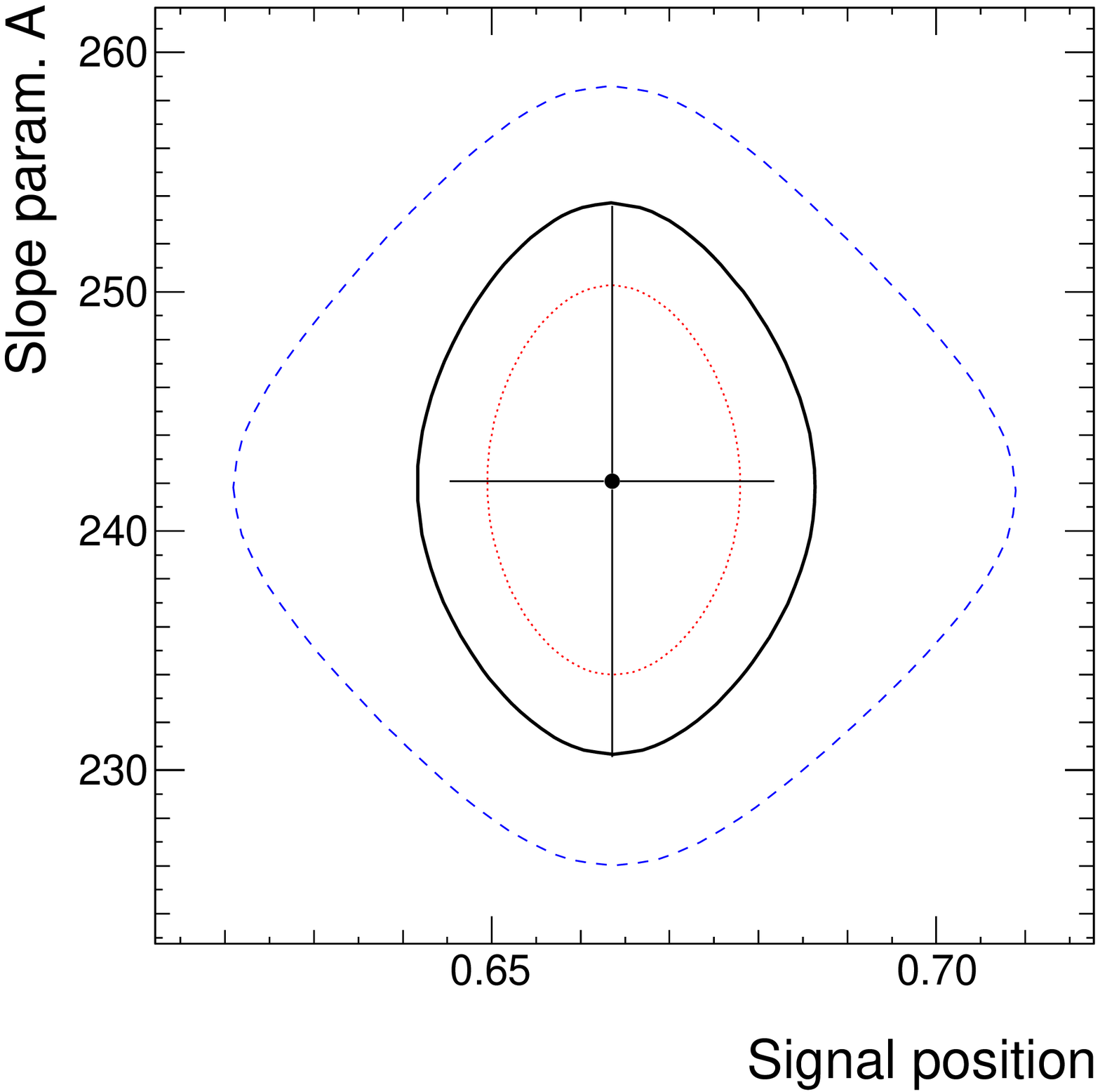}
\label{fig:figures/dataset_10/contourAE.eps}
} \\
\caption{\label{fig:exampleDiscovery} \subref{fig:figures/dataset_10/dataAndFitAndDifference.eps}:  The data of dataset 10, with the result of fitting eq.~\ref{eq:BanffBkg} as described in paragraph~\ref{sec:omitAnomaly}.  The bottom of the figure compares the data ($D$) to the background ($B$) in each bin, using the $\frac{D-B}{\sqrt{B}}$ approximation of significance.  The blue vertical lines show the most discrepant bump found, namely the central window of the local hypothesis test which yielded the smallest \pval.
 \subref{fig:figures/dataset_10/fittedWithGaussian.eps}:  The fit of eq.~\ref{eq:bkgAndSig} to the data.
 \subref{fig:figures/dataset_10/nullStatistic.eps}: The distribution of the \bh statistic in 690 pseudo-experiments ($t$) generated to follow the distribution obtained by the fit in \subref{fig:figures/dataset_10/dataAndFitAndDifference.eps}.  The observed \bh statistic ($t_o$) is marked by the blue arrow.
 \subref{fig:figures/dataset_10/contourDE.eps}: The 2-dimensional 0.5$\sigma$ (red), 1$\sigma$ (black), and 2$\sigma$ (blue) confidence contour for the signal position and amount.  The black marker and the error bars correspond to the most likely values and the uncertainty returned by {\tt TF1::GetParError}.
\subref{fig:figures/dataset_10/contourAE.eps}: Same as \subref{fig:figures/dataset_10/contourDE.eps}, but showing the signal position and slope parameter $A$.
}
\end{figure}

As an example where we claim discovery, we present dataset 10, the first dataset where discovery is claimed.  Fig.~\ref{fig:exampleDiscovery} summarizes the information extracted from this dataset.  

For this dataset, we estimate the most likely \pval to be $\frac{0}{690}=0$.  With the 690 pseudo-experiments generated, and assuming a flat prior in $[0,1]$, we infer that the \pval is less than 0.01 with probability about 0.99904.

The signal mean is estimated at $E = 0.664 \pm 0.018$.  Similarly, $D = 0.13 \pm 0.07$, and $A = 242 \pm 12$.  It should be reminded that each bin was width 0.025, and $242/0.025 = 9680$, which is comparable to what is known about $A$, i.e.\ that it is a random variable around $10^4$.  Similarly, $0.13/0.025  = 5.2$, which is comparable to the number of events one can identify as signal in Fig.~\ref{fig:exampleDiscovery}\subref{fig:figures/dataset_10/fittedWithGaussian.eps}.

\subsection{A non-discovery example}

\begin{figure}[p]
\centering
\subfigure[]{
\includegraphics[width=0.48\textwidth]{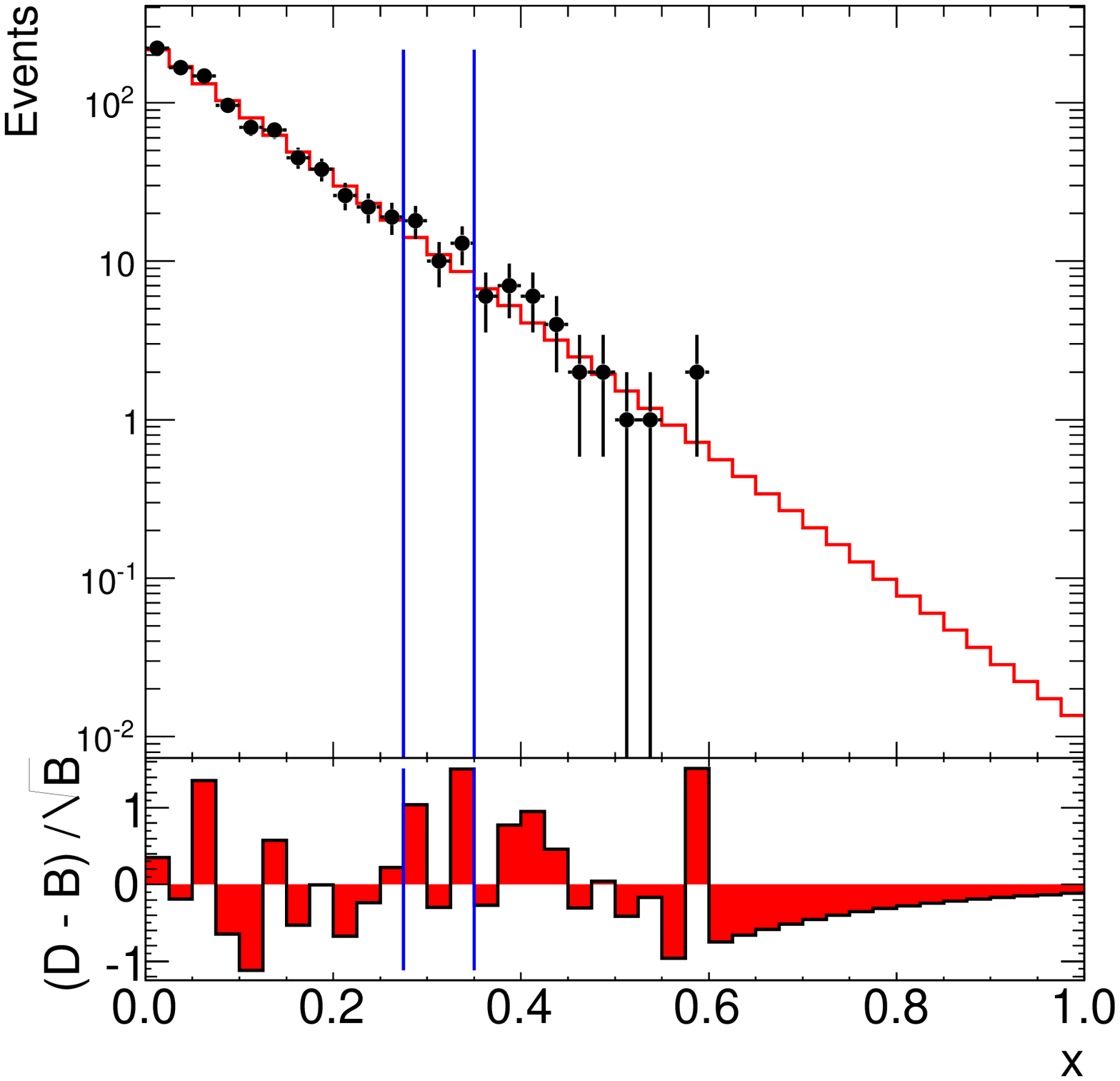}
\label{fig:figures/dataset_0/dataAndFitAndDifference.eps}
}
\subfigure[]{
\includegraphics[width=0.48\textwidth]{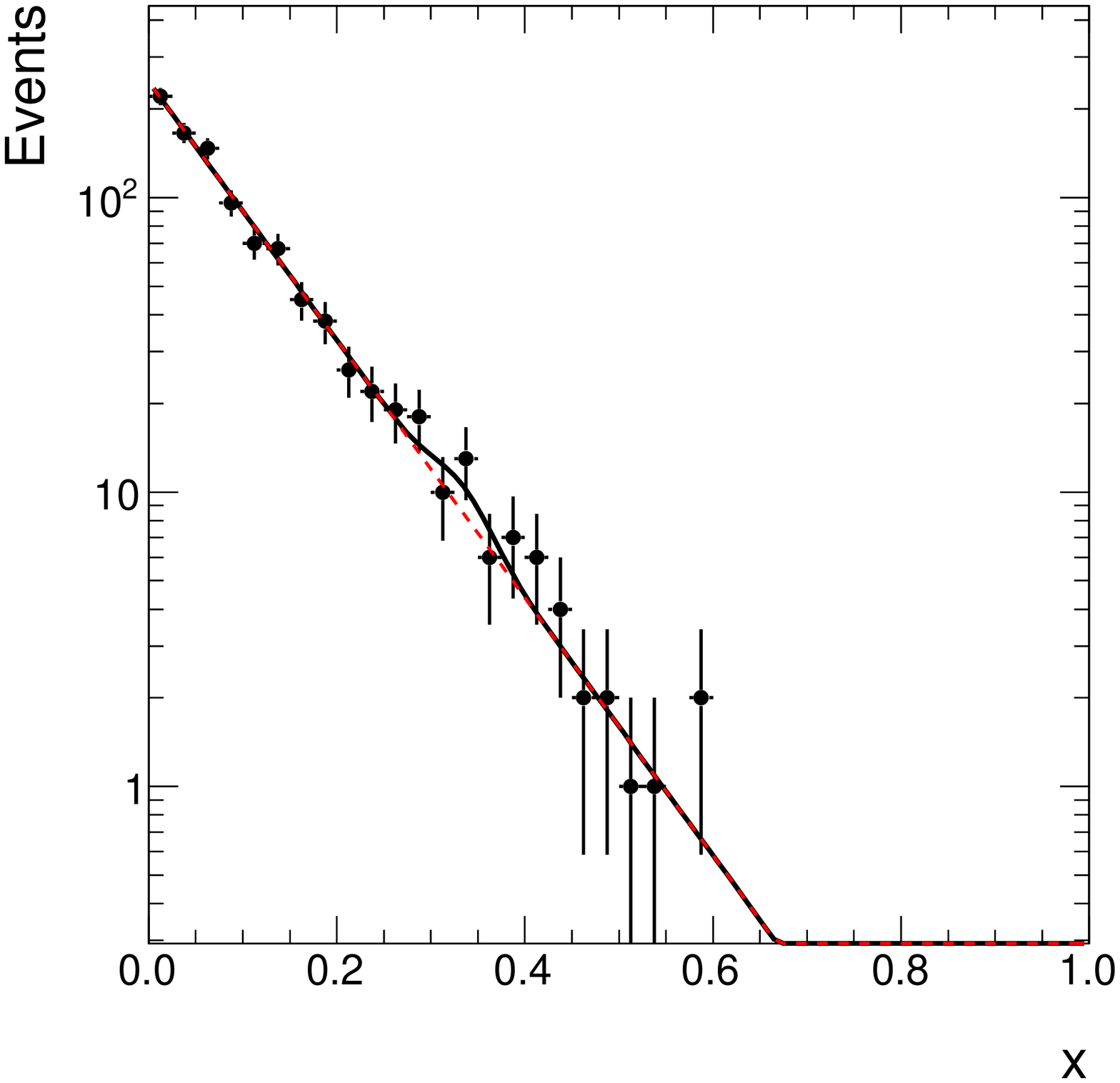}
\label{fig:figures/dataset_0/fittedWithGaussian.eps}
}
\\
\subfigure[]{
\includegraphics[width=0.3\textwidth]{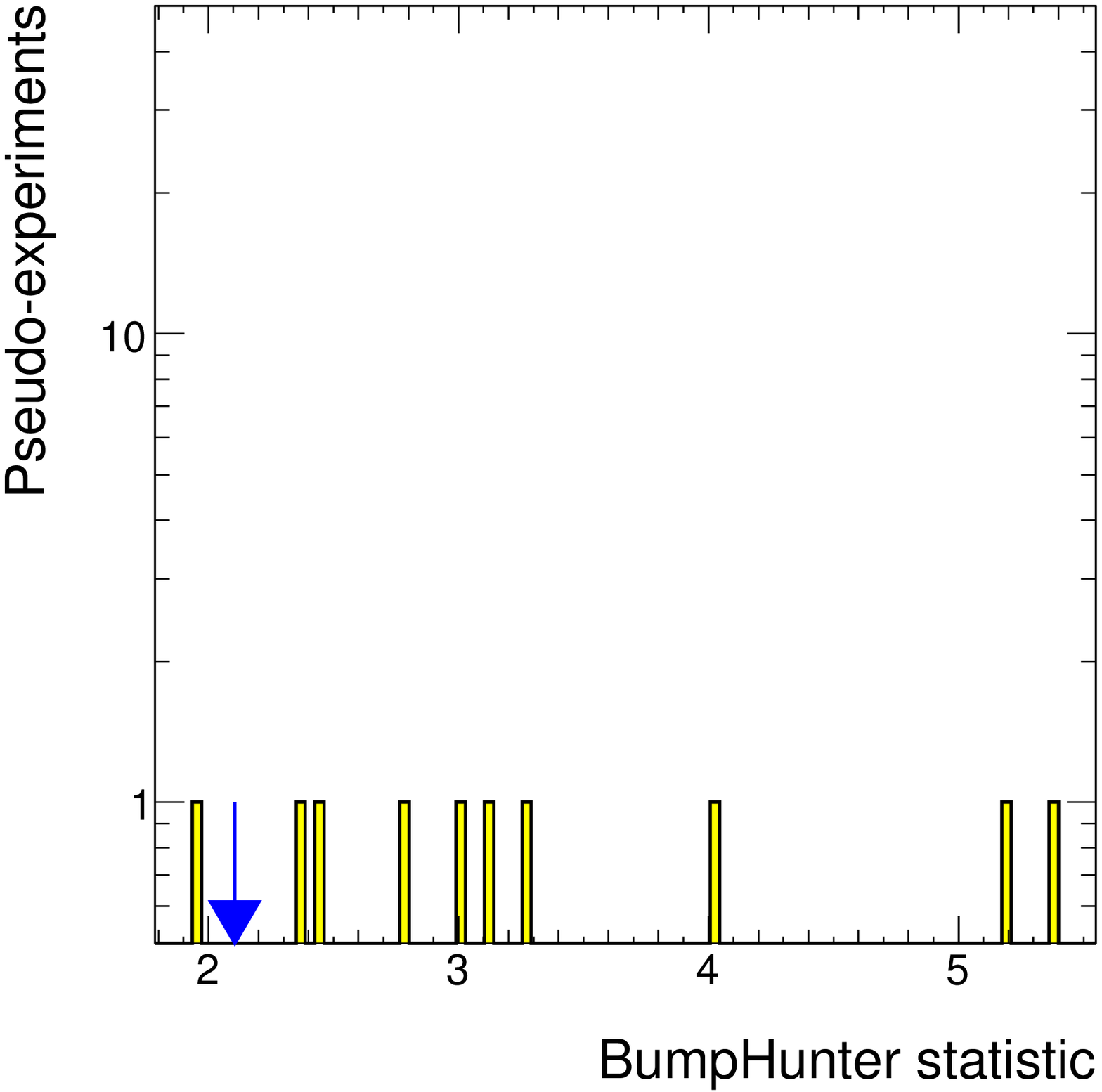}
\label{fig:figures/dataset_0/nullStatistic.eps}
}
\subfigure[]{
\includegraphics[width=0.3\textwidth]{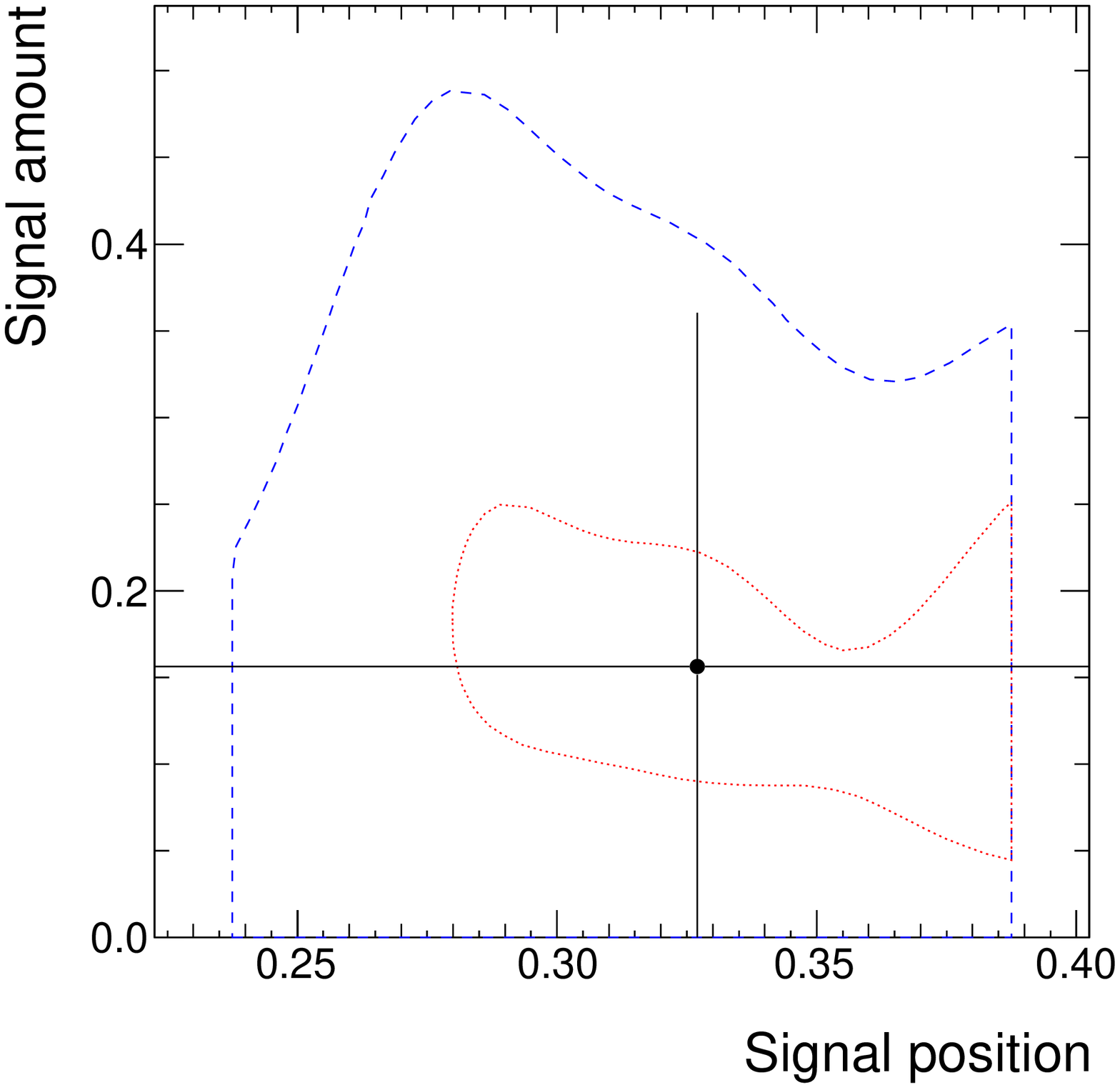}
\label{fig:figures/dataset_0/contourDE.eps}
}
\subfigure[]{
\includegraphics[width=0.3\textwidth]{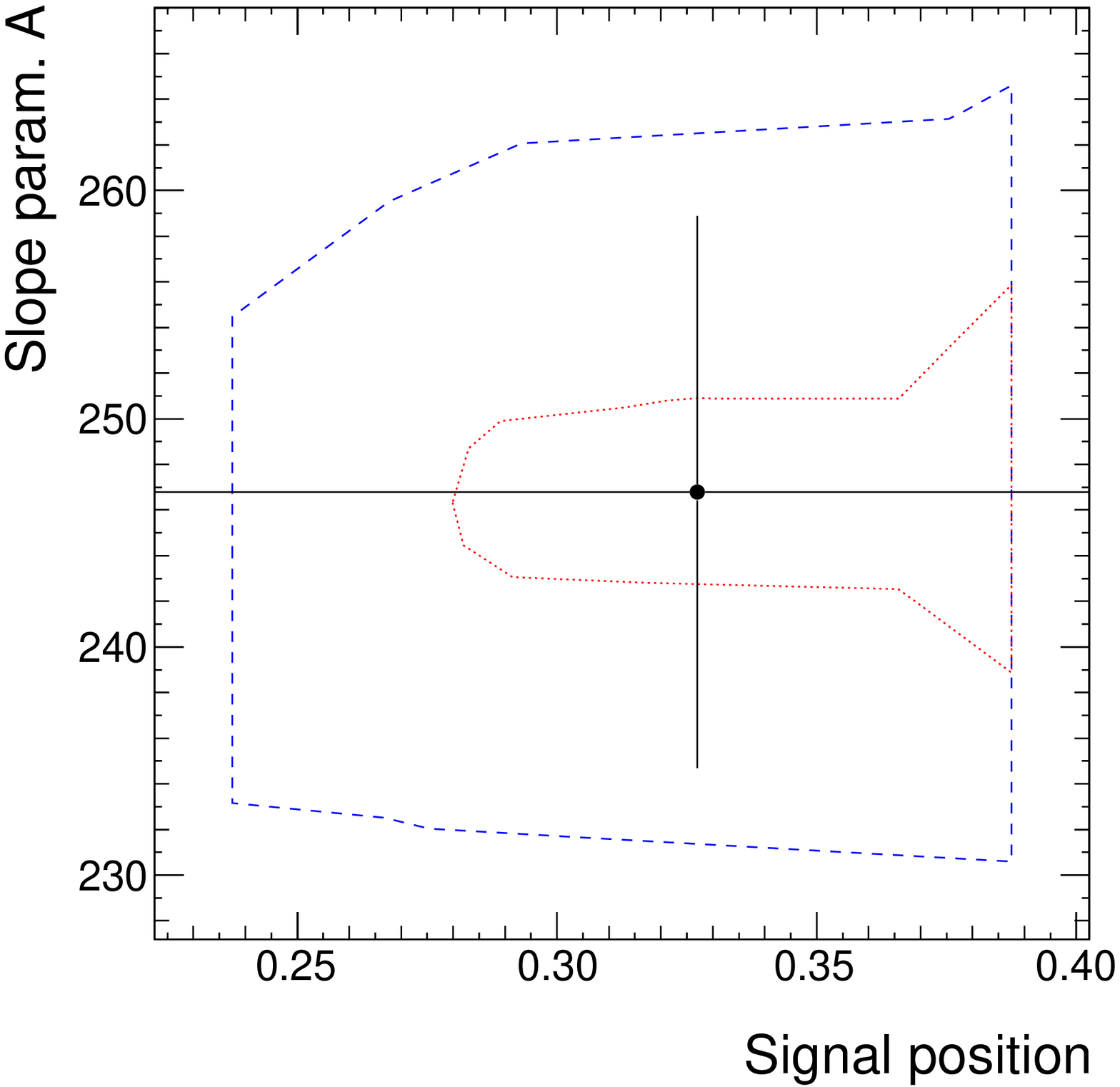}
\label{fig:figures/dataset_0/contourAE.eps}
} \\
\caption{\label{fig:exampleNonDiscovery} Same as Fig.~\ref{fig:exampleDiscovery}, but for dataset 0, where most likely there is no signal.  An obvious difference is that only 10 pseudo-experiments as generated, 9 of which have a bigger \bh statistic than observed, as shown in \subref{fig:figures/dataset_0/nullStatistic.eps}.
}
\end{figure}

As an example where we do not claim discovery, we present dataset 0.  Fig.~\ref{fig:exampleNonDiscovery} summarizes the information extracted from this dataset.  

For this dataset, we estimate the most likely \pval to be $\frac{9}{10}$.  Of course, this number is not so useful, because it reflects only 10 pseudo-experiments.  The useful inference from those 10 pseudo-experiments, though, is that the \pval is greater than 0.01 with probability indistinguishably close to 100\%.  

\subsection{Summary of datasets}
\label{sec:summaryOfDatasets}
Of the 20000 datasets, we found 1819 where the most likely \pval was estimated to be $\le 0.01$.
Of the 20000 datasets, there are 107 datasets where it was decided to stop producing pseudo-experiments, because we ran out of time.  For those 107 datasets, 57 have estimated $\pval \le 0.01$, and 50 have $\pval > 0.01$.  The reason it took too long to conclude was that the \pval is very close to 0.01, so many trials are required to discern, with 0.999 credibility, on which side of 0.01 the \pval is.   However, of those 107 datasets where 0.999 credibility was not attained, 64 concluded with credibility less than 0.99, 38 concluded with credibility less than 0.9, and just 2 with credibility less than 0.5.  Indicatively, these 2 datasets estimated the most likely \pval to be $\frac{1000}{100010}\simeq0.009999$.

\begin{figure}[p]
\hspace{-1cm}
\begin{tabular}{cccc}
\includegraphics[width=0.25\textwidth]{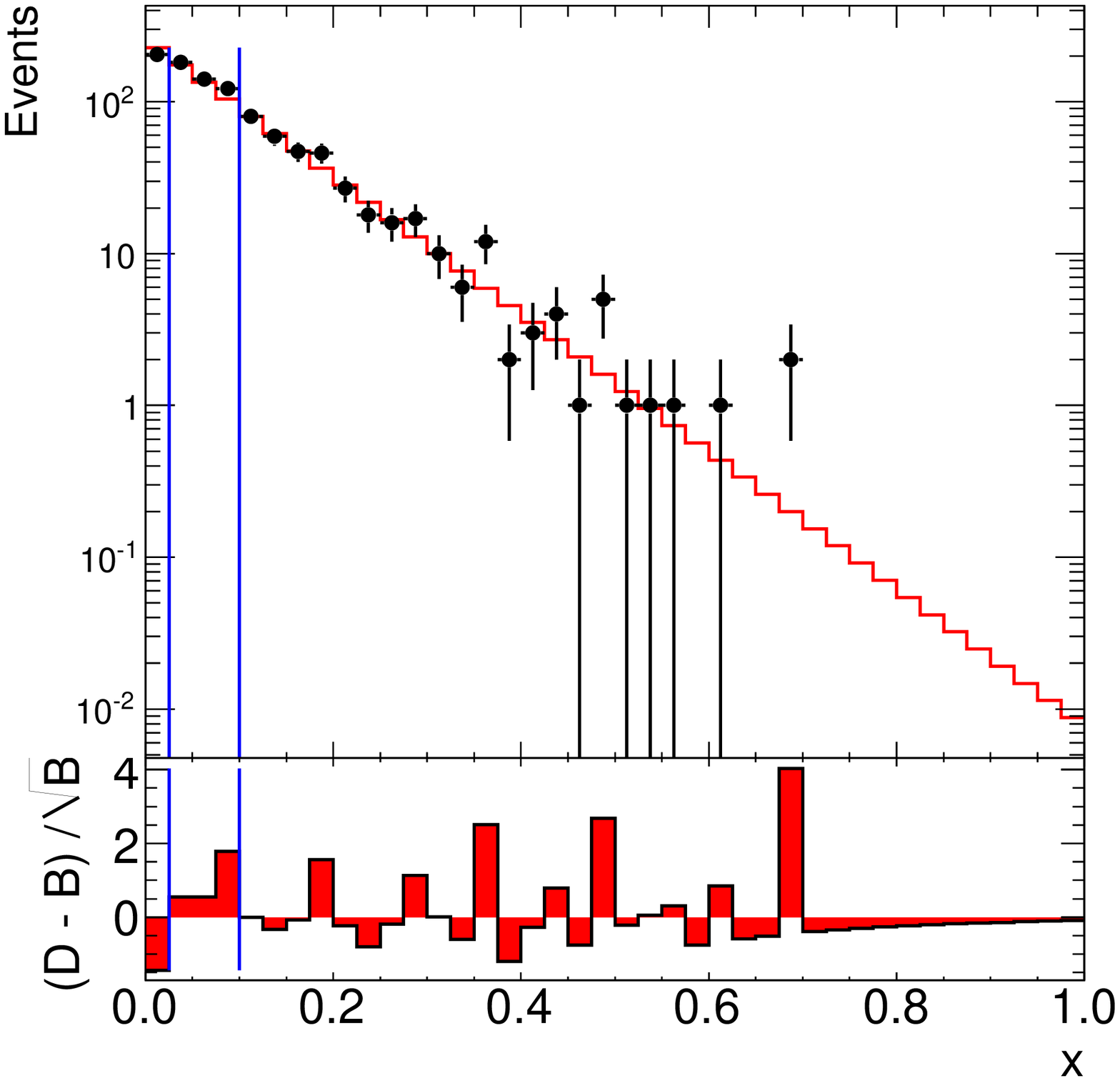} &
\includegraphics[width=0.25\textwidth]{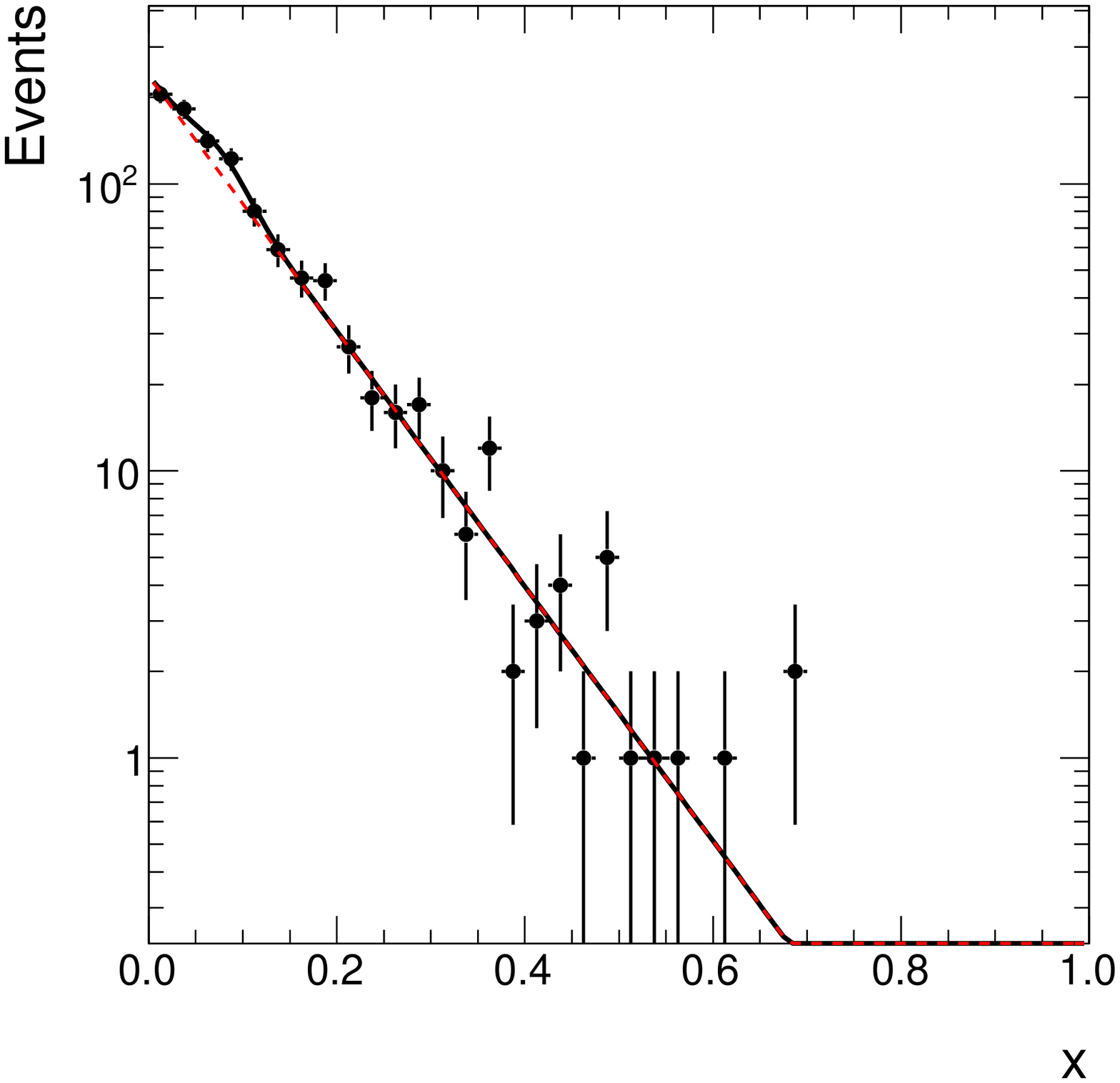} &
\includegraphics[width=0.25\textwidth]{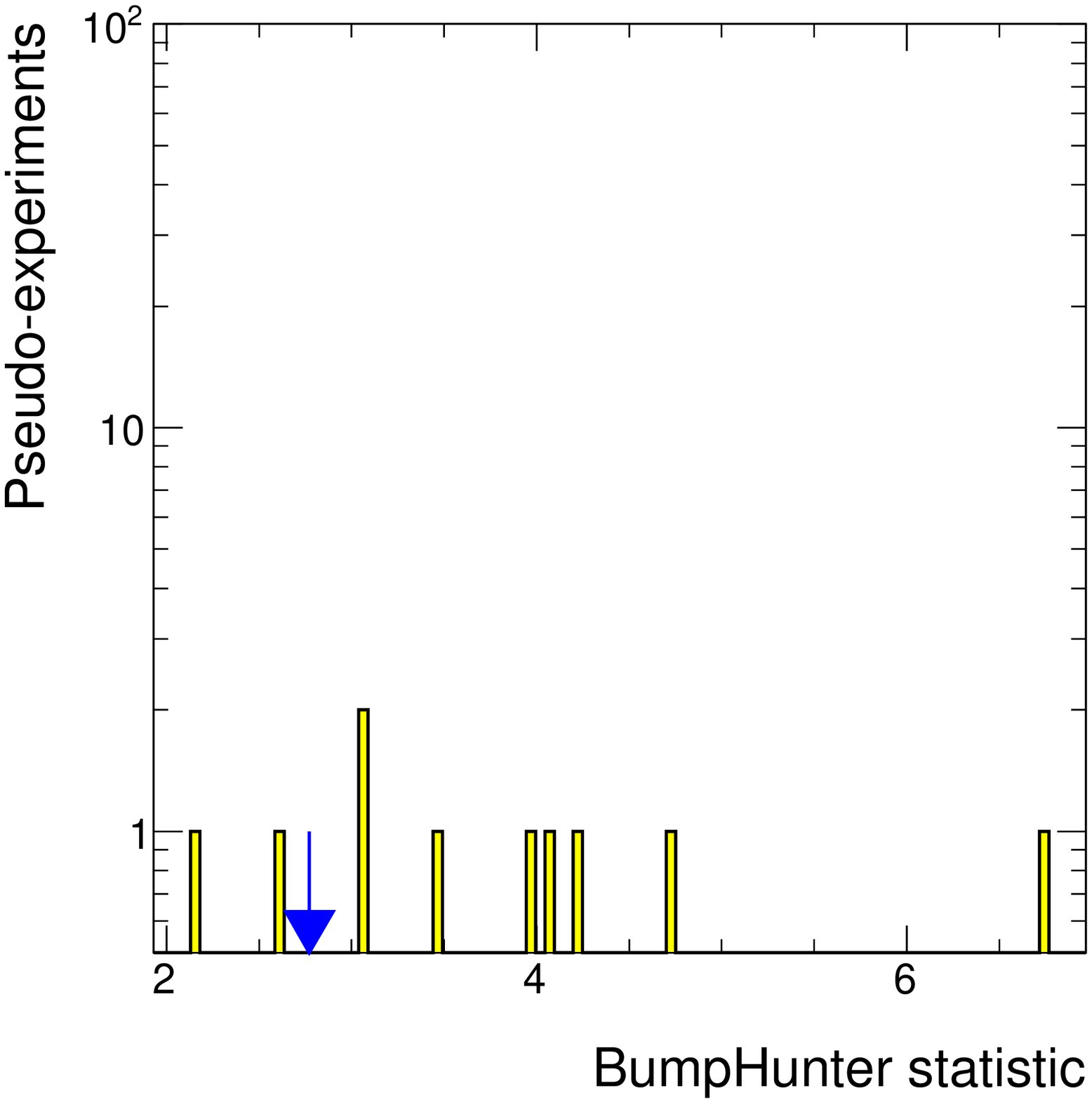} &
\includegraphics[width=0.25\textwidth]{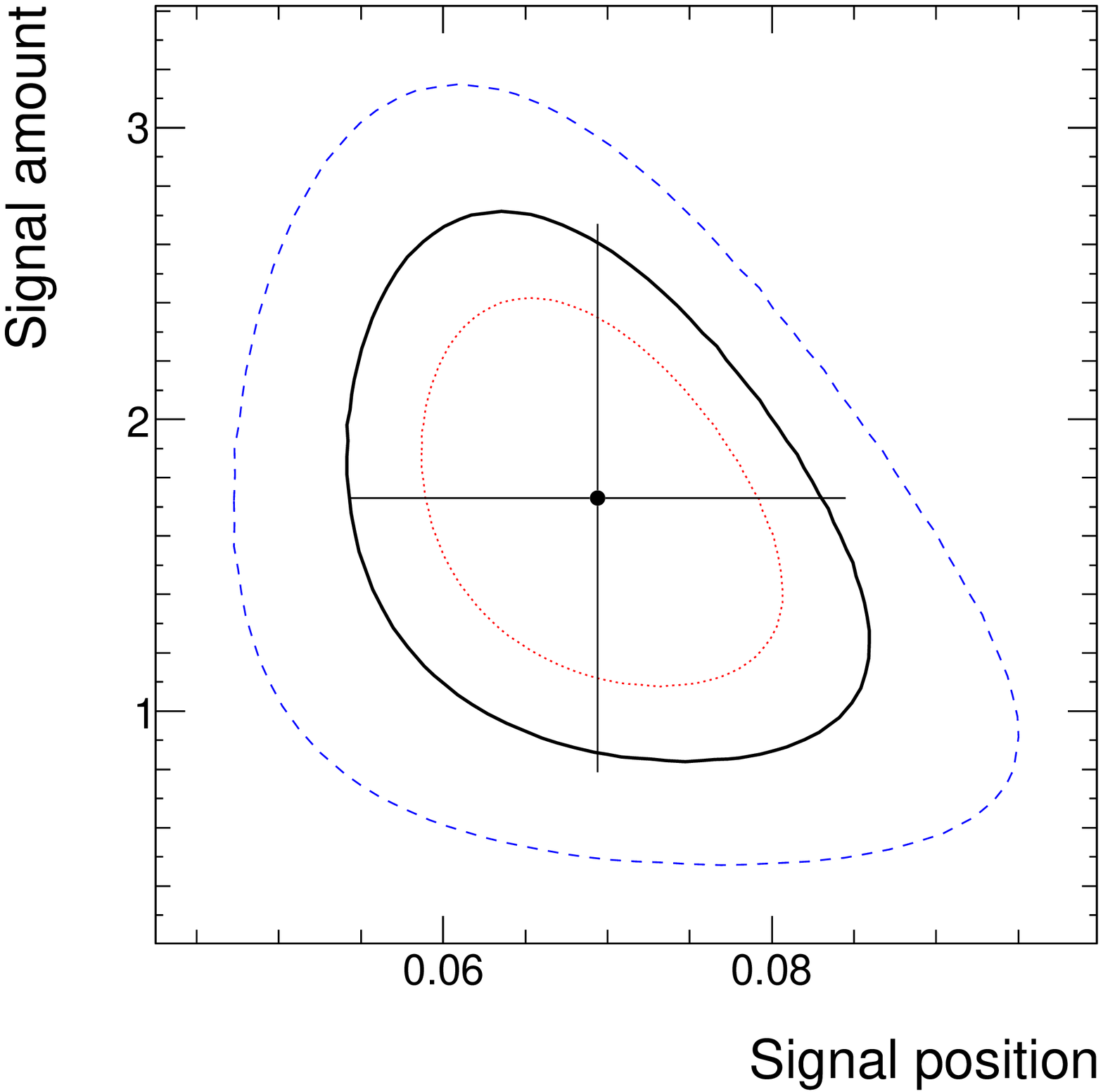} \\
\includegraphics[width=0.25\textwidth]{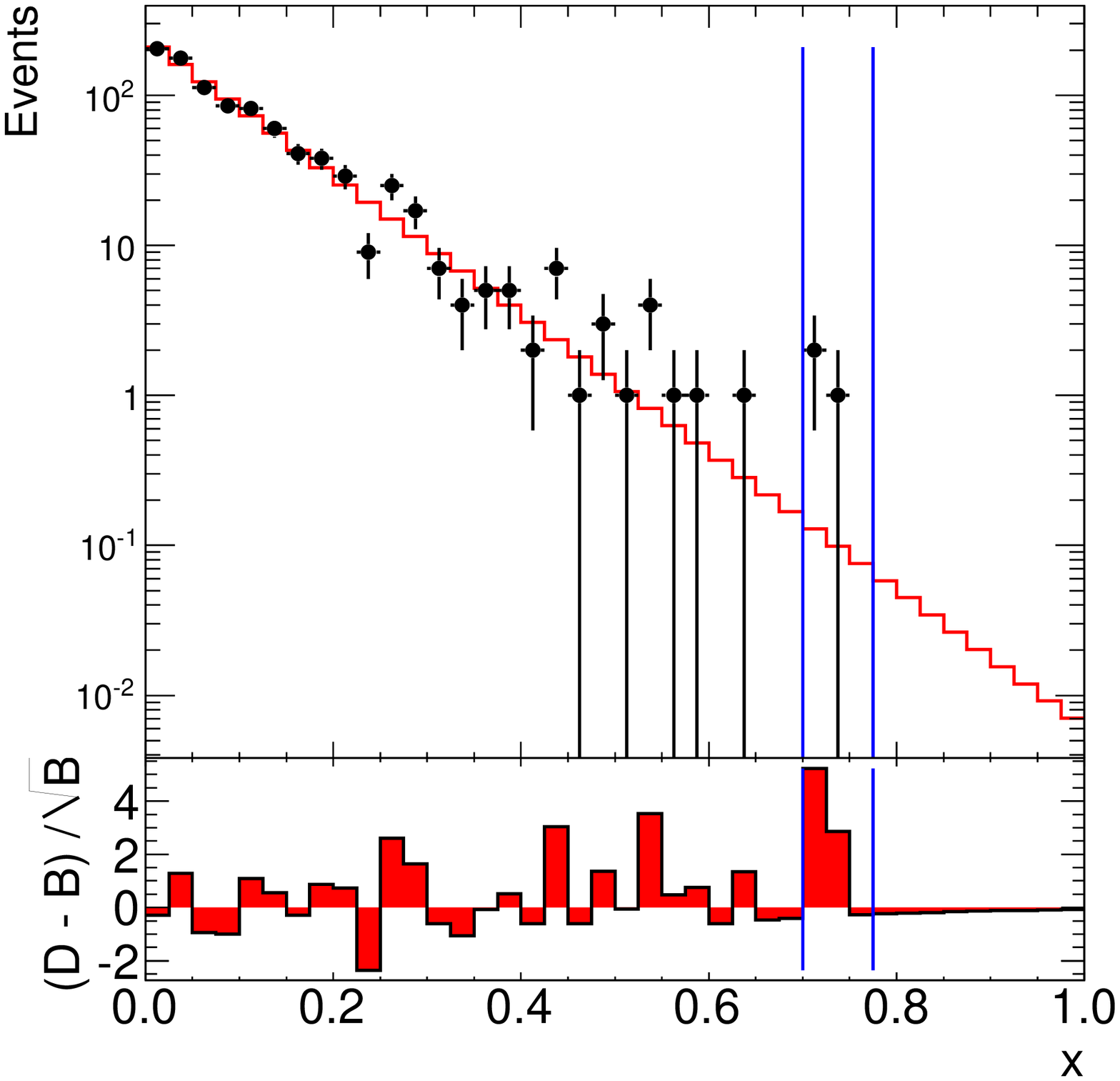}&
\includegraphics[width=0.25\textwidth]{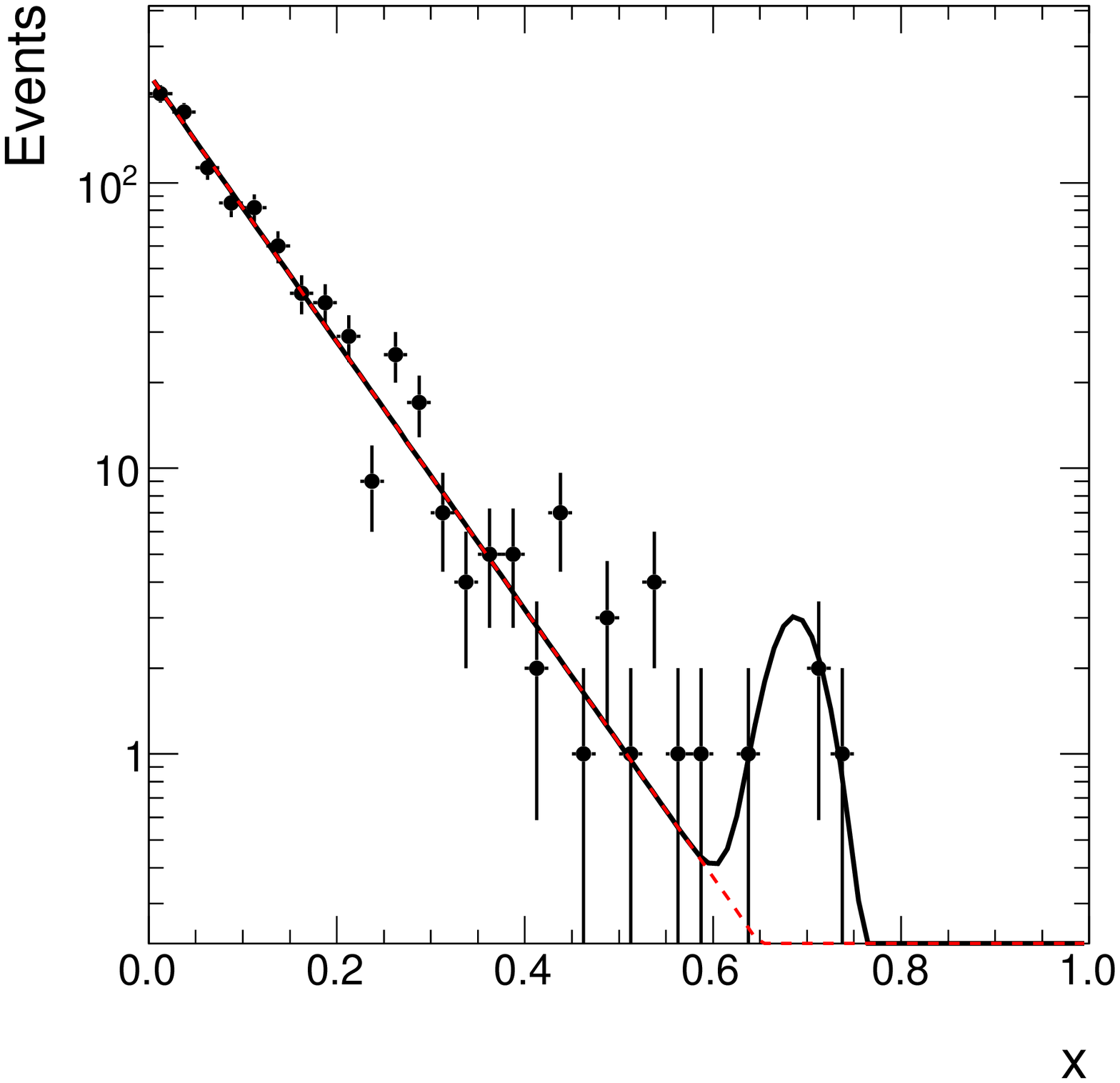} &
\includegraphics[width=0.25\textwidth]{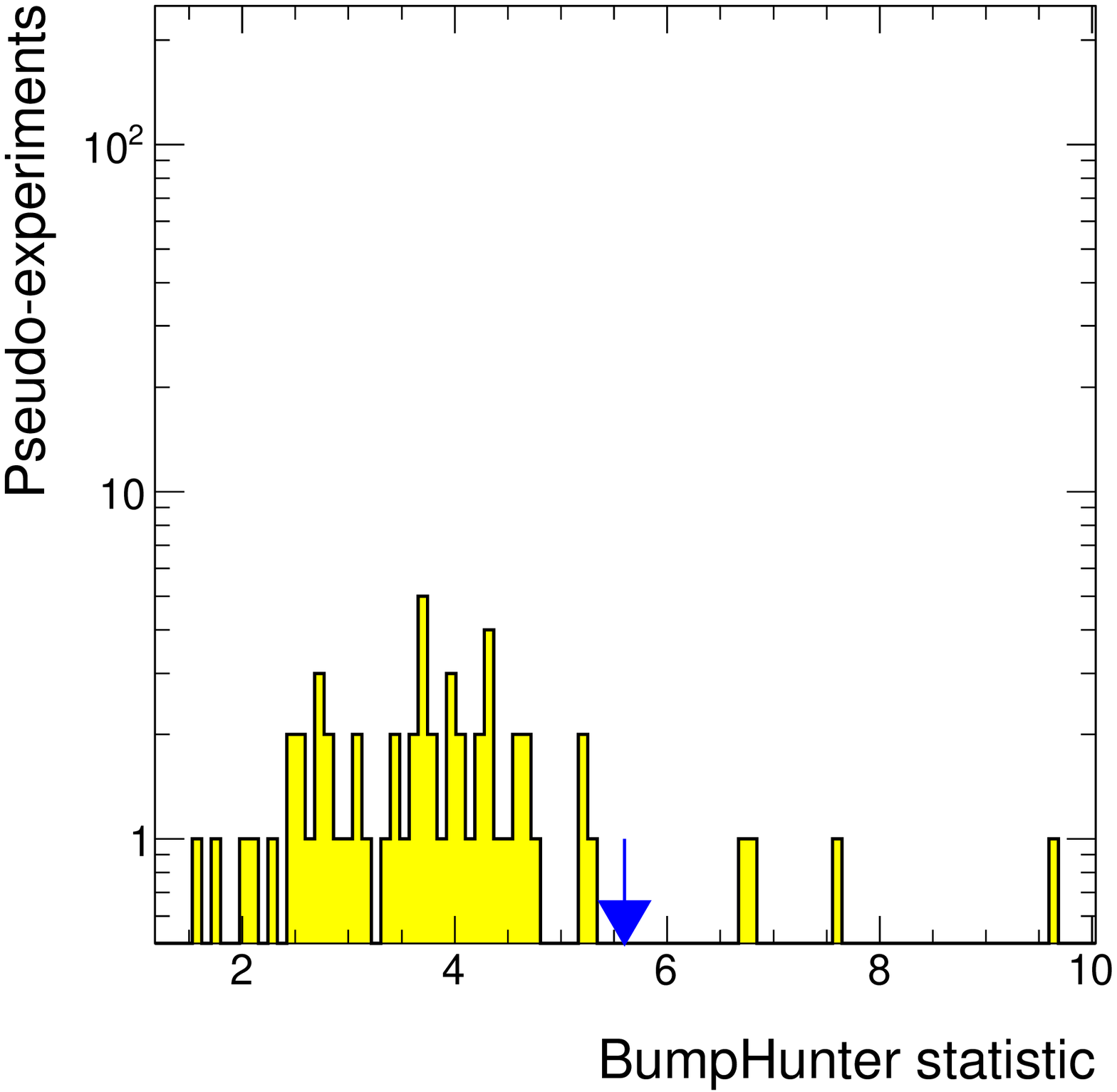}&
\includegraphics[width=0.25\textwidth]{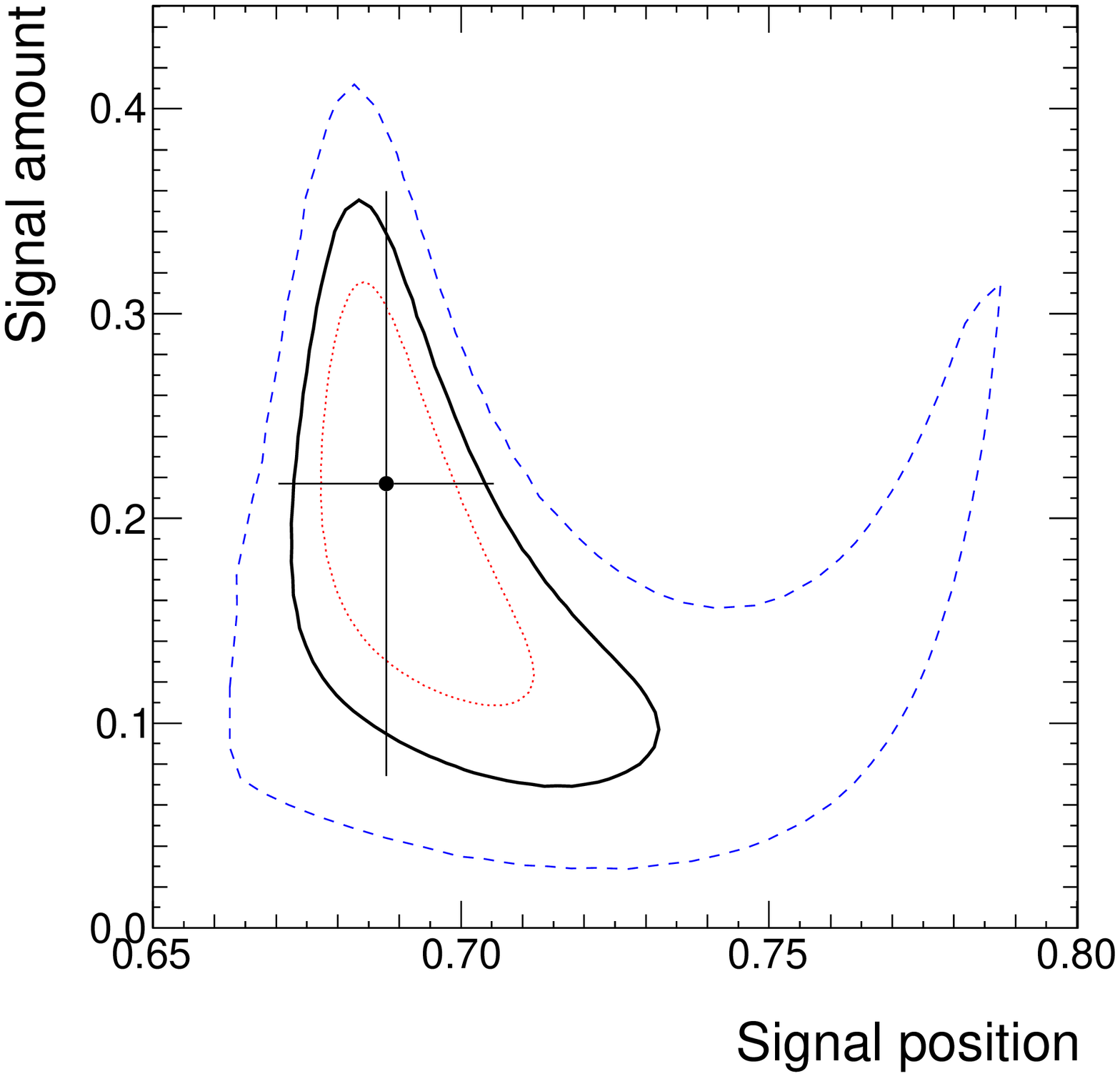} \\ 
\includegraphics[width=0.25\textwidth]{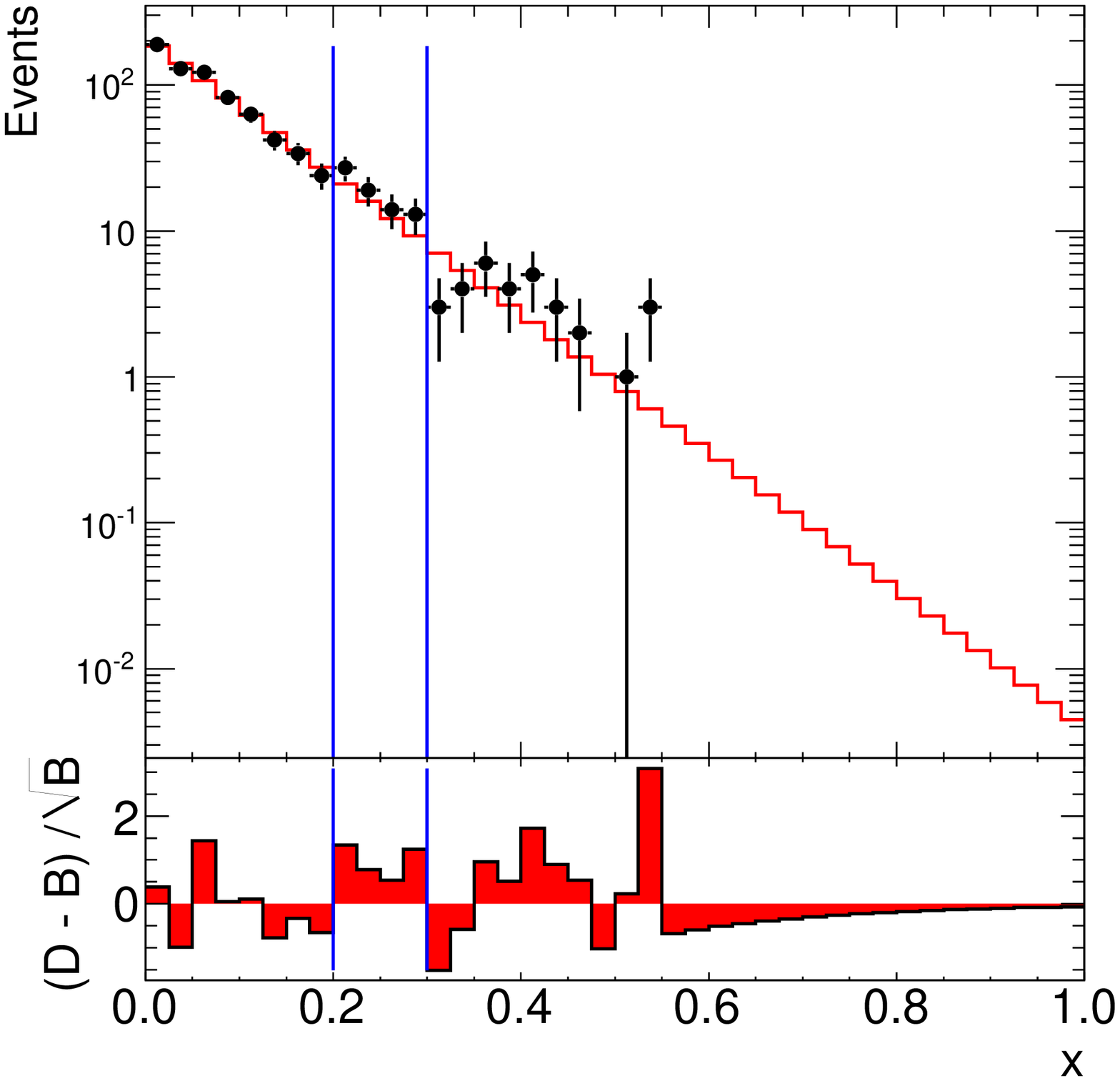}&
\includegraphics[width=0.25\textwidth]{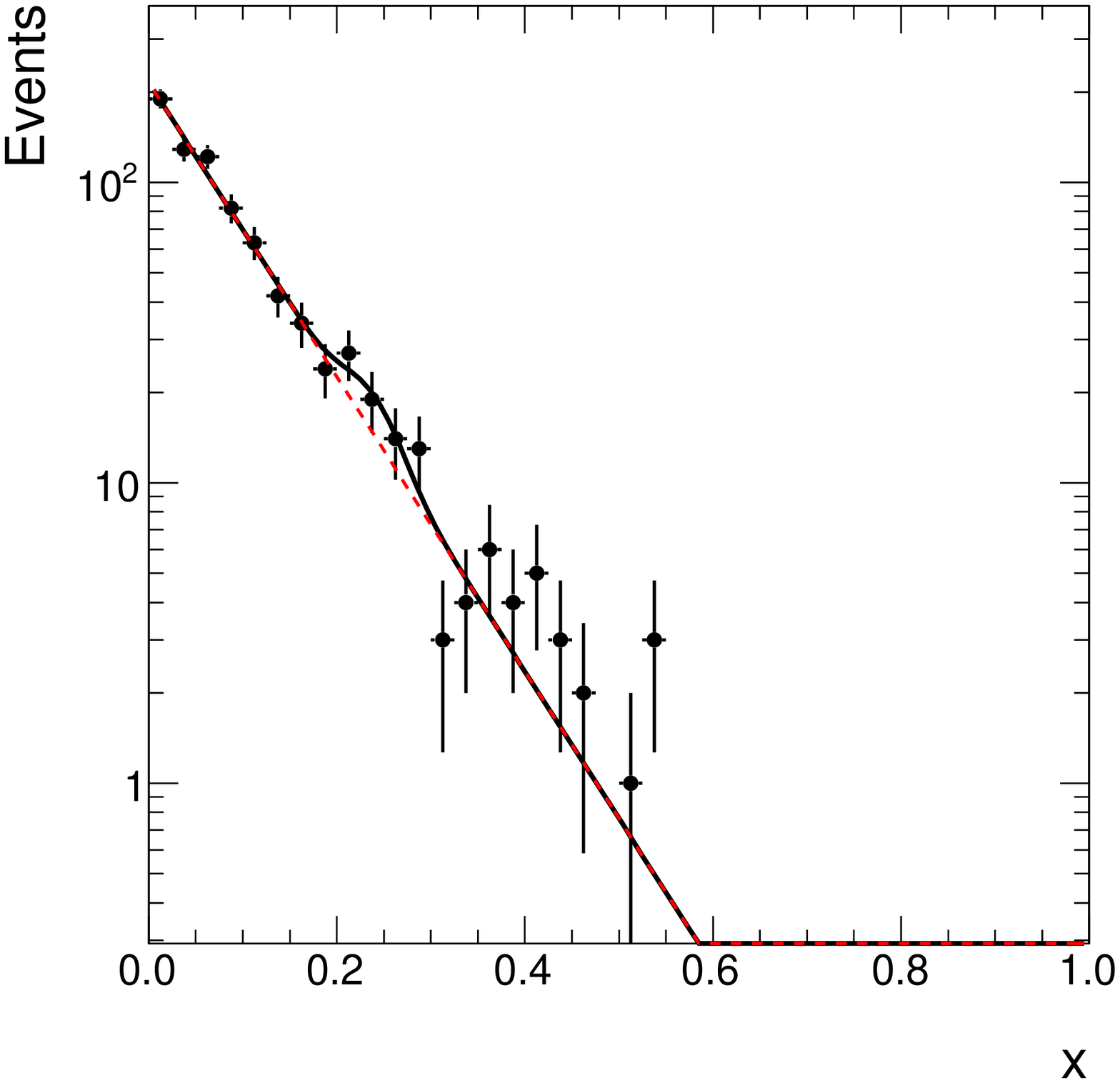} &
\includegraphics[width=0.25\textwidth]{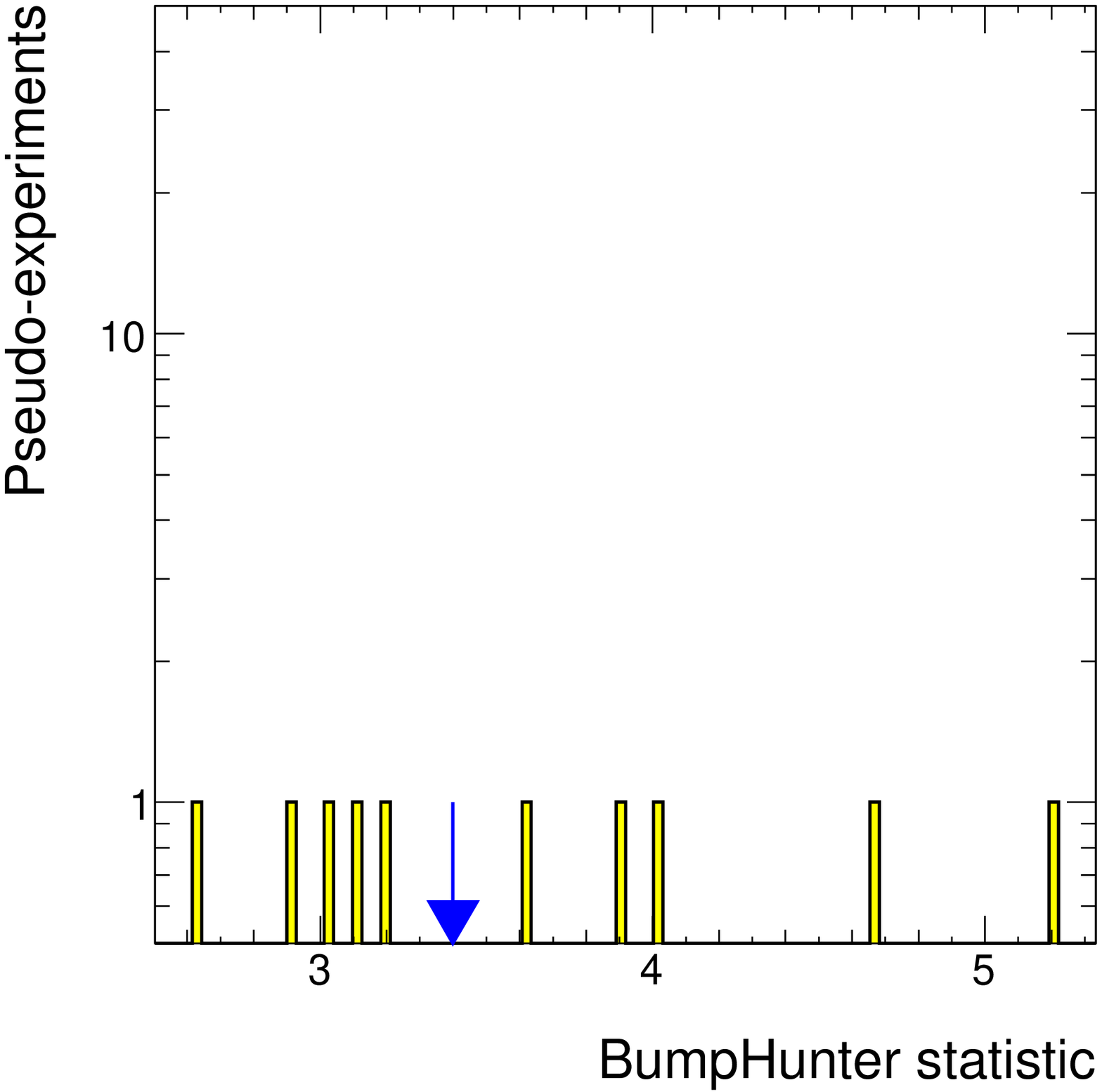}&
\includegraphics[width=0.25\textwidth]{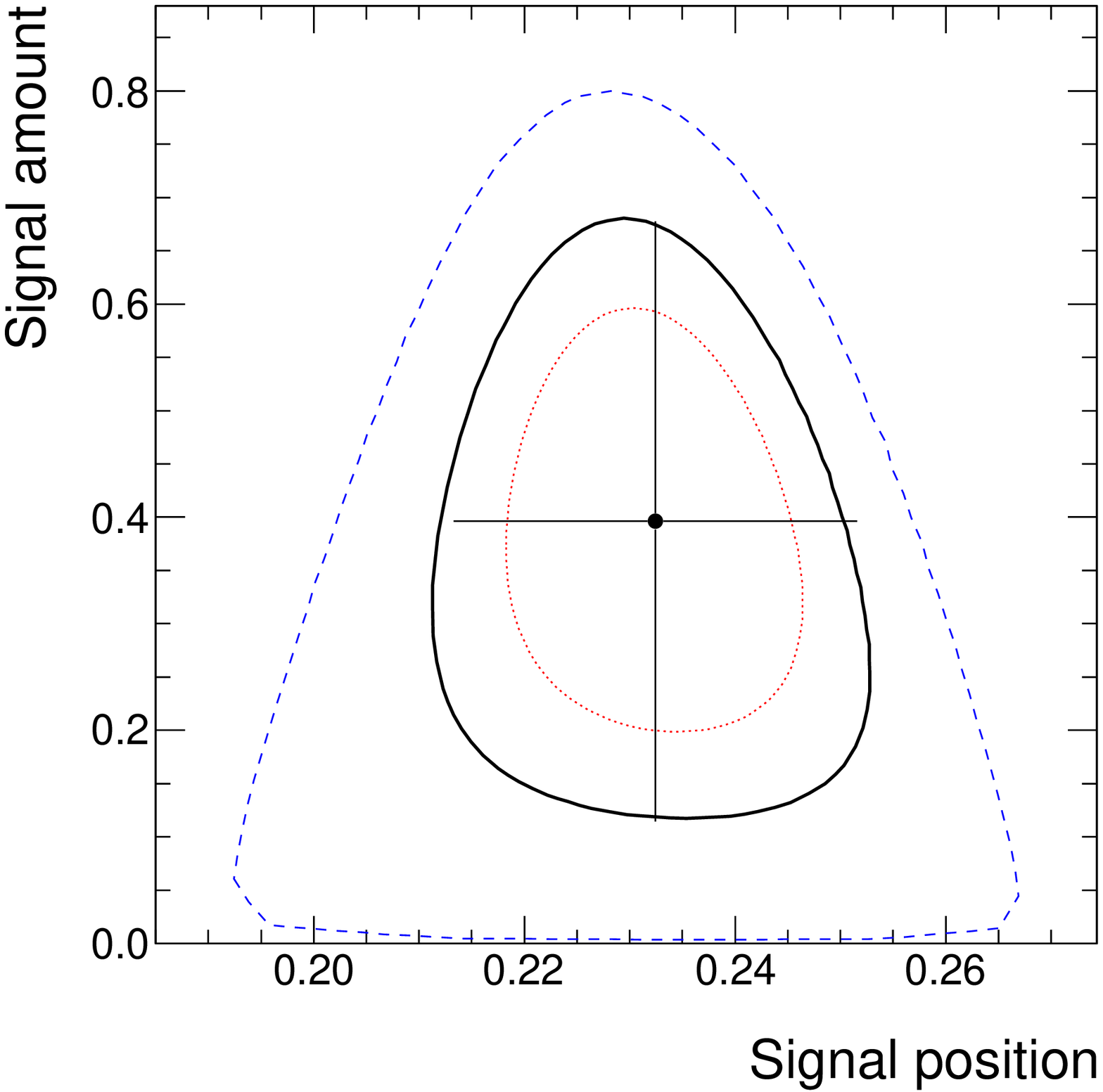} \\ 
\includegraphics[width=0.25\textwidth]{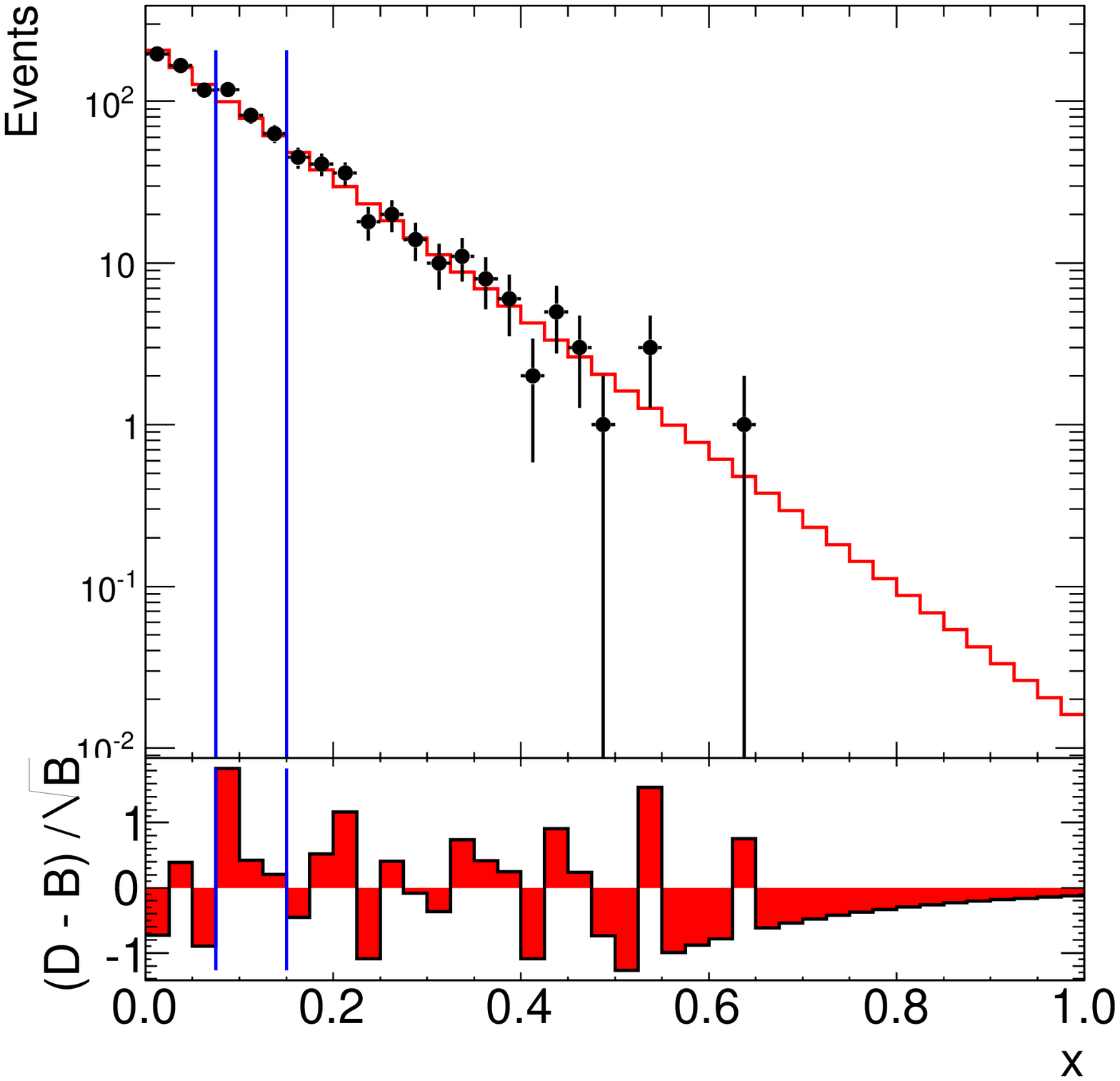} &
\includegraphics[width=0.25\textwidth]{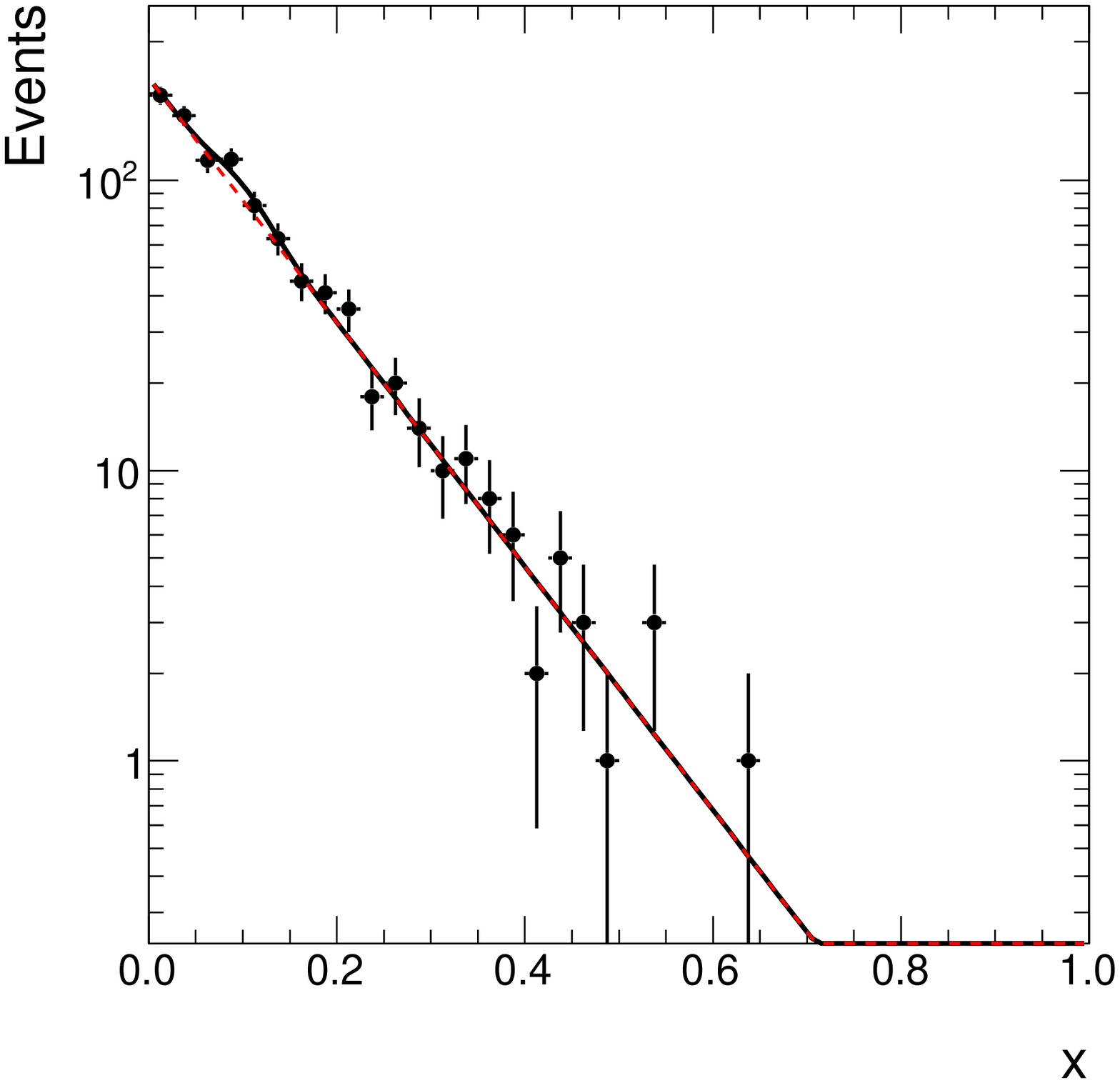} &
\includegraphics[width=0.25\textwidth]{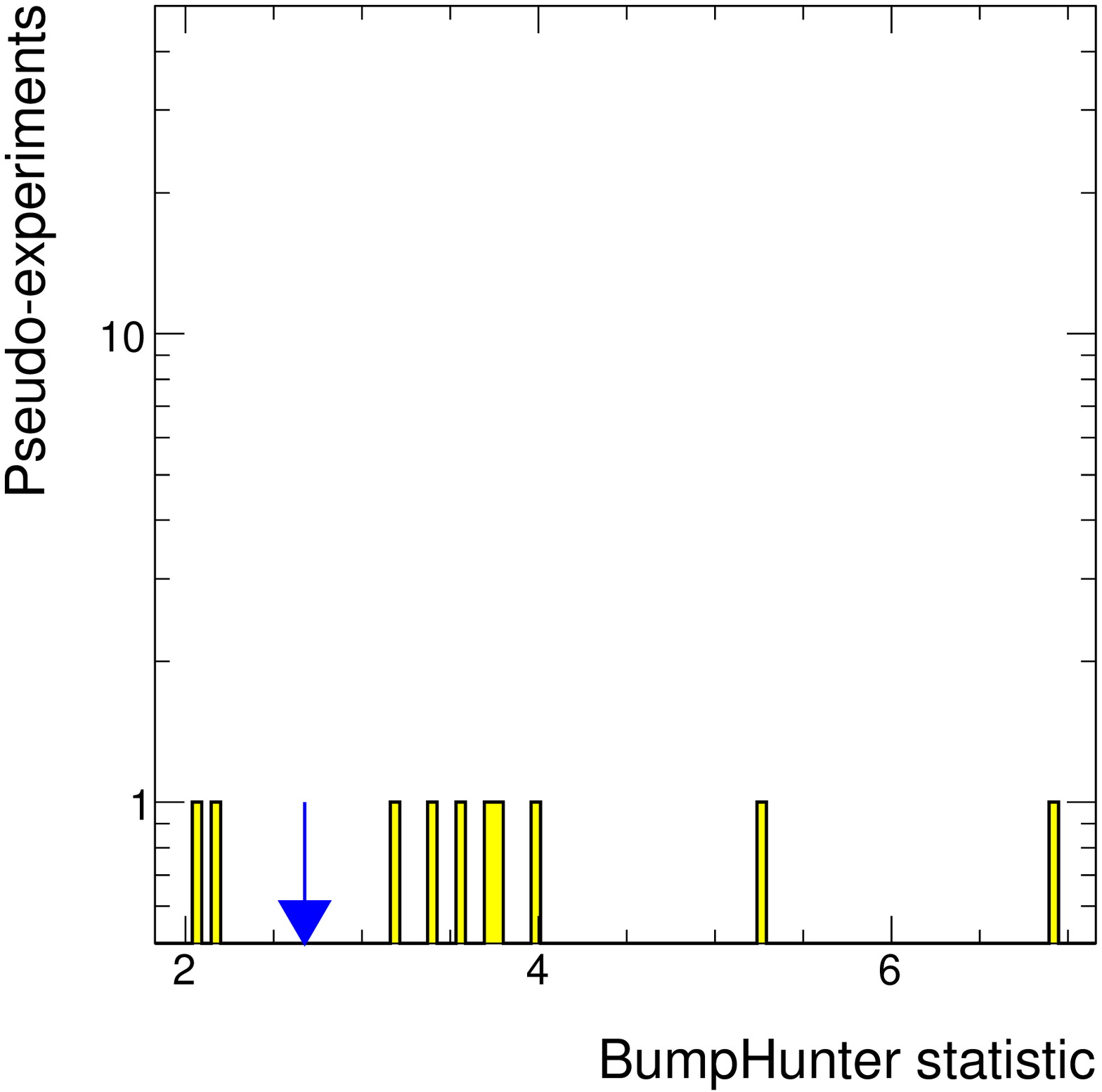}&
\includegraphics[width=0.25\textwidth]{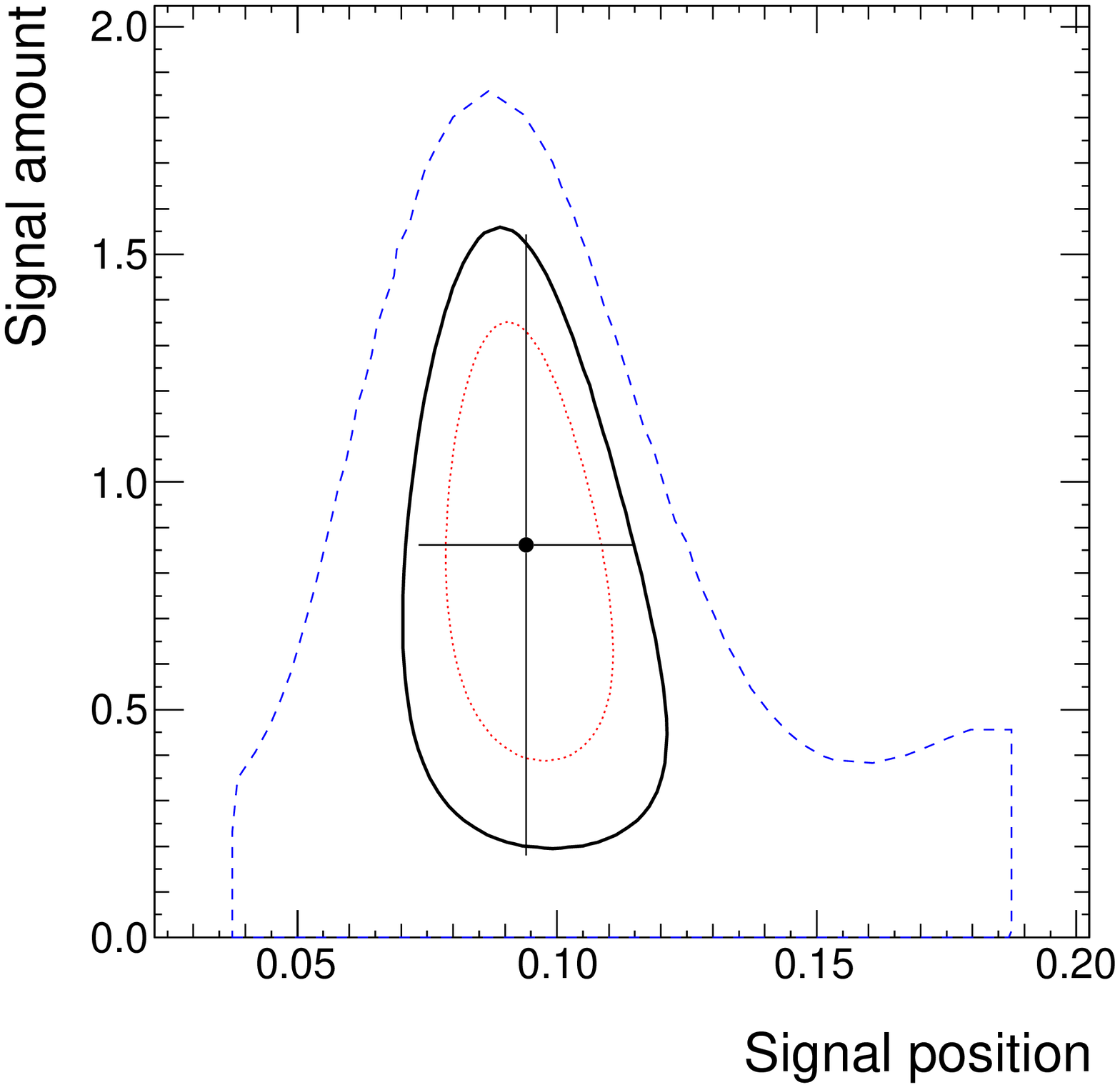} \\ 
\includegraphics[width=0.25\textwidth]{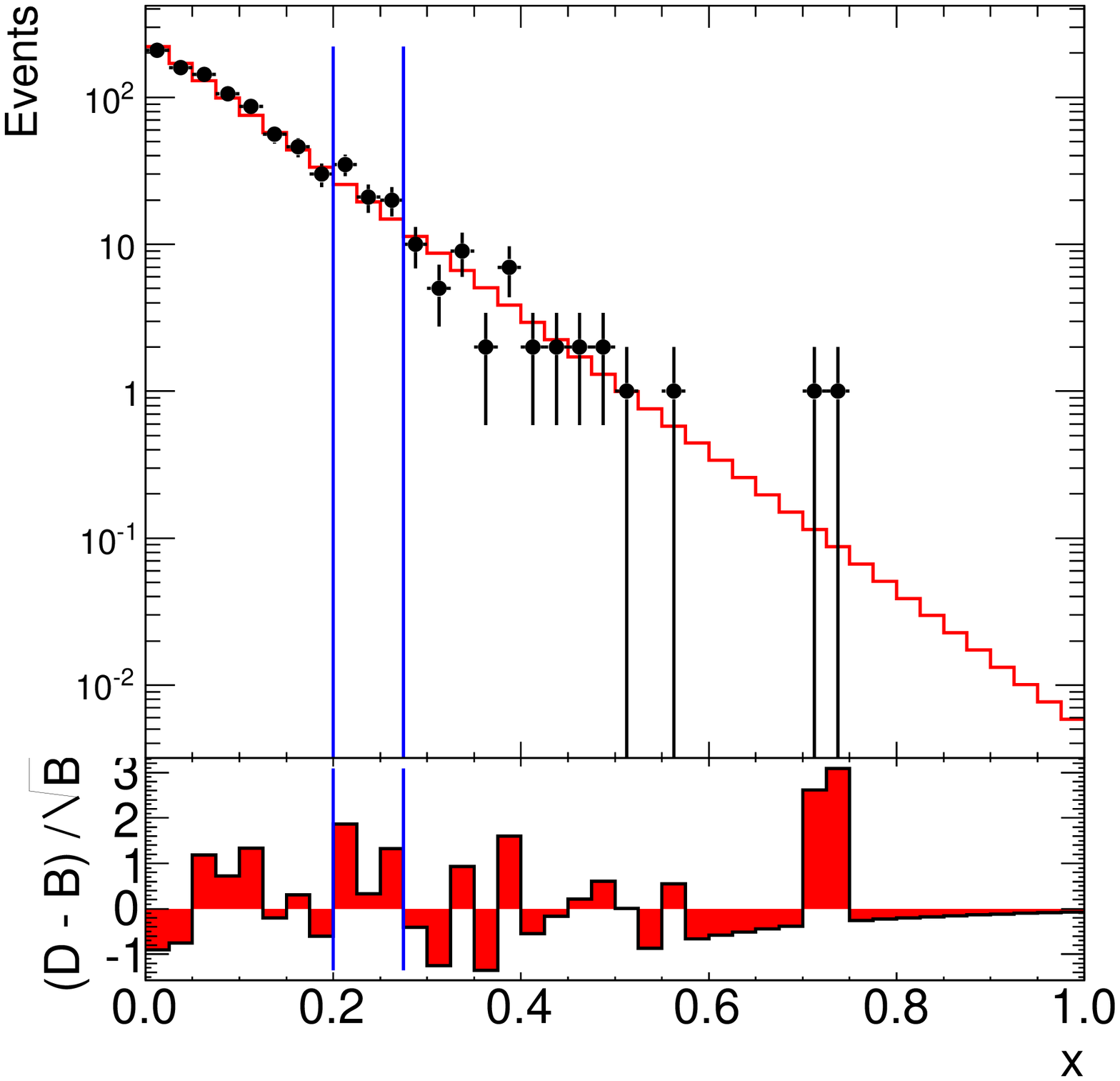}&
\includegraphics[width=0.25\textwidth]{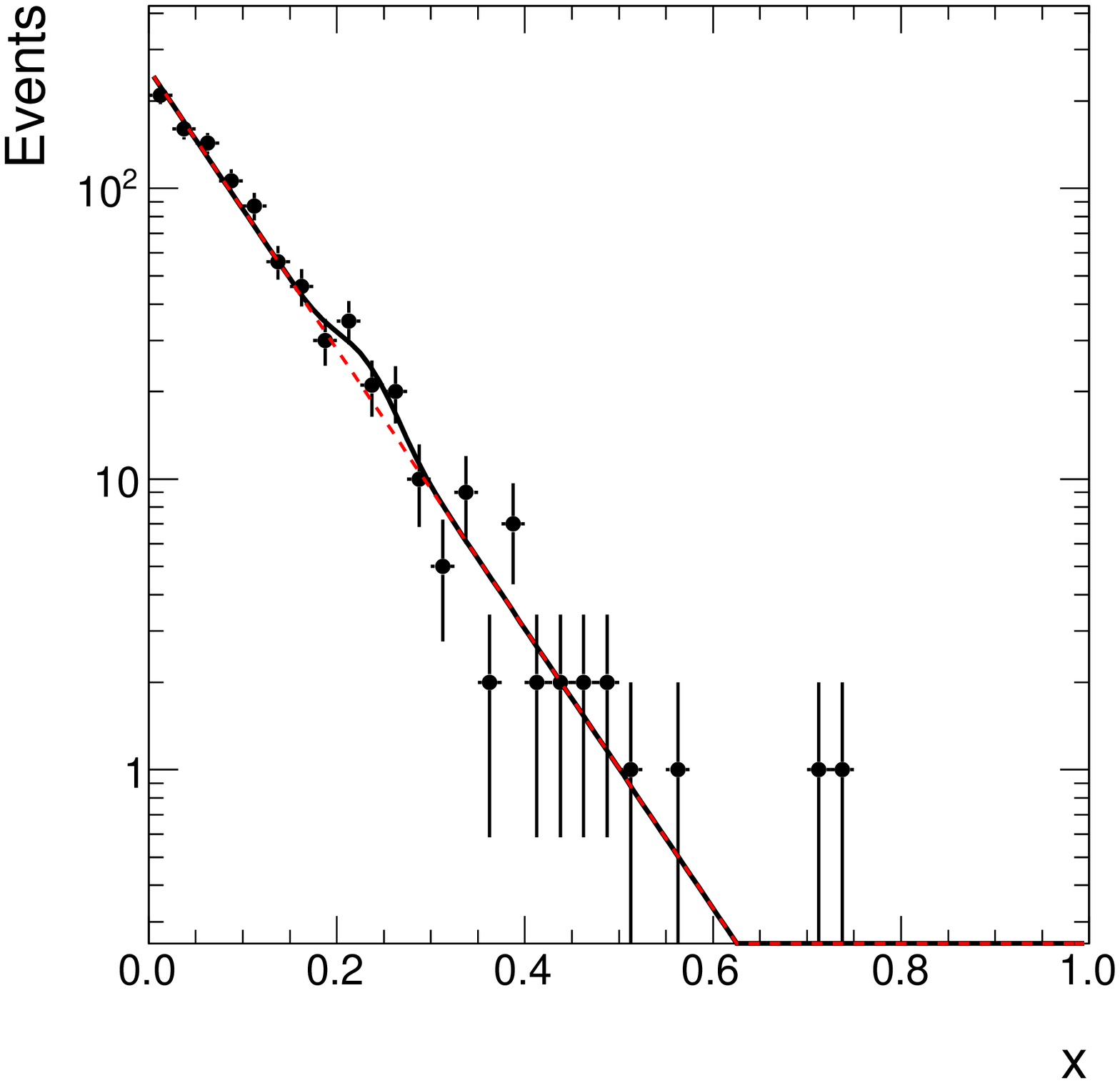} &
\includegraphics[width=0.25\textwidth]{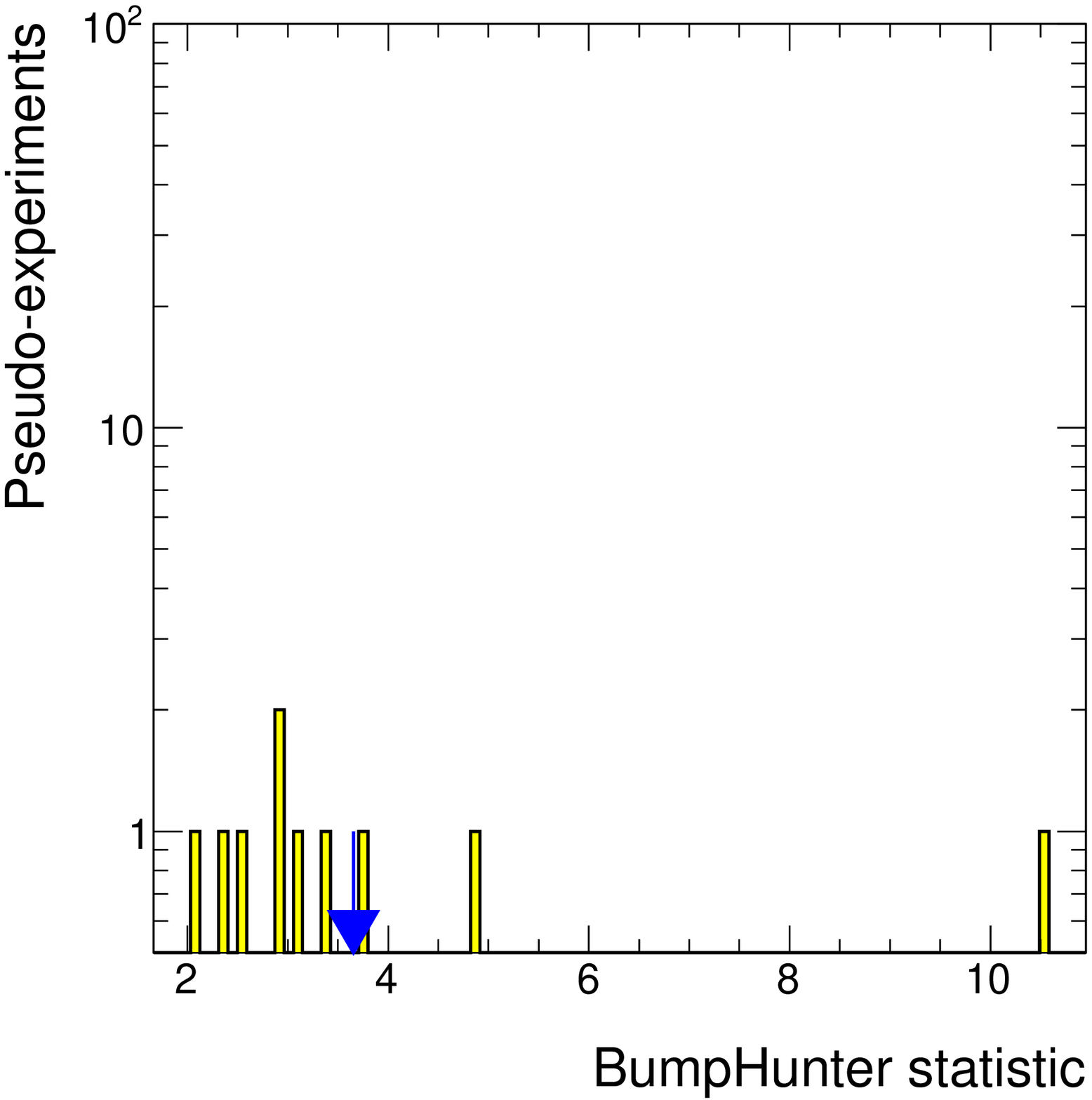}&
\includegraphics[width=0.25\textwidth]{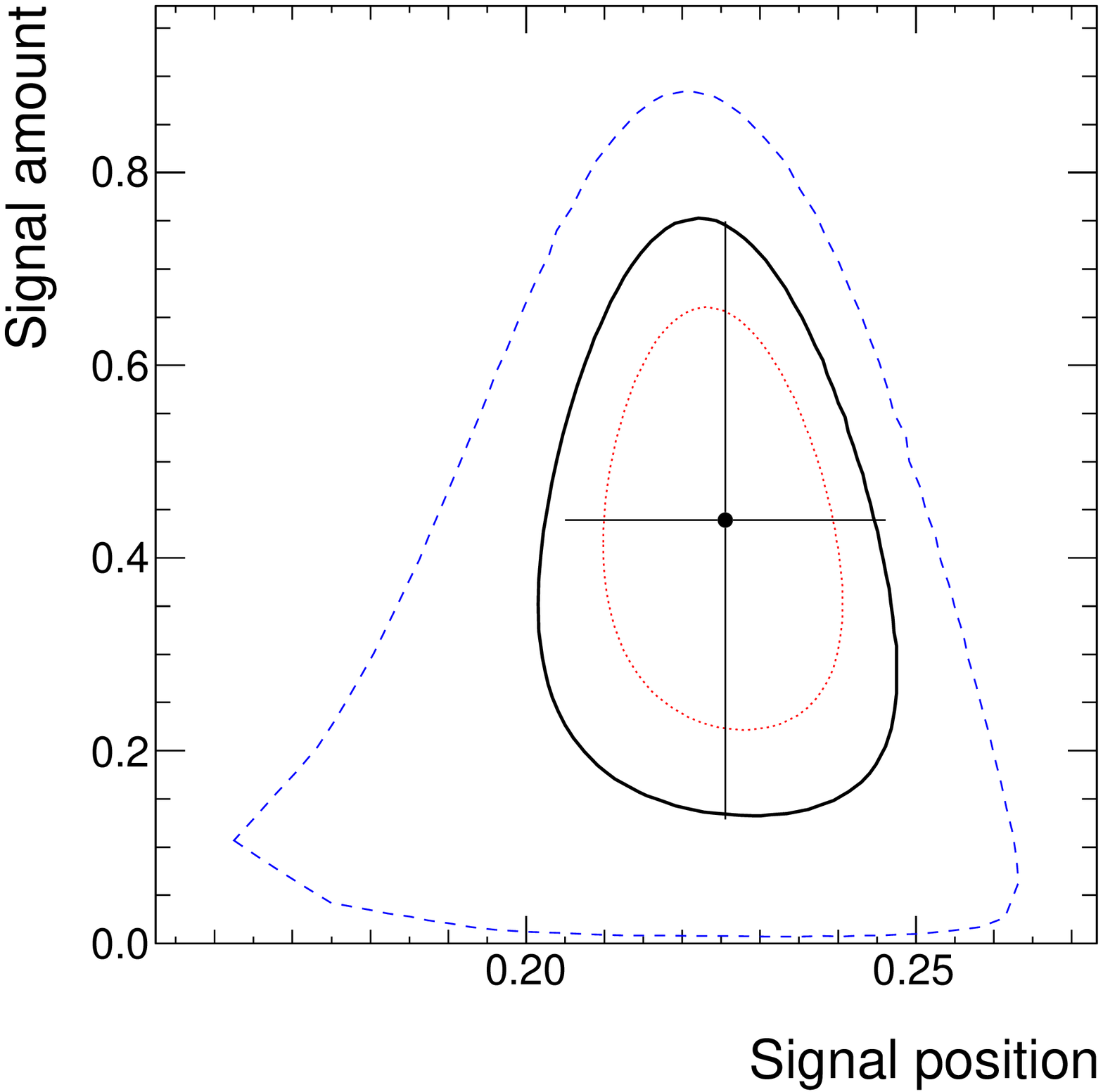} \\ 
\end{tabular}
\caption{\label{fig:appendix1} Summary of the results from 5 Banff Challenge Problem 1 datasets, where no discovery was claimed.  The datasets are \{100, 400, 500, 700, 800\}, and one row of figures corresponds to each respectively.  We see from the 2-dimensional contours that parameters $D$ and $E$ are poorly constrained, because there is not significant signal in the data to constrain them.  The corresponding most likely \pvals are: $\{\frac{8}{10}, \frac{4}{60}, \frac{5}{10}, \frac{8}{10}, \frac{3}{10} \}$
}
\end{figure}

\begin{figure}[p]
\hspace{-1cm}
\begin{tabular}{cccc}
\includegraphics[width=0.25\textwidth]{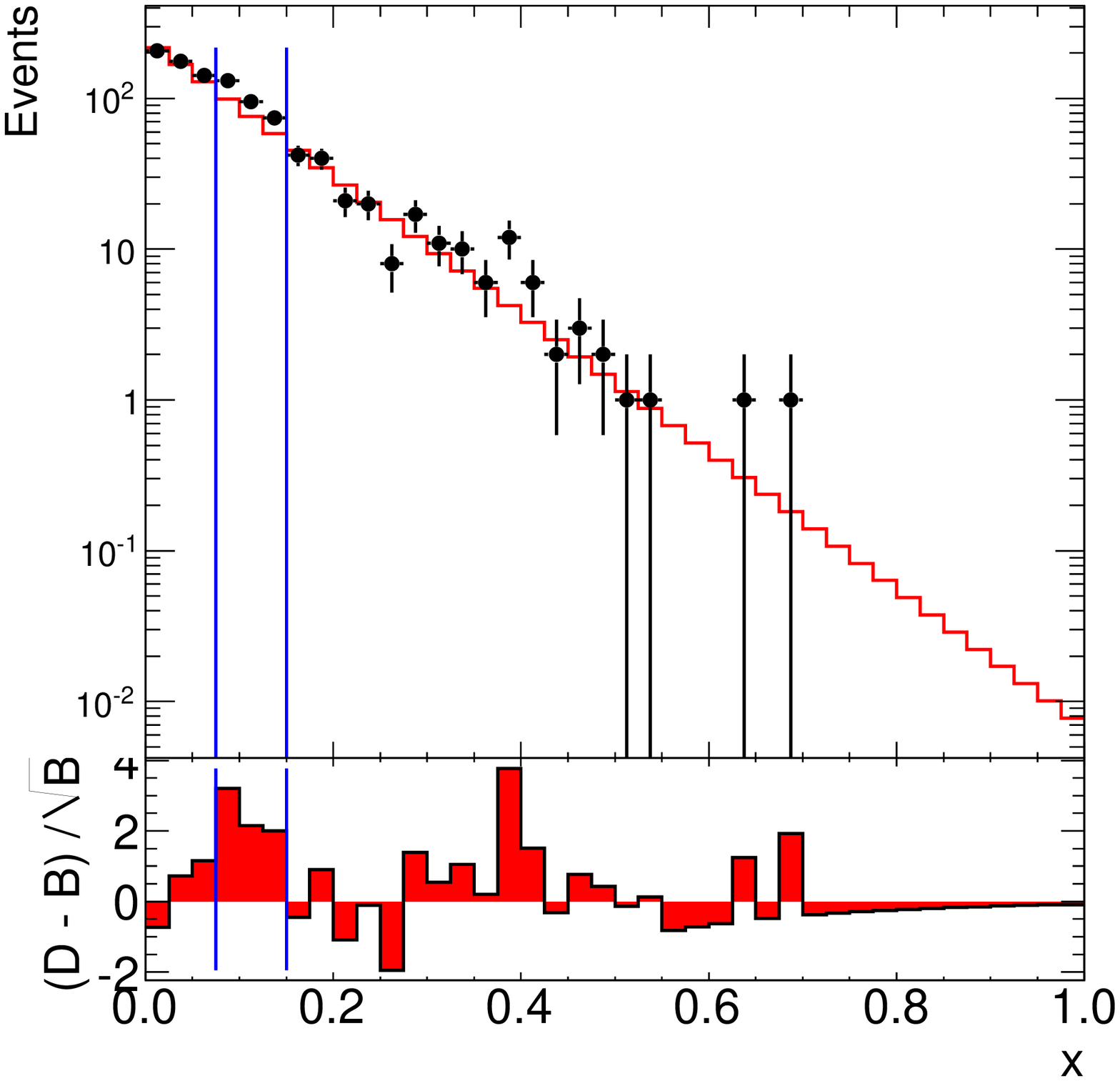} &
\includegraphics[width=0.25\textwidth]{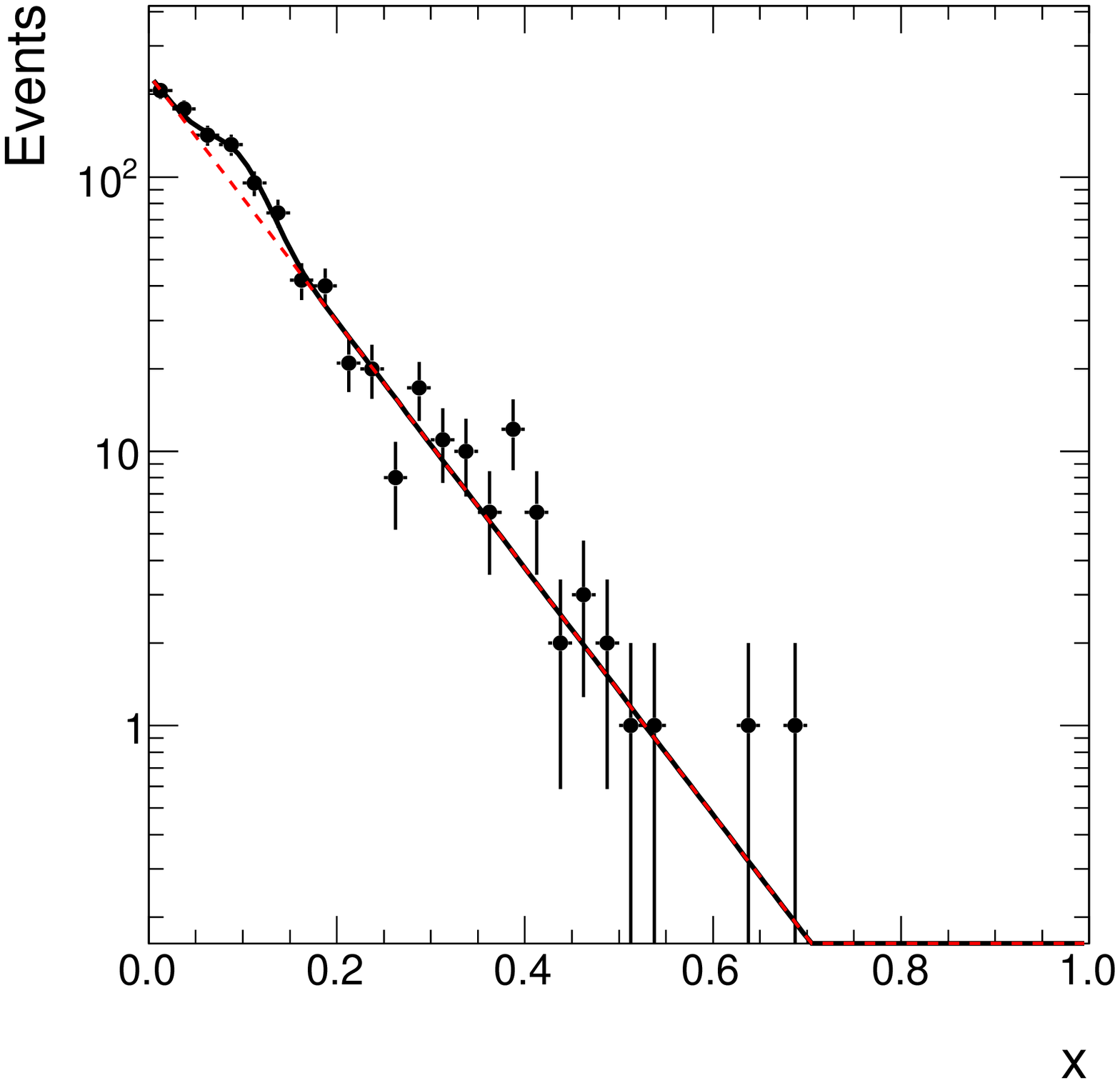} &
\includegraphics[width=0.25\textwidth]{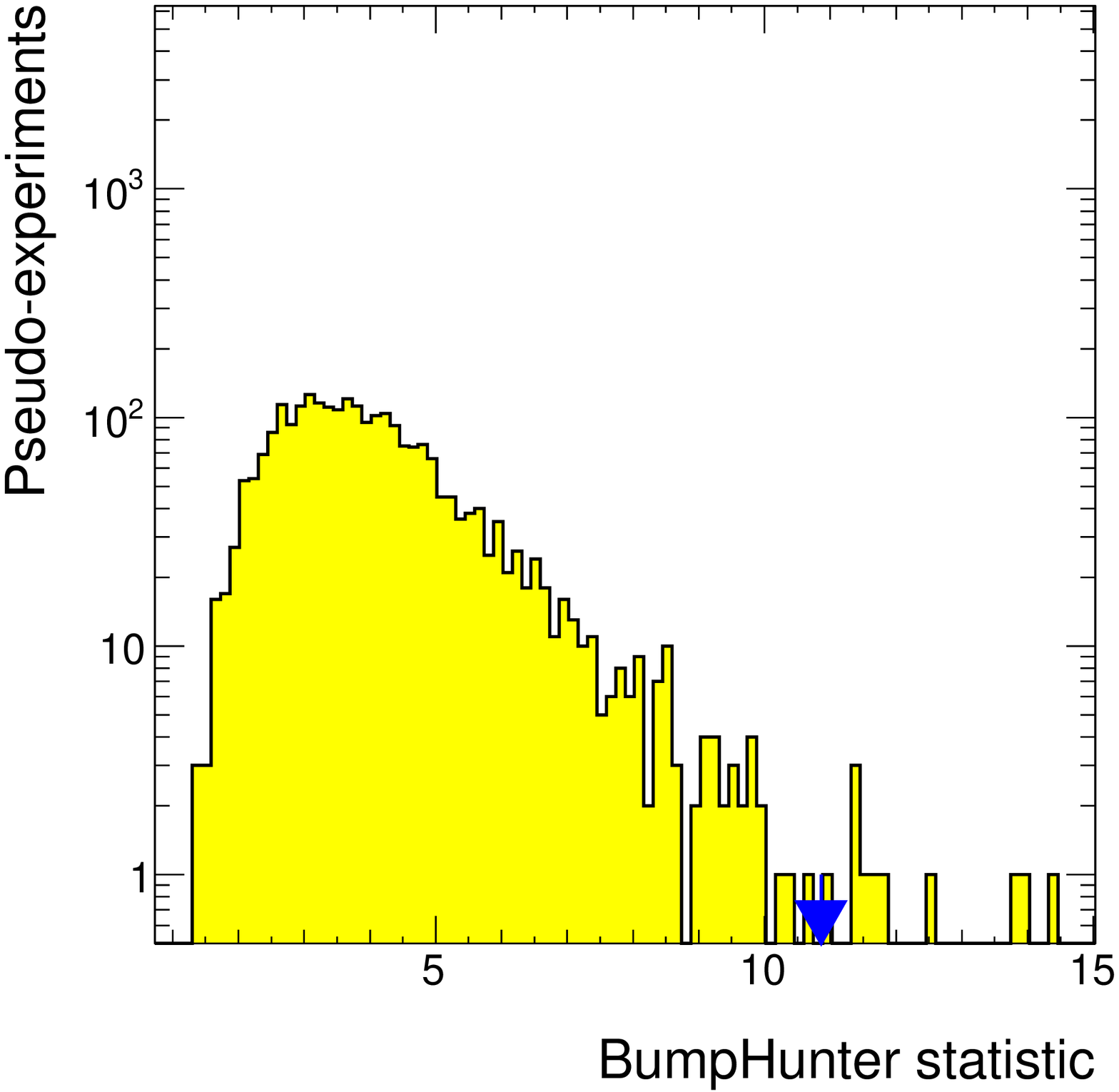} &
\includegraphics[width=0.25\textwidth]{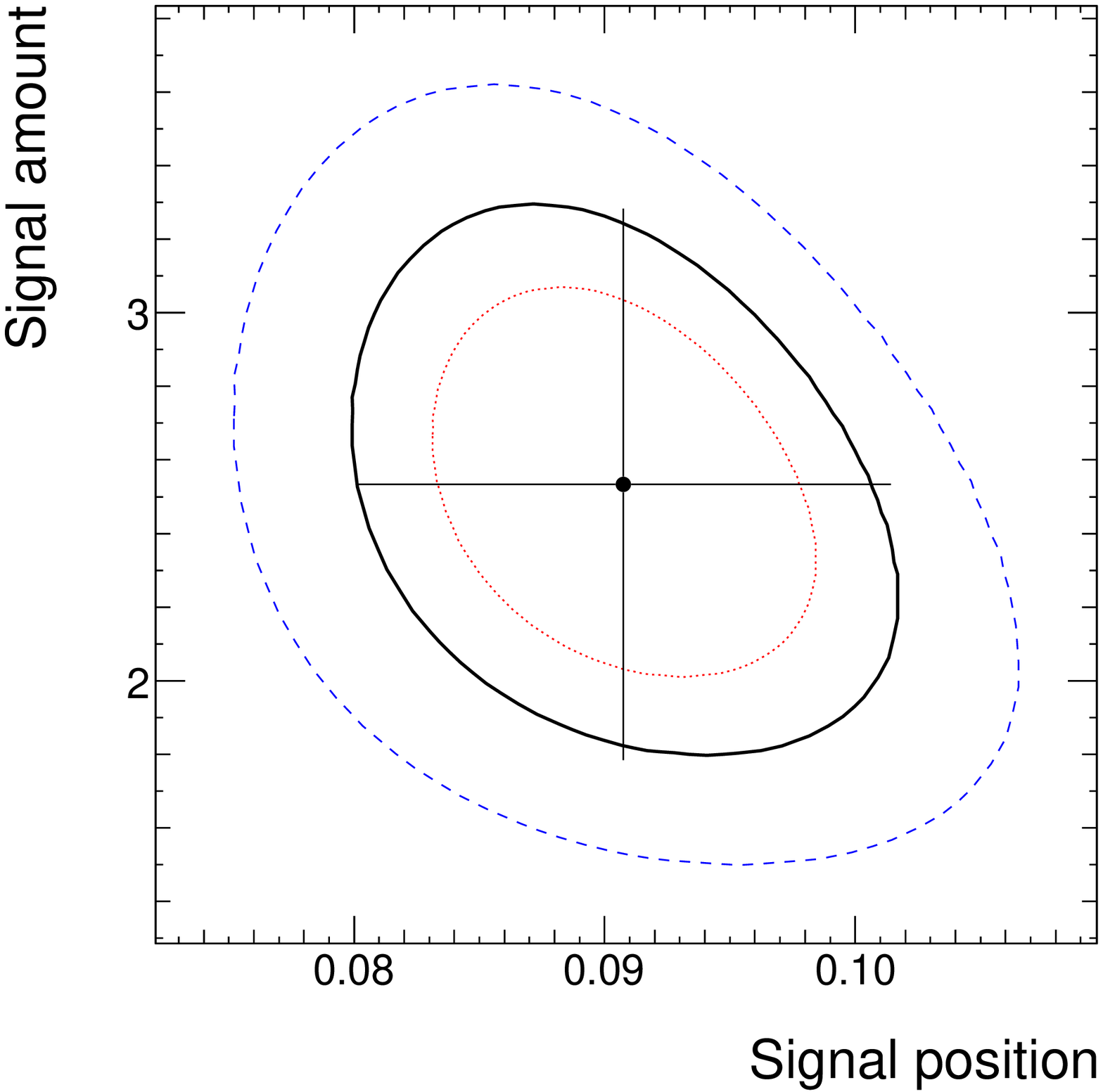} \\
\includegraphics[width=0.25\textwidth]{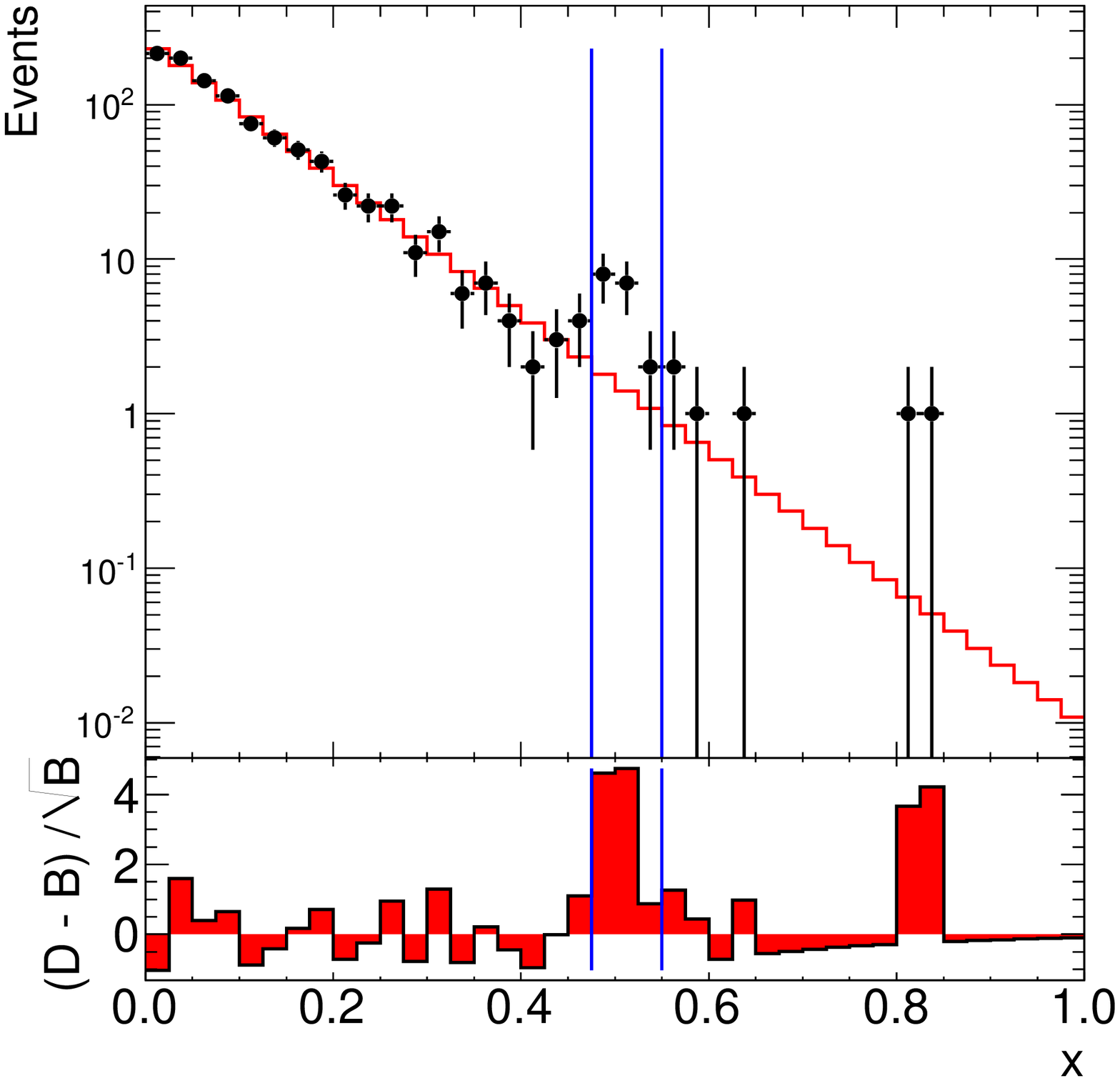} &
\includegraphics[width=0.25\textwidth]{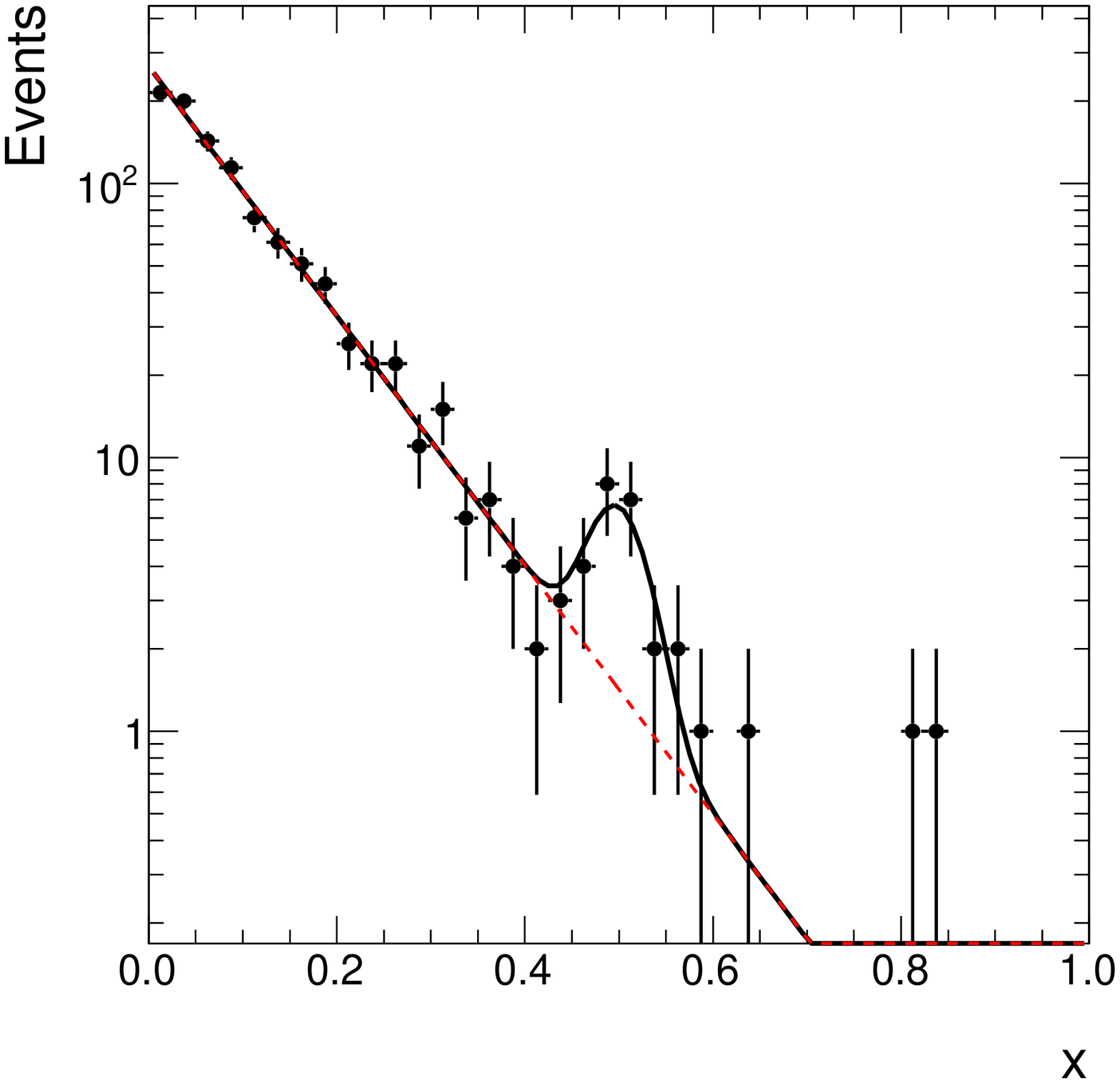} &
\includegraphics[width=0.25\textwidth]{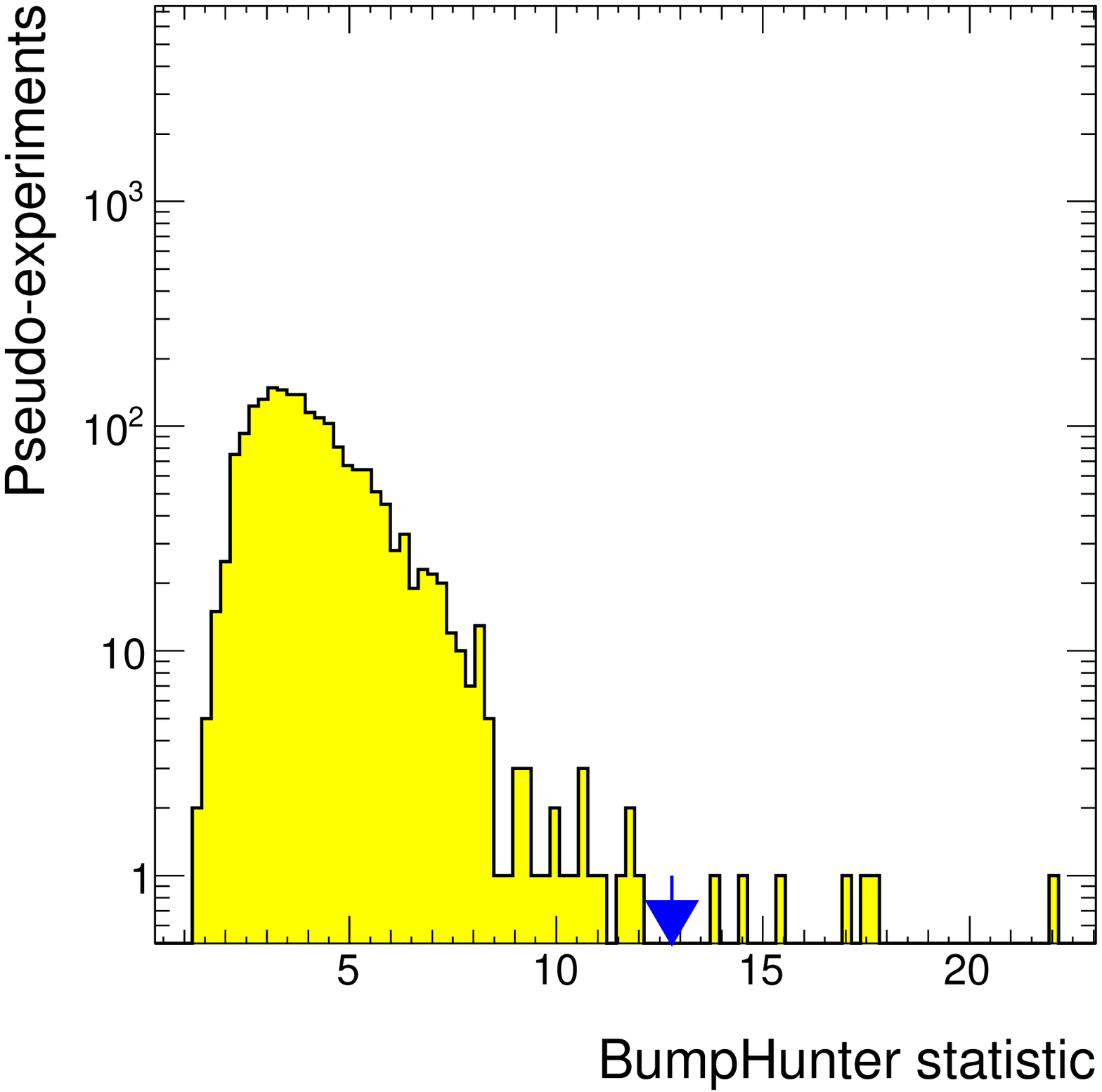} &
\includegraphics[width=0.25\textwidth]{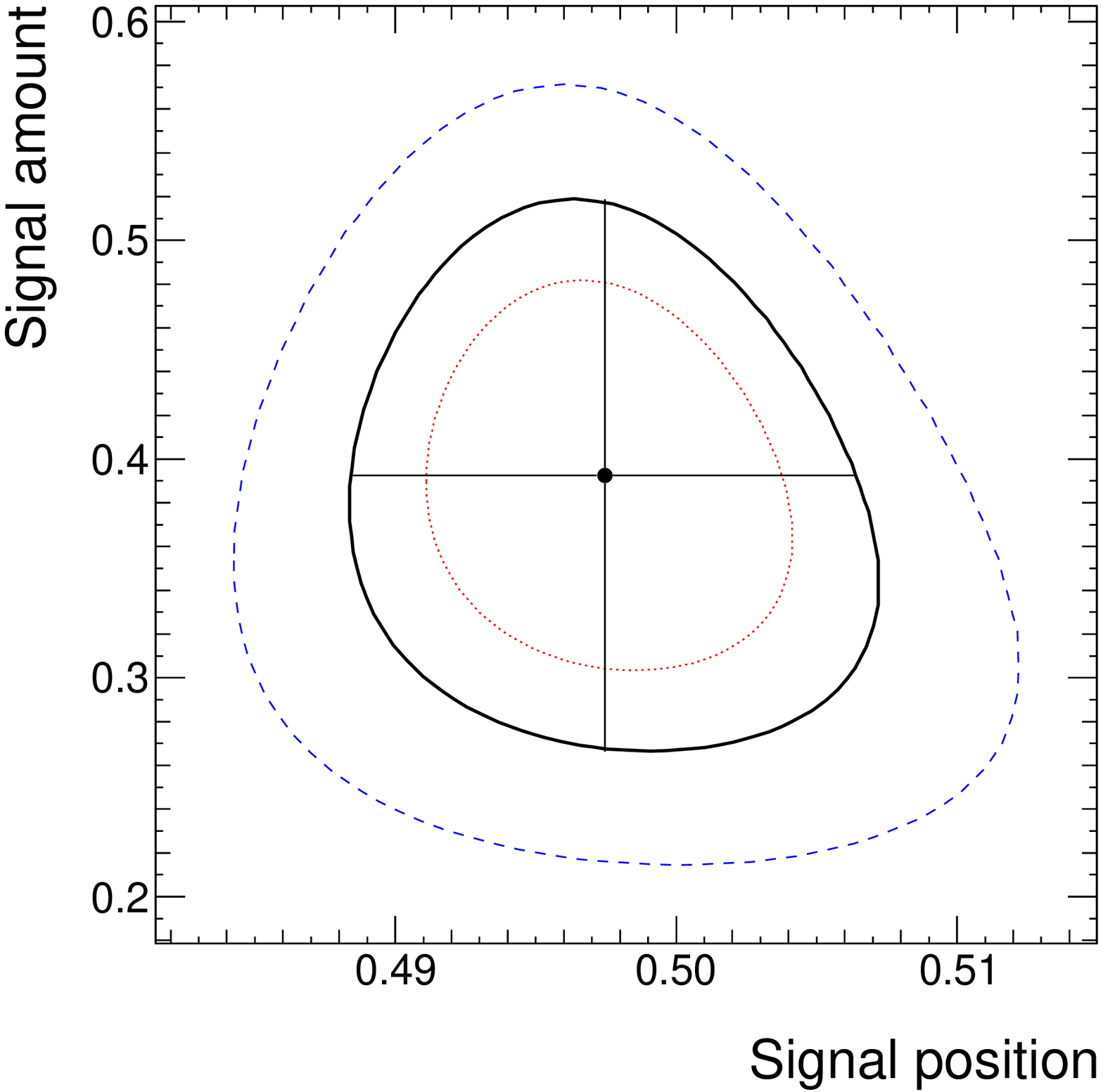} \\
\includegraphics[width=0.25\textwidth]{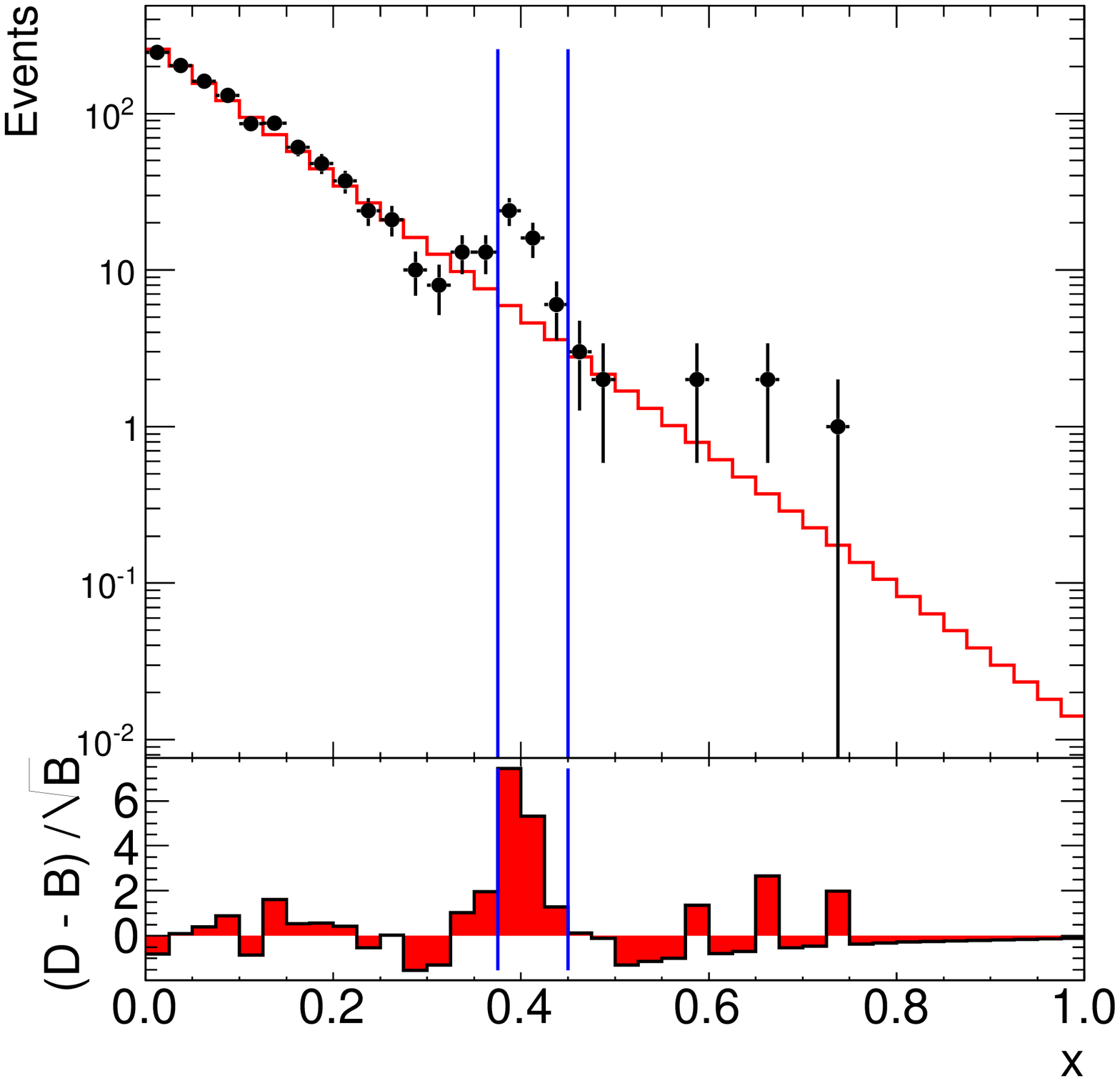} &
\includegraphics[width=0.25\textwidth]{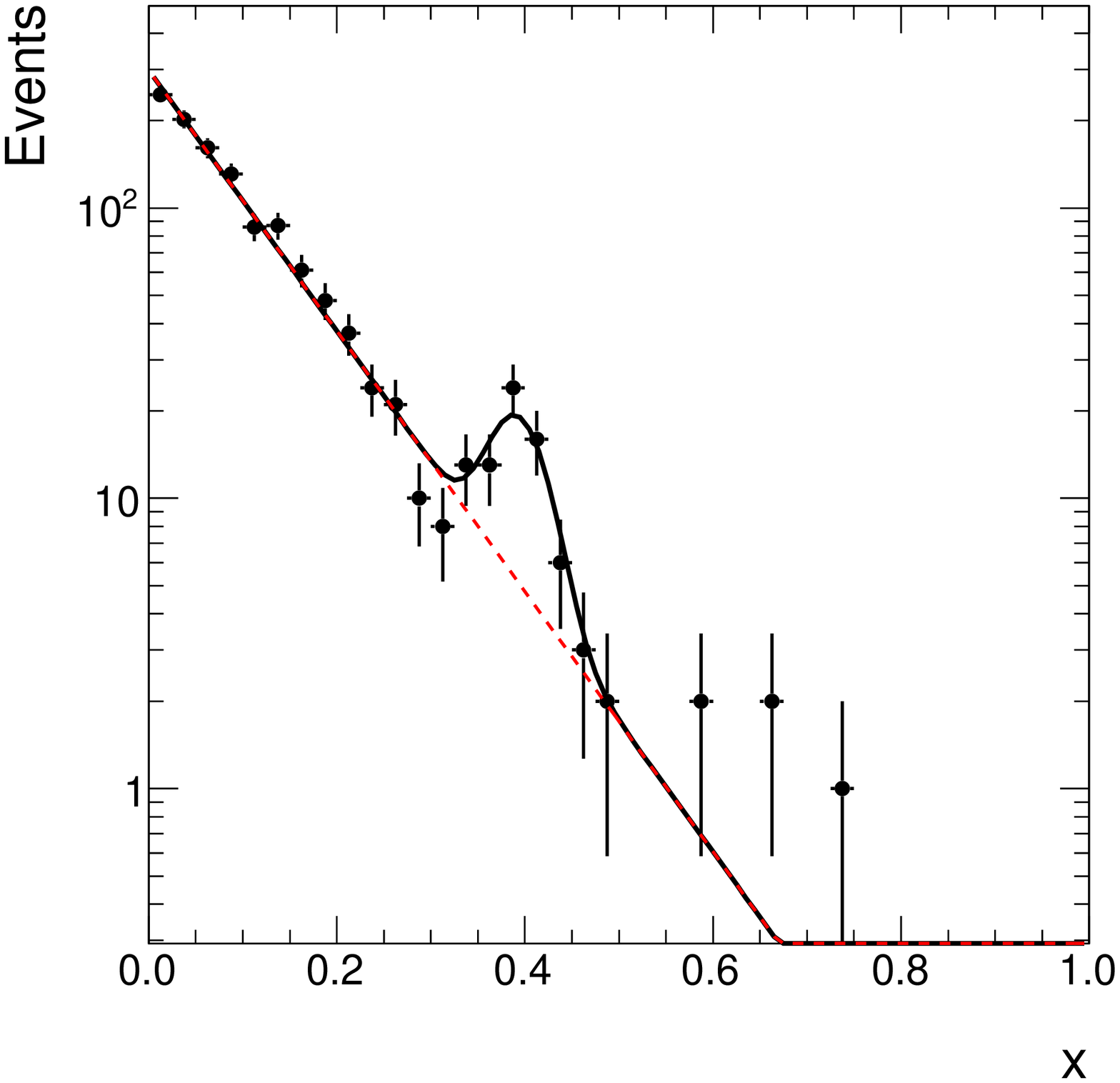} &
\includegraphics[width=0.25\textwidth]{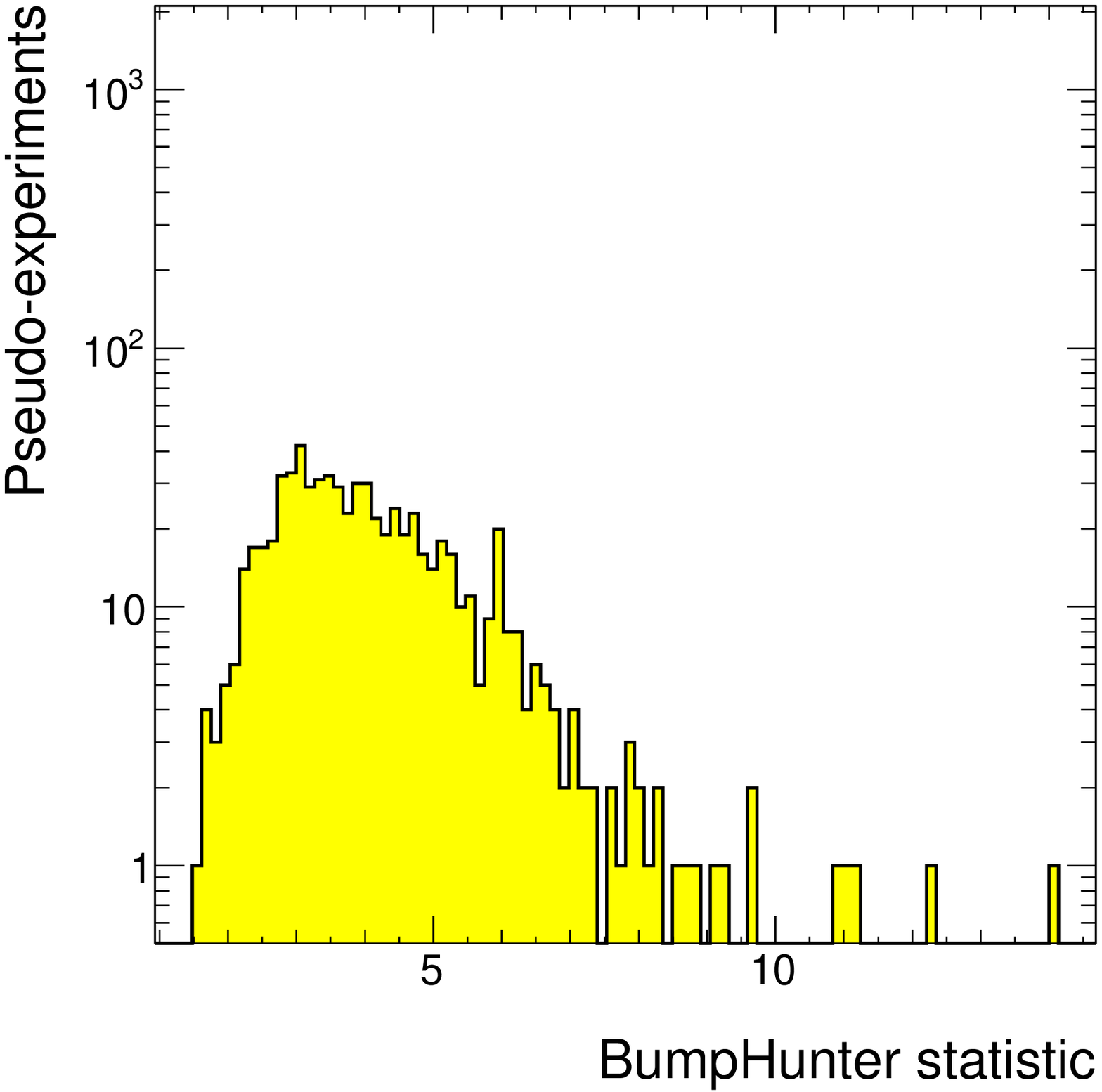}&
\includegraphics[width=0.25\textwidth]{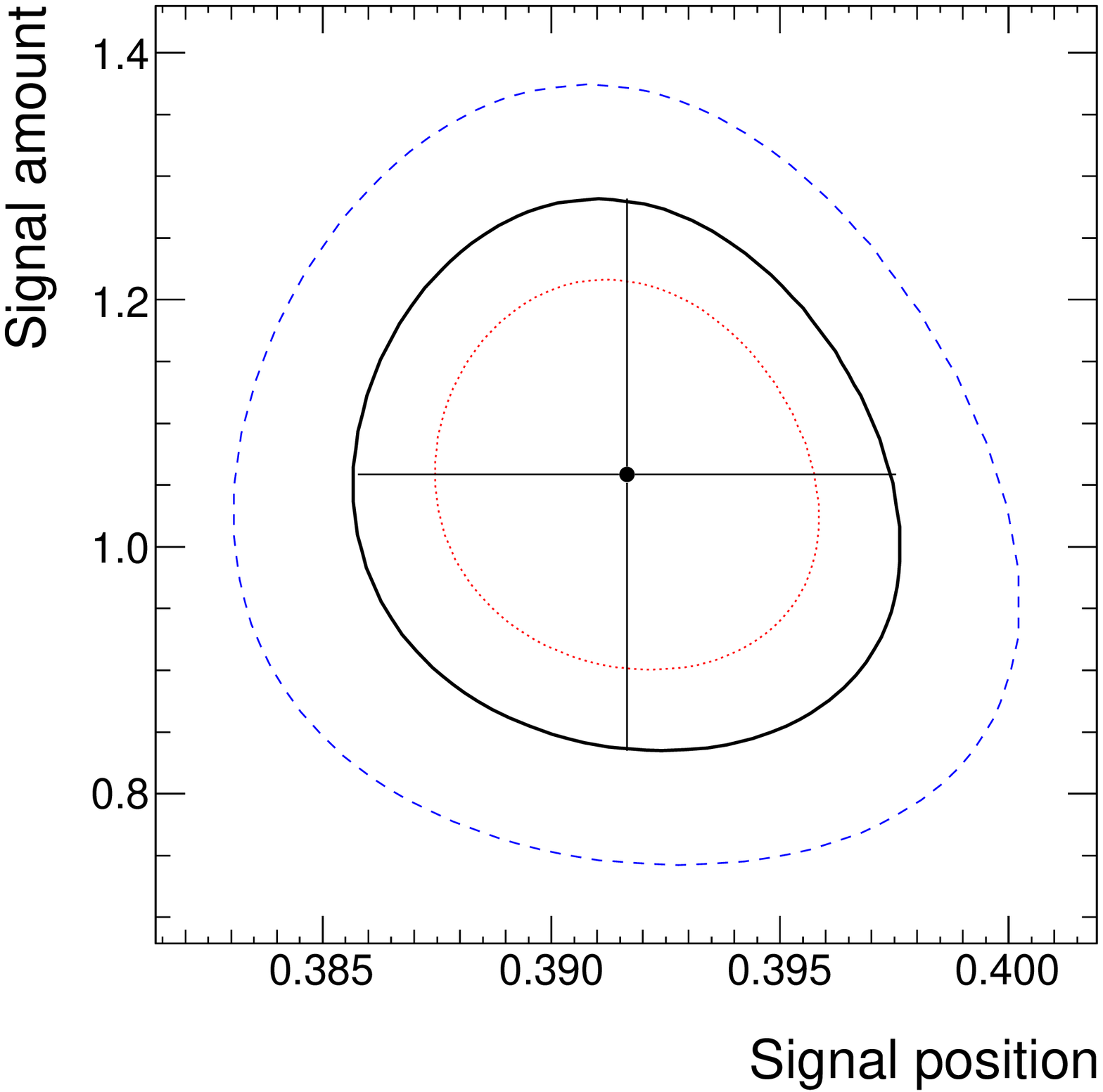} \\ 
\includegraphics[width=0.25\textwidth]{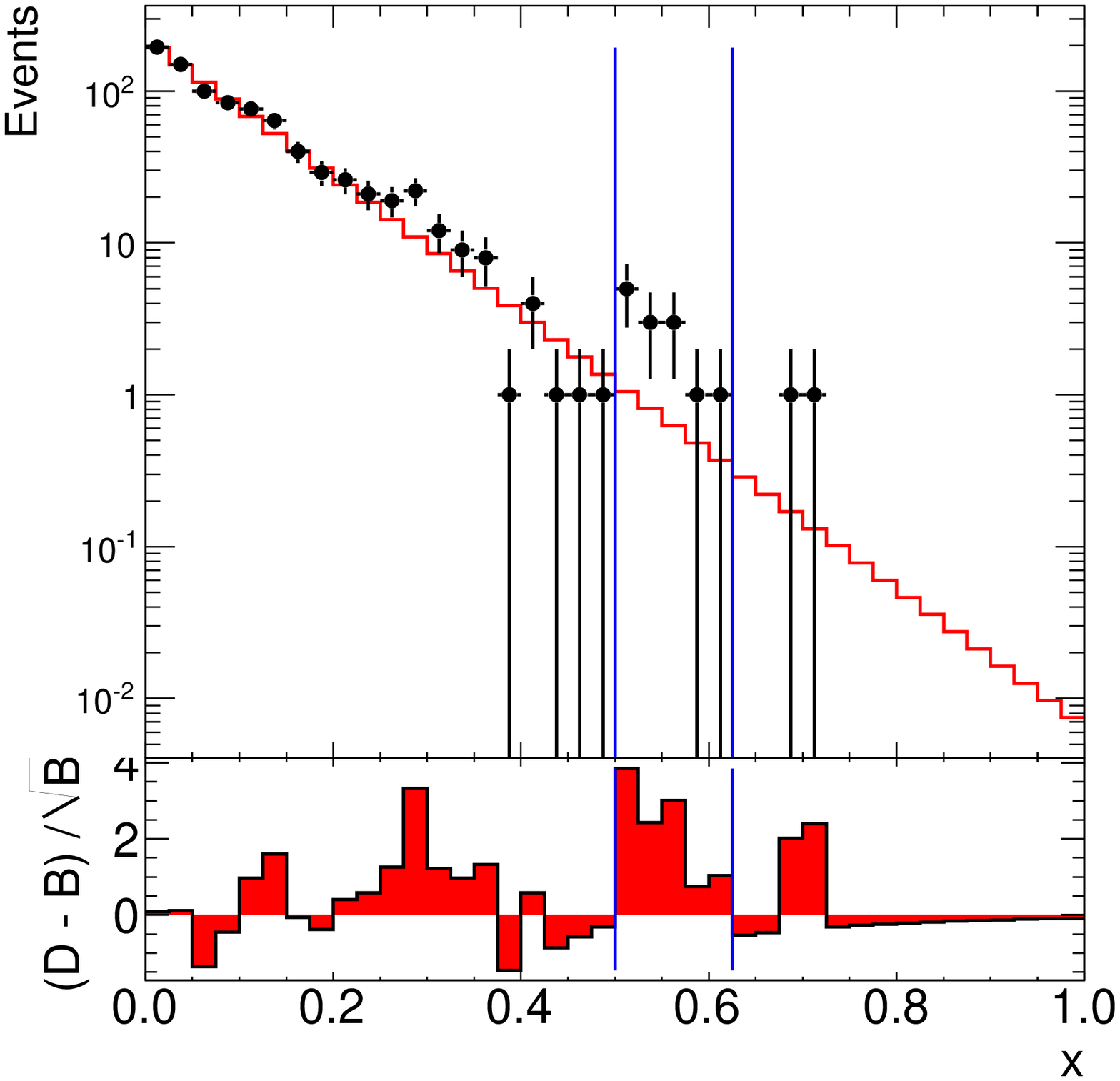}&
\includegraphics[width=0.25\textwidth]{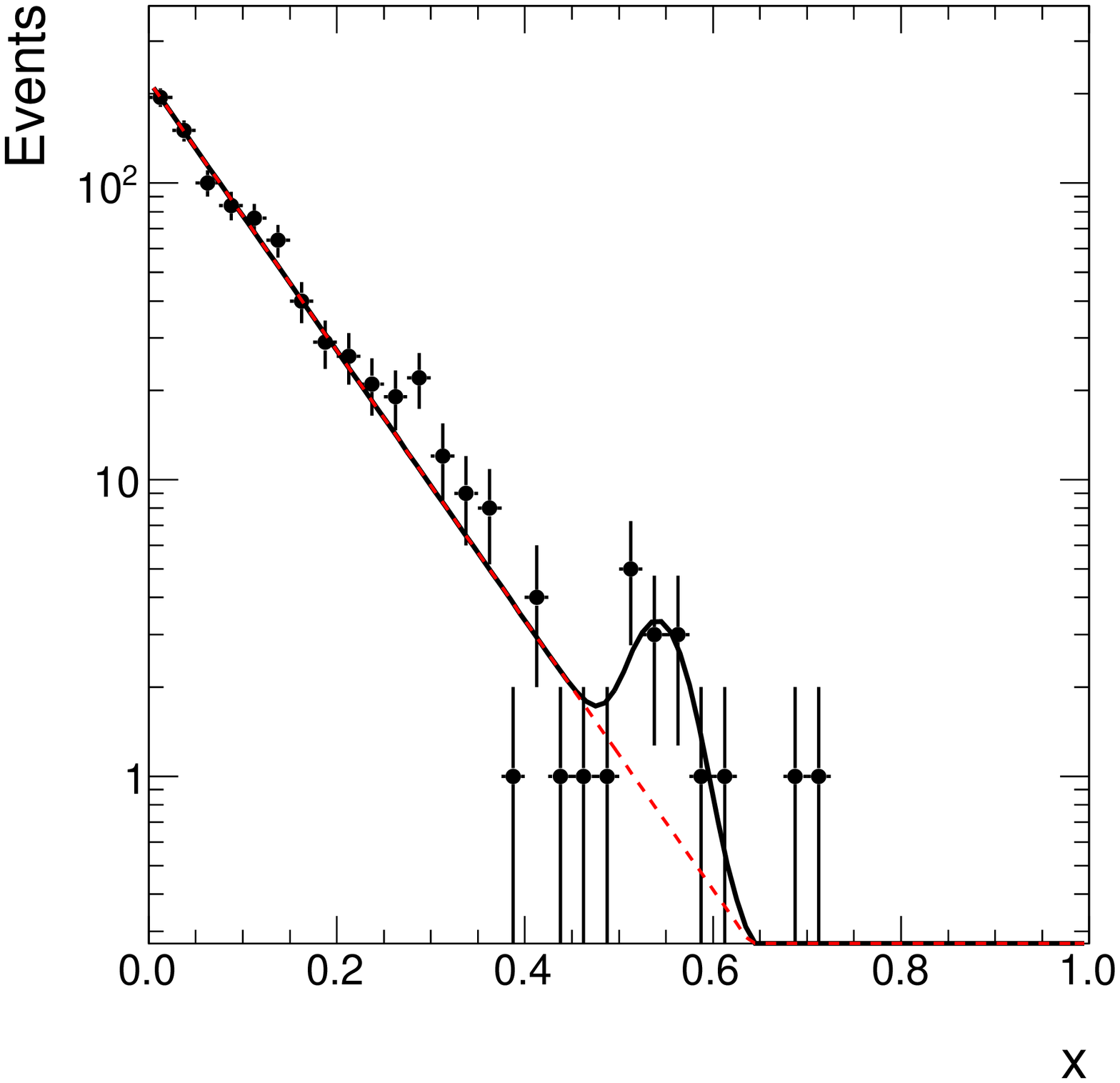} &
\includegraphics[width=0.25\textwidth]{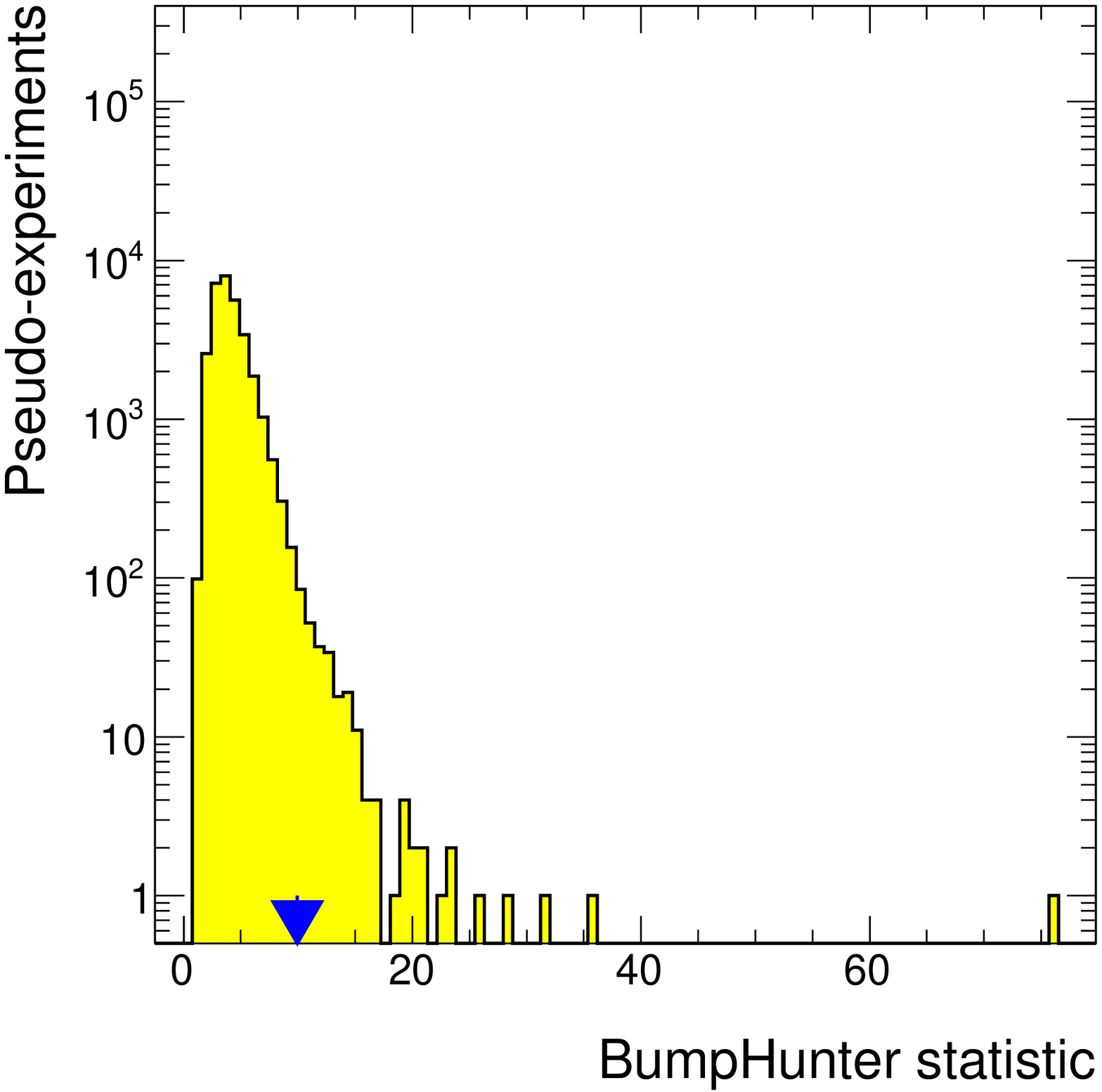}&
\includegraphics[width=0.25\textwidth]{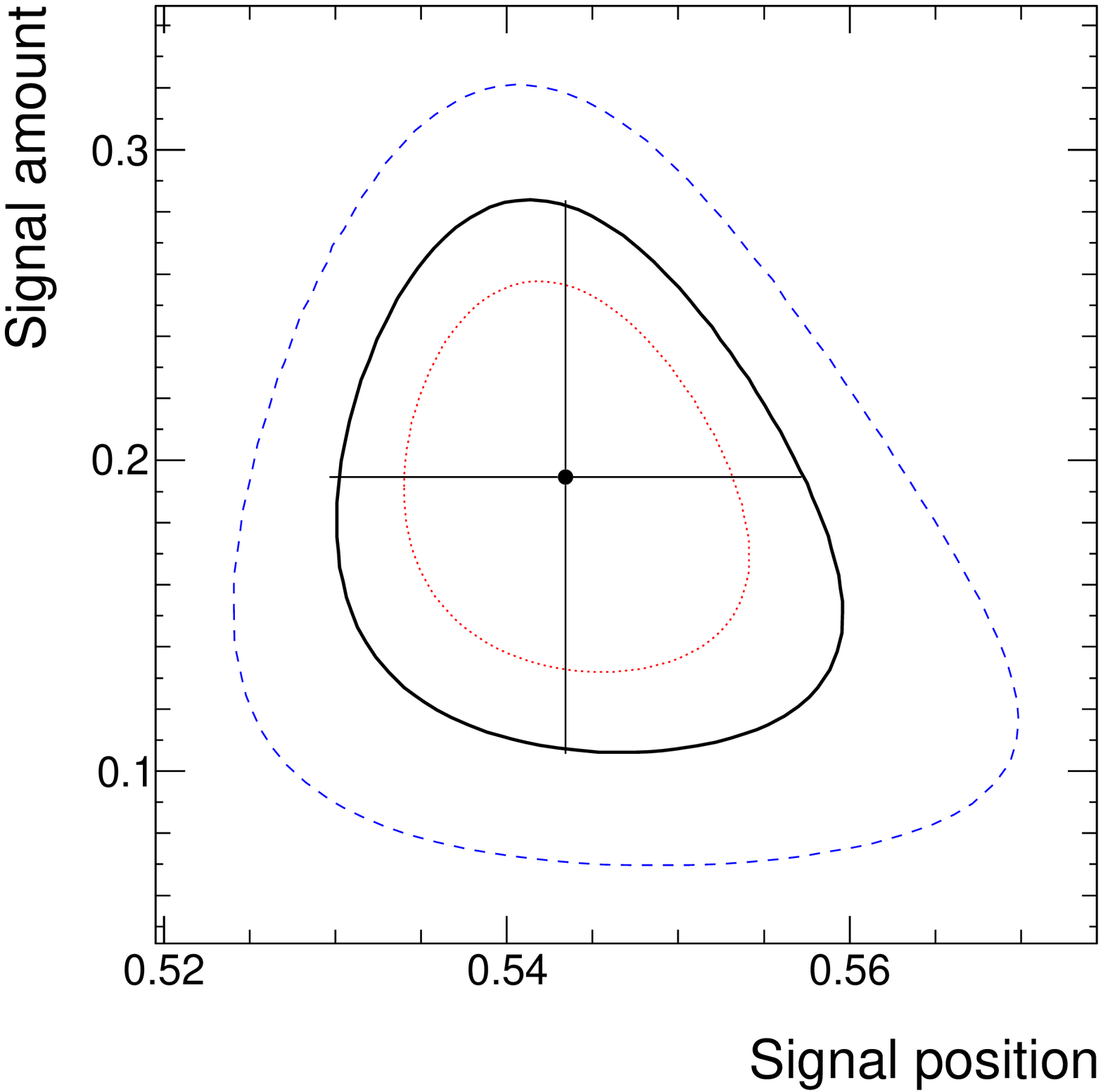} \\ 
\includegraphics[width=0.25\textwidth]{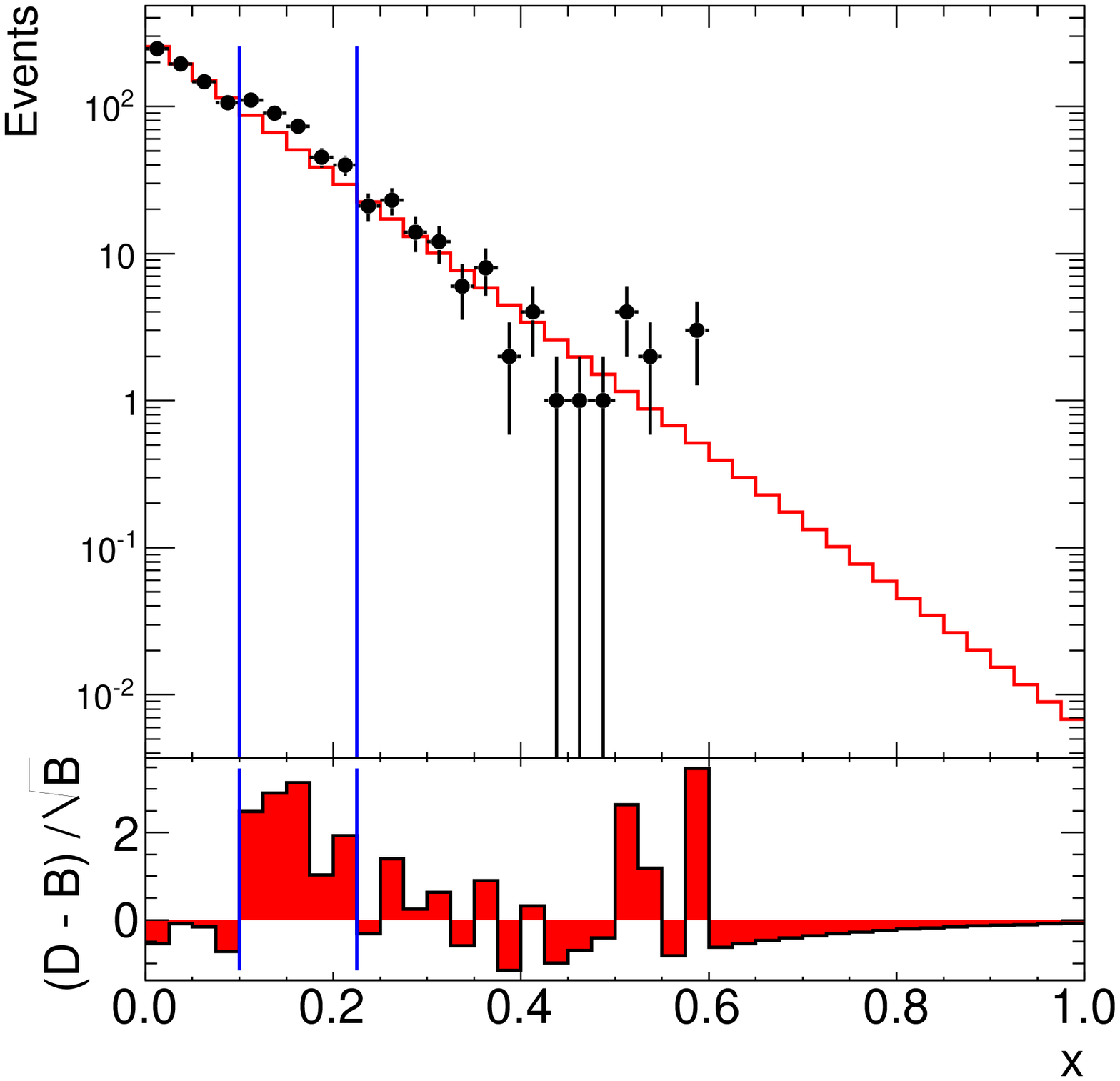}&
\includegraphics[width=0.25\textwidth]{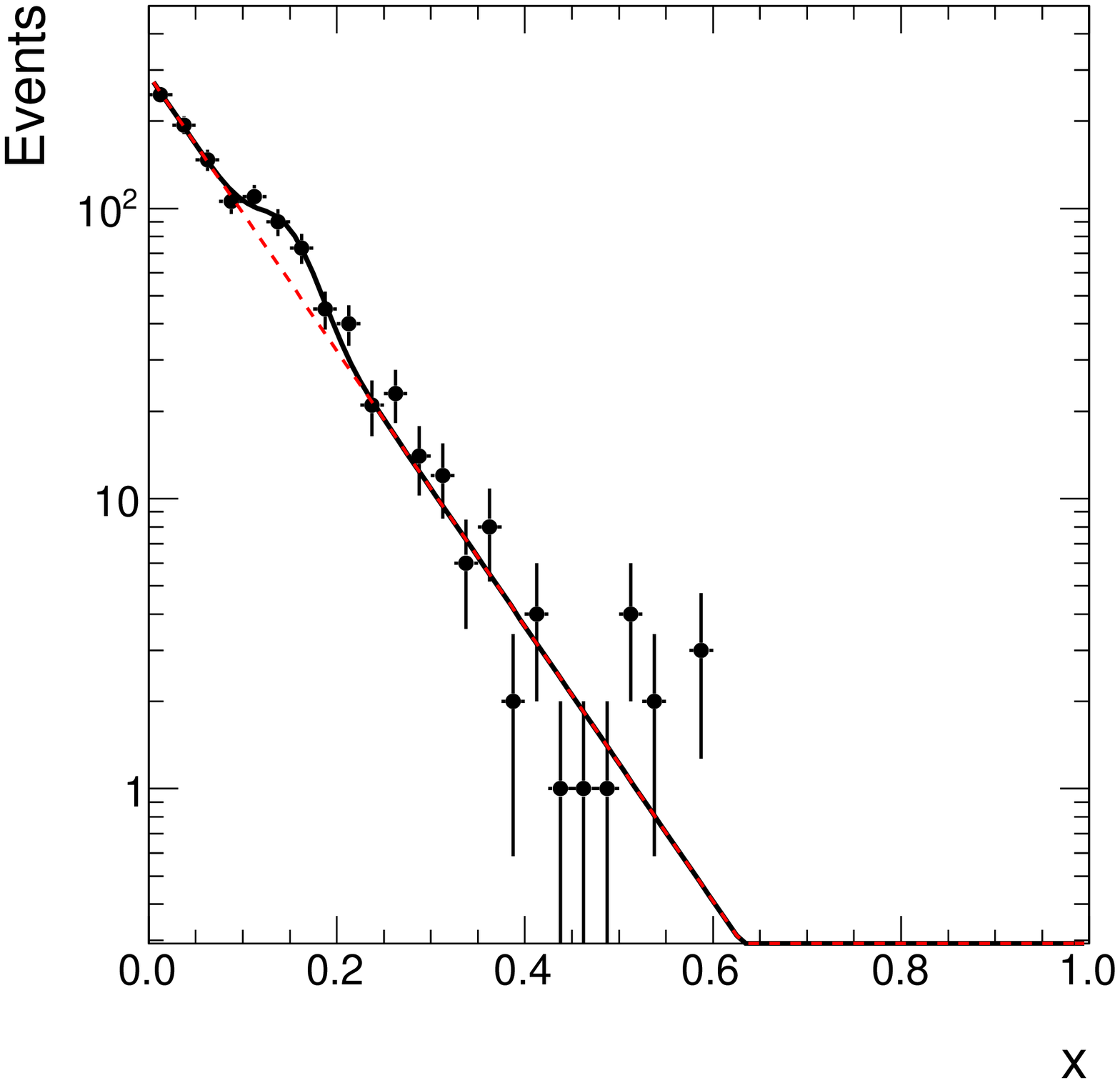} &
\includegraphics[width=0.25\textwidth]{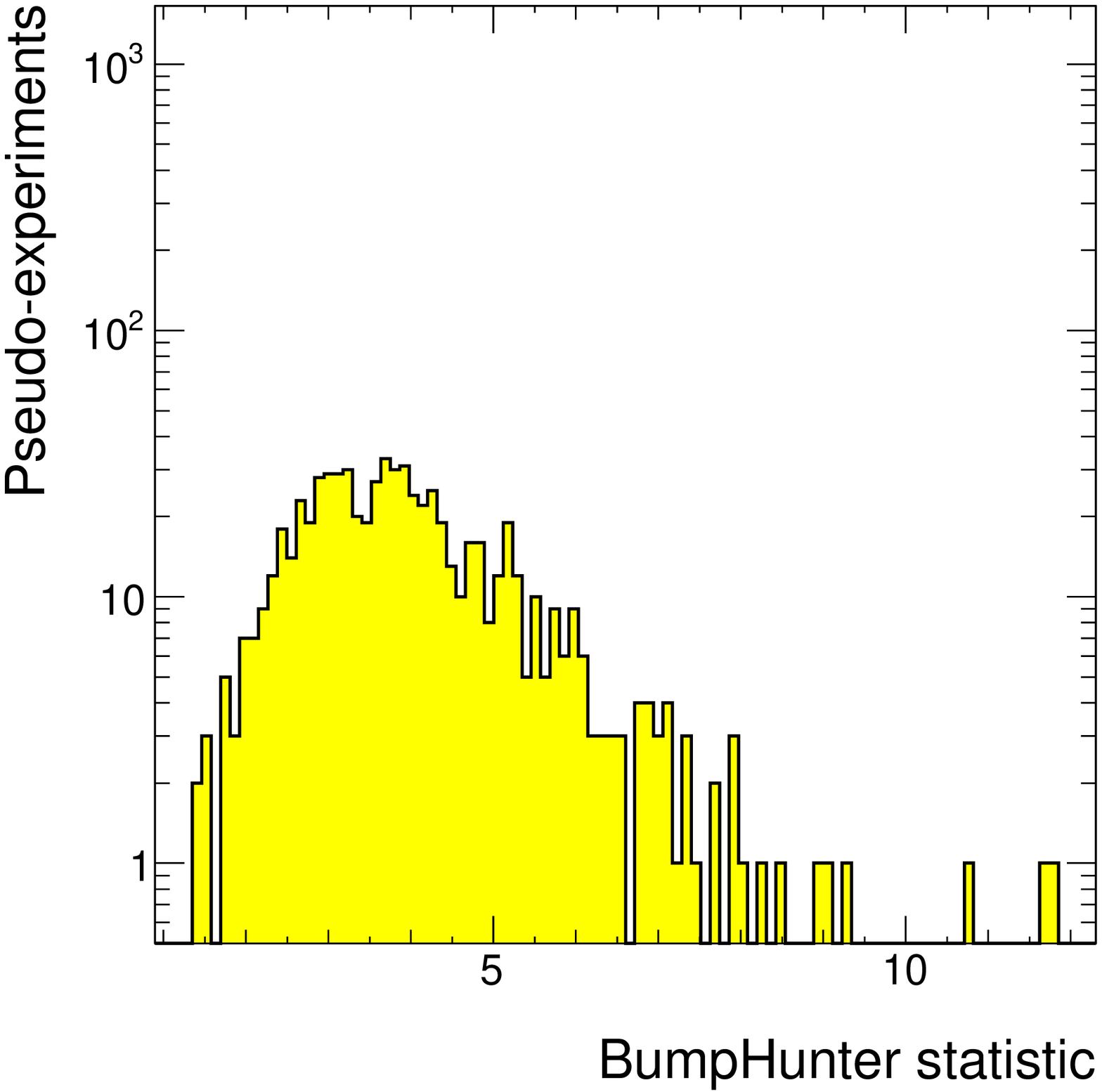}&
\includegraphics[width=0.25\textwidth]{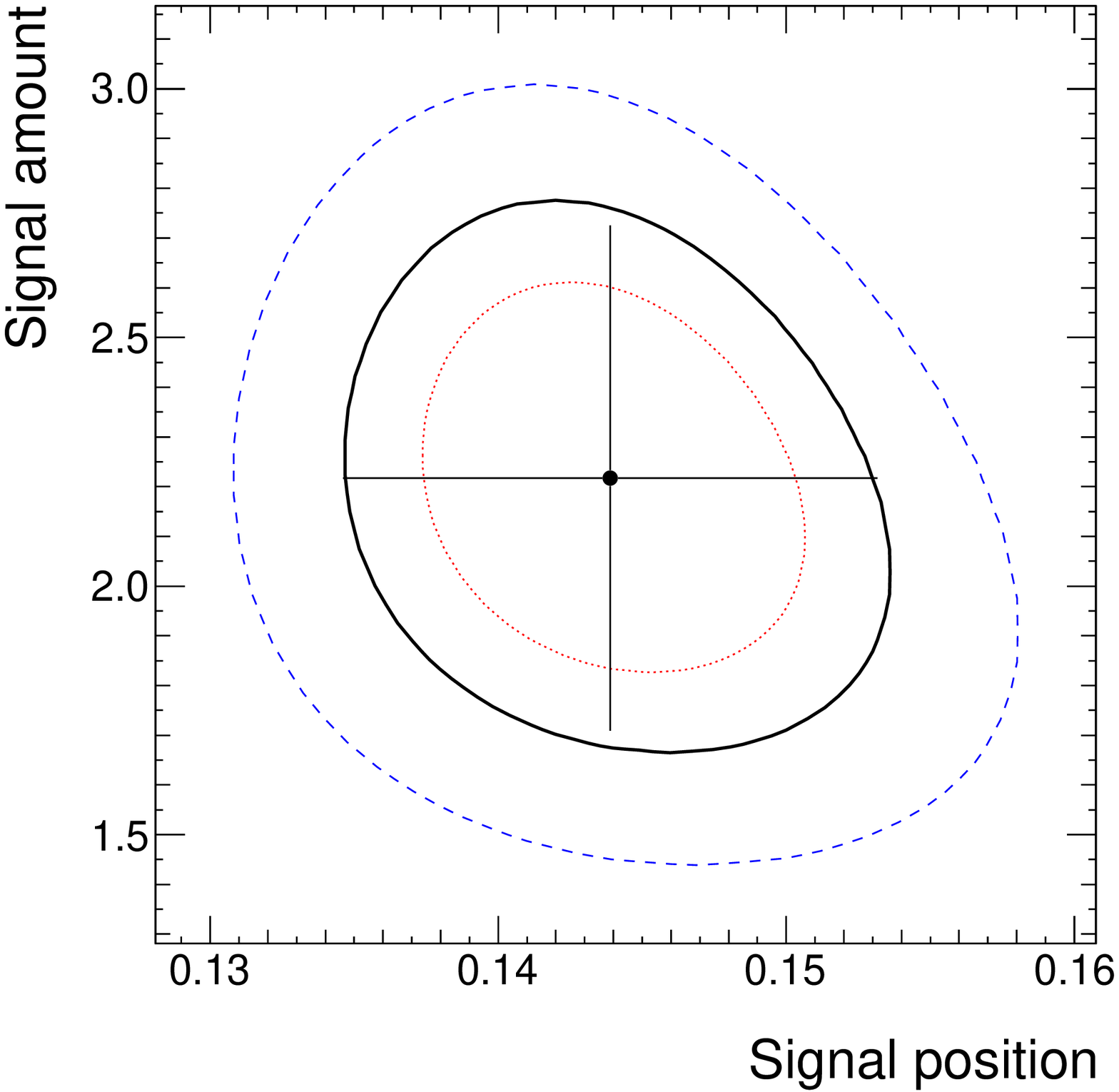} \\ 
\end{tabular}
\caption{\label{fig:appendixWithDiscovery} Summary of results from 5 Banff Challenge Problem 1 datasets, where a discovery was claimed.  The datasets are \{22, 25, 35, 41, 42\}. One row of figures corresponds to each.  In the 3$^{\rm rd}$ row, 3$^{\rm rd}$ column, the blue arrow is missing because the observed \bh statistic $t_o = 24.9$ is outside the plotted range.  The same happens in dataset 42, last row, with $t_o = 14.9$.  The corresponding most likely \pvals are: $\{\frac{11}{2250}, \frac{7}{1960}, \frac{0}{690}, \frac{266}{31080}, \frac{0}{690} \}$.}
\end{figure}

Fig.~\ref{fig:appendix1} and Fig.~\ref{fig:appendixWithDiscovery} show 10 more examples of datasets (5 with a discovery claim and 5 without).

\begin{figure}[p]
\centering
\subfigure[]{
\includegraphics[width=0.45\textwidth]{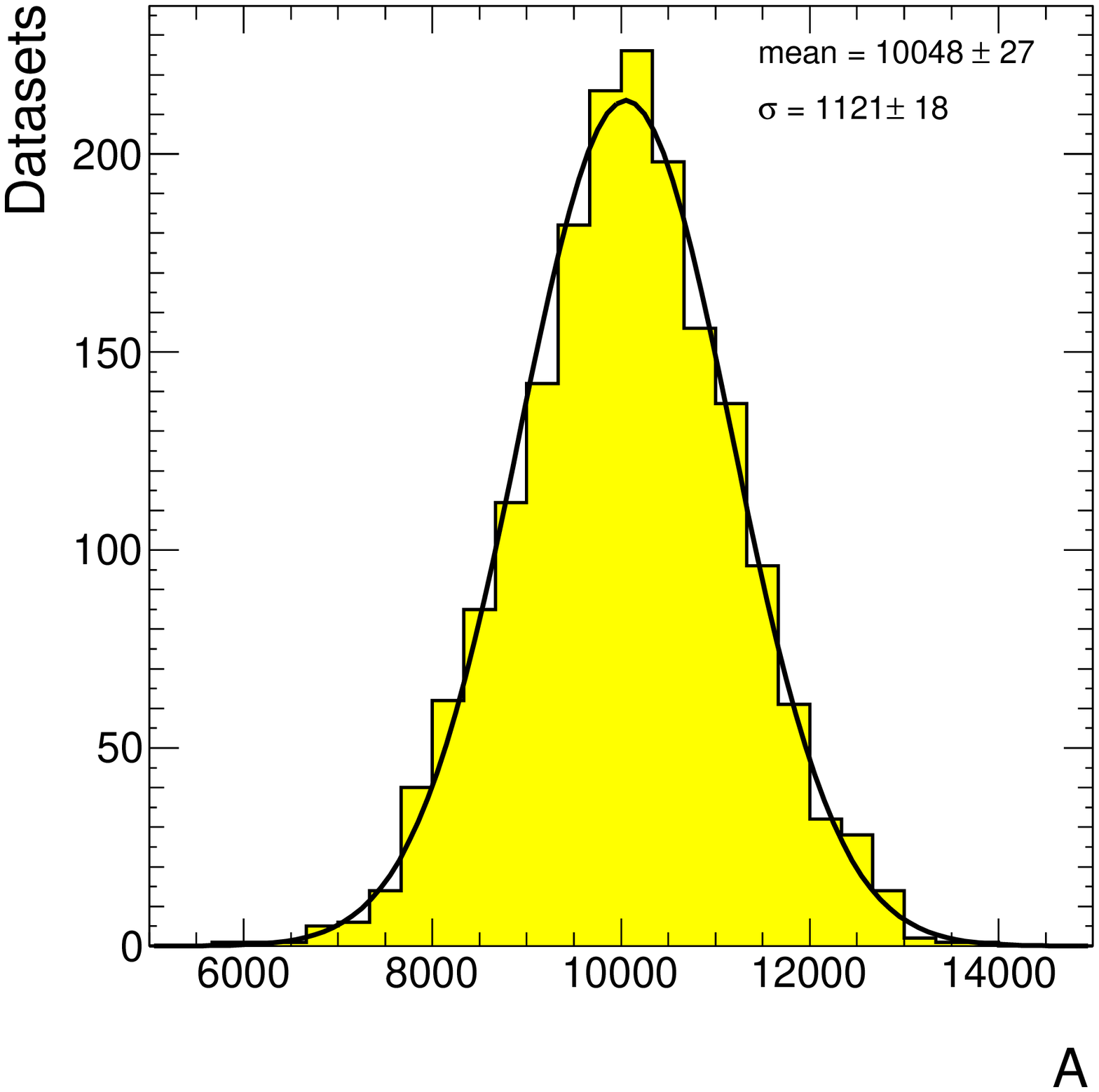}
\label{fig:figures/A.eps}
}
\subfigure[]{
\includegraphics[width=0.45\textwidth]{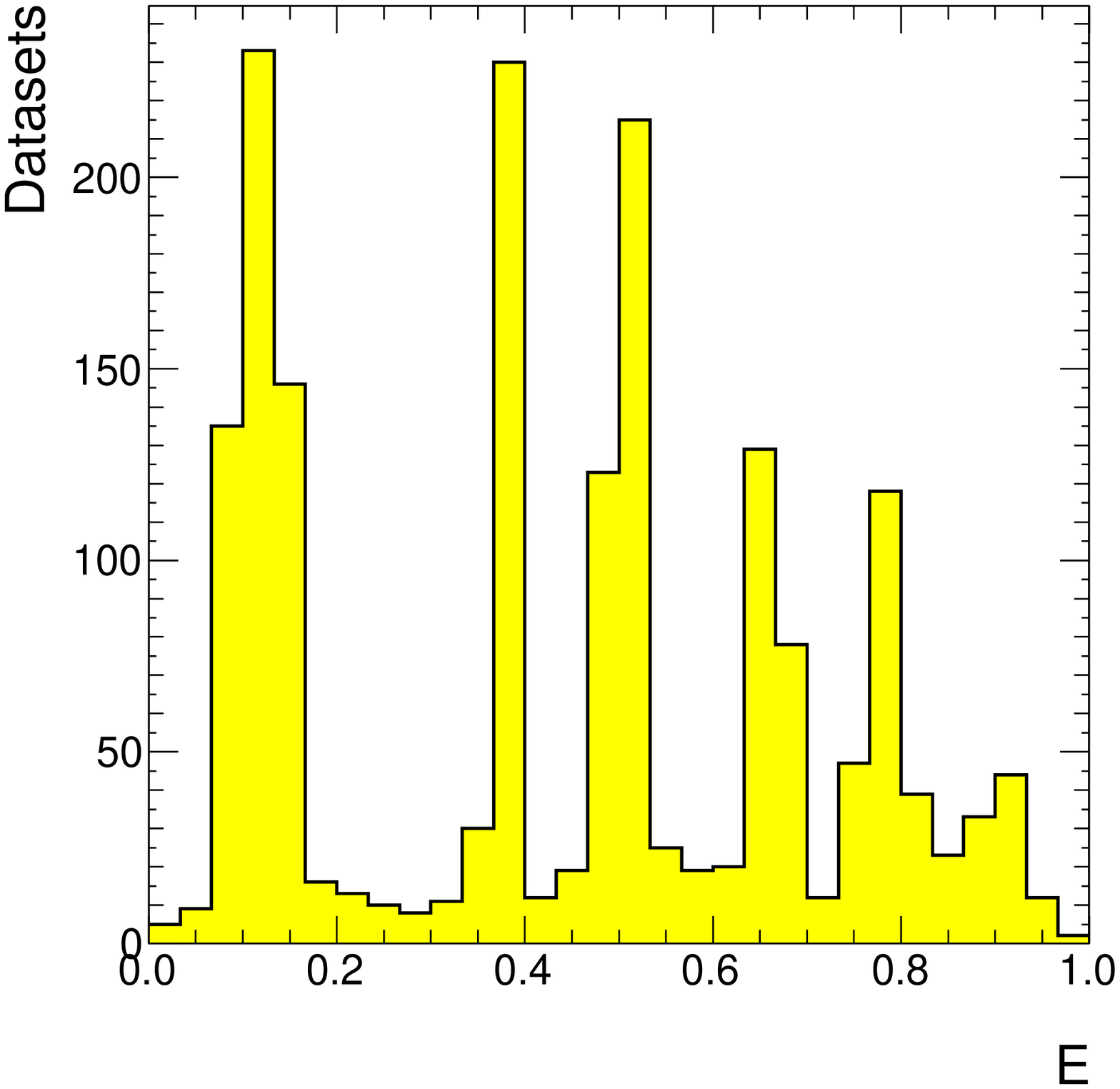}
\label{fig:figures/E.eps}
}
\subfigure[]{
\includegraphics[width=0.45\textwidth]{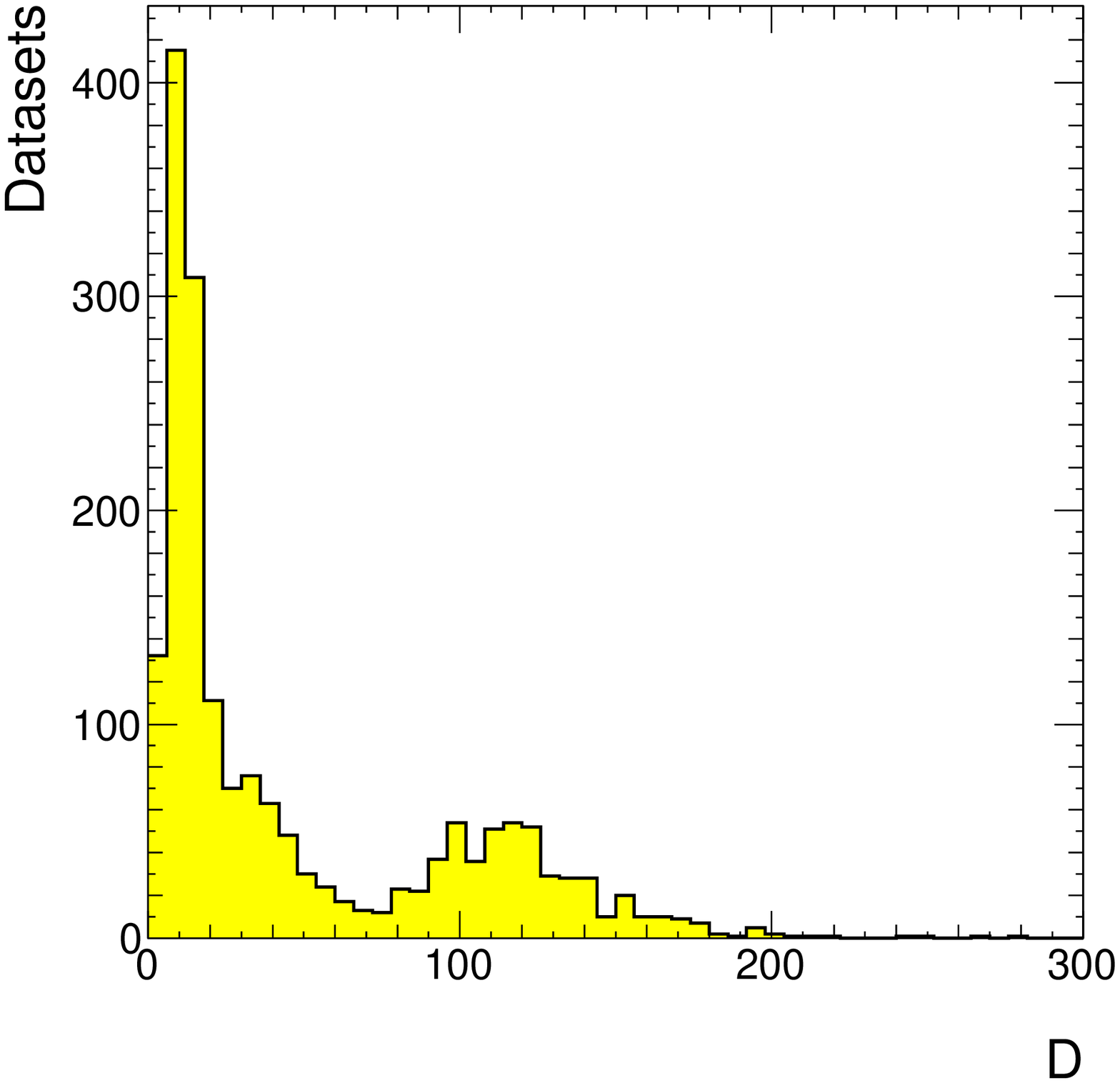}
\label{fig:figures/D.eps}
}
\subfigure[]{
\includegraphics[width=0.45\textwidth]{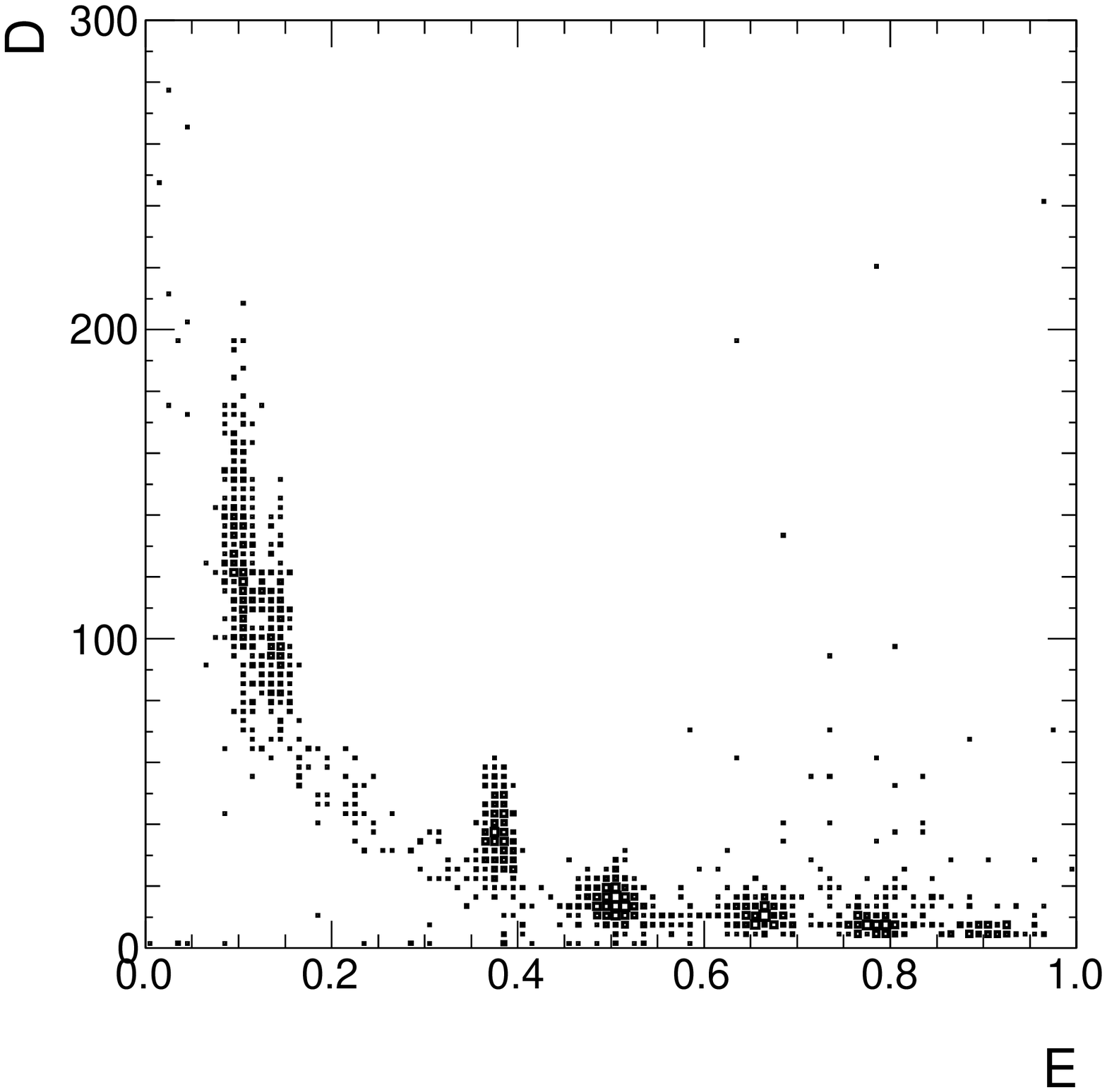}
\label{fig:figures/DE.eps}
}
\caption{\label{fig:parameters}  For the 1819 datasets where $\pval$ was estimated to be $\le 0.01$, the distributions of fitted parameters $\{A,E,D\}$ are shown, as well as the joint distribution of $D$ and $E$ in \subref{fig:figures/DE.eps}.  To remove the effect of binning, $A$ and $D$ are shown after dividing the fitted values by the bin size (0.025).  We have the information that $A$ is actually following a Gaussian of mean = $10^4$ and standard deviation = 1000, and we see that the $A$ we obtain with our procedure is distributed in a very similar way, even though this population of datasets is the subset where signal exists, therefore the estimation of $A$ becomes more challenging.}
\end{figure}

Fig.~\ref{fig:parameters} summarizes the best fitting values of $A$, $D$ and $E$ in just those 1711 pseudo-experiments where a discovery was claimed at the level of 0.01 Type-I error probability.

\section{Sensitivity}

\subsection{The Banff Challenge sensitivity tests}
\label{sec:tests}

The sensitivity of the \bh is measured in three signal cases, as required by the Banff Challenge.  ``Sensitivity'' means the probability of observing a $\pval \le 0.01$ in the presence of a specific amount and kind of signal.  In all signal cases, the signal is injected in the nominal background distribution, which comes from integrating $10^4 e^{-10 x}$ in each $x$ bin.  In all cases, the signal is given by a function $D e^{-\frac{(x-E)^2}{2\cdot 0.03^2}}$.

\begin{figure}[p]
\centering
\subfigure[]{
\includegraphics[width=0.3\textwidth]{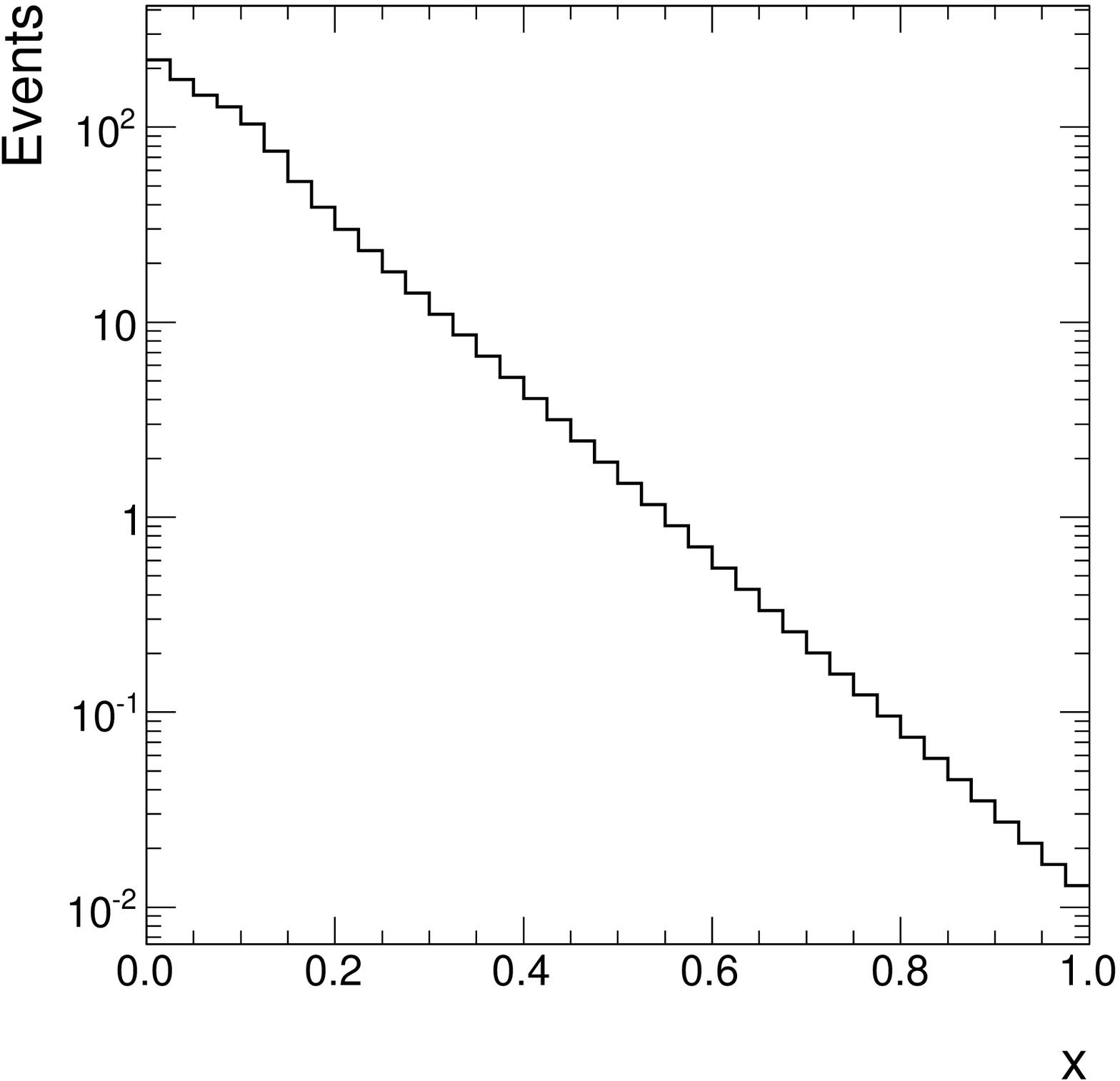}
\label{fig:distr1}
}
\subfigure[]{
\includegraphics[width=0.3\textwidth]{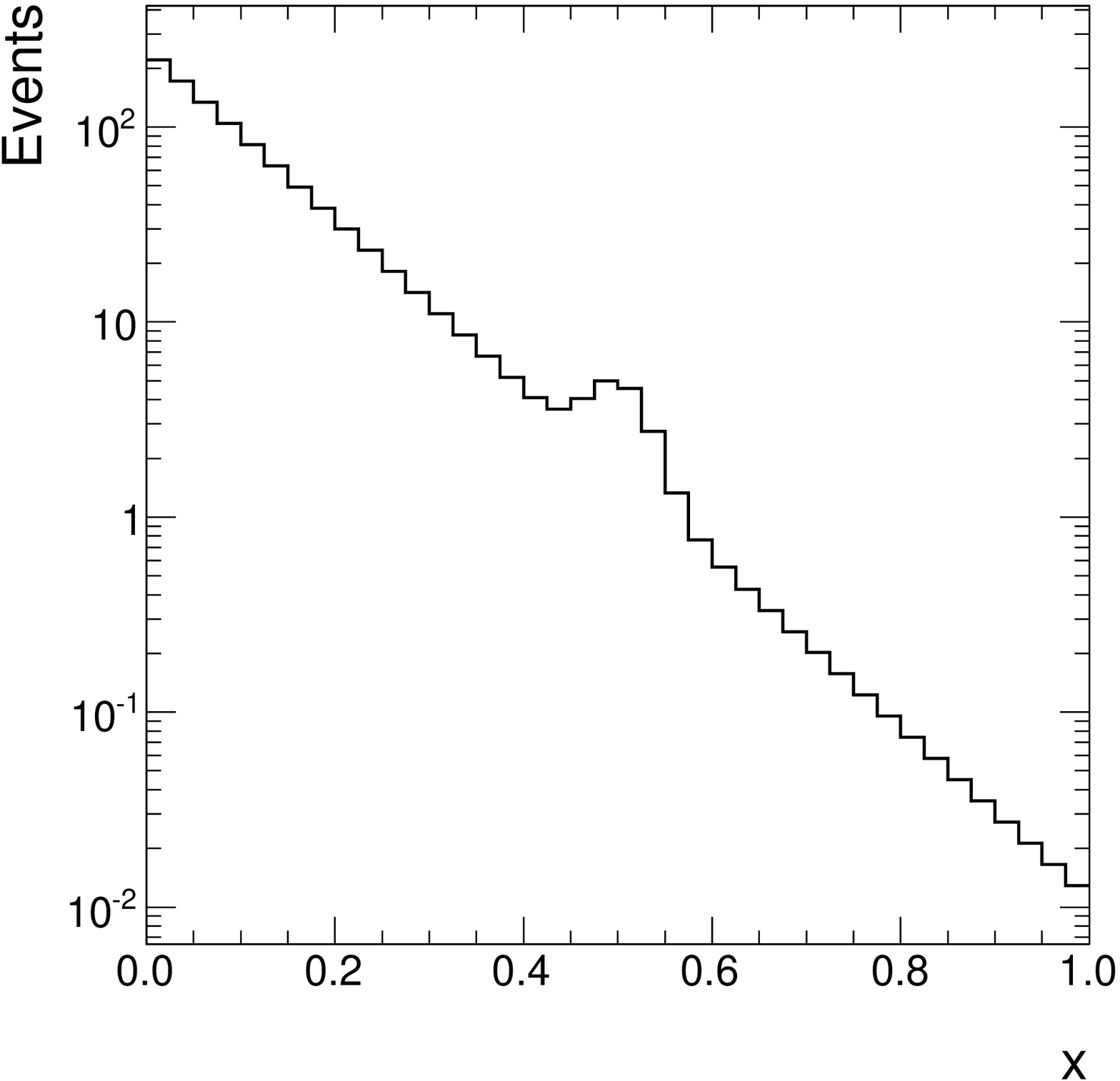}
\label{fig:distr2}
}
\subfigure[]{
\includegraphics[width=0.3\textwidth]{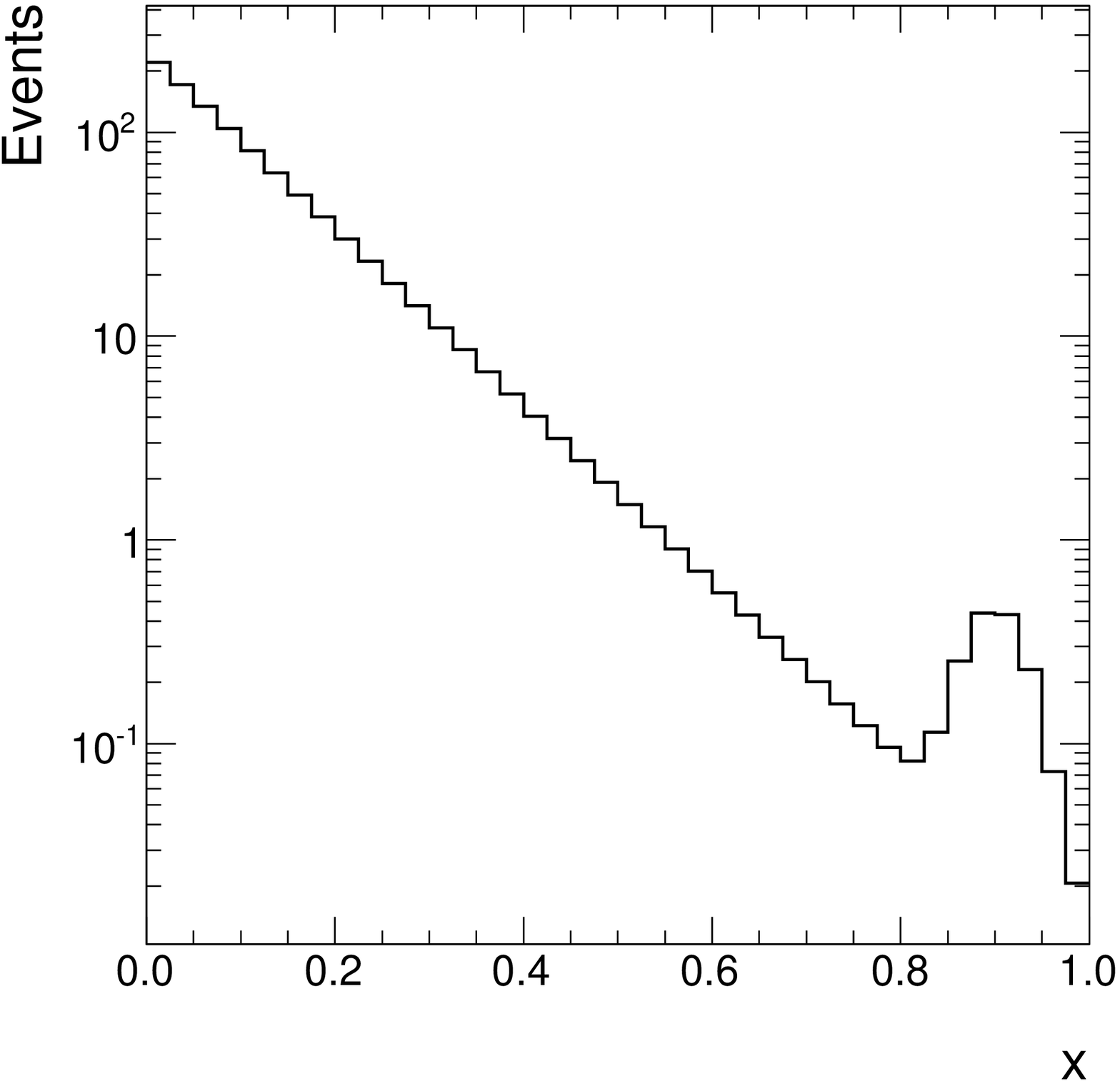}
\label{fig:distr3}
}\\
\subfigure[]{
\includegraphics[width=0.3\textwidth]{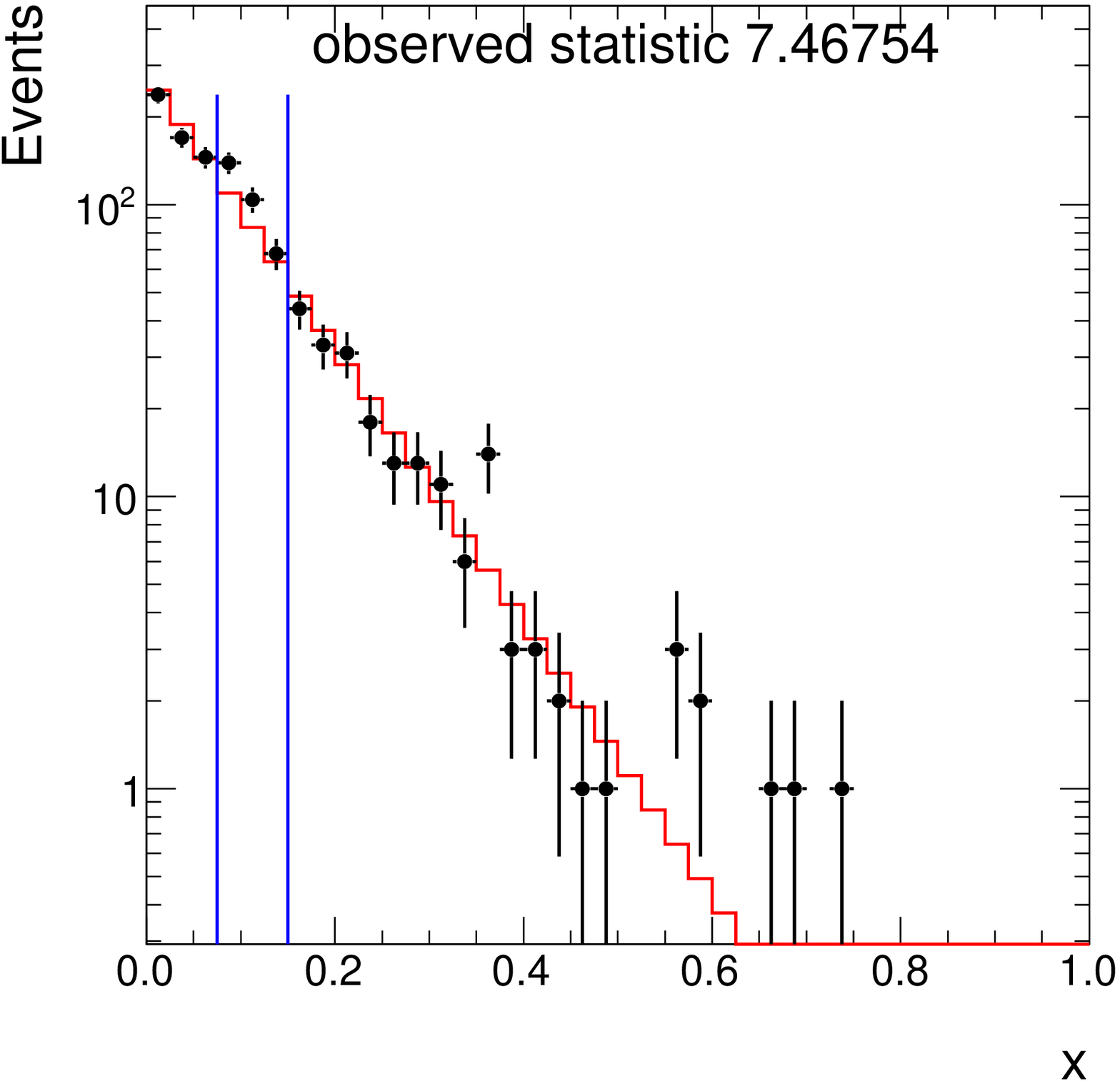}
\label{fig:example1}
}
\subfigure[]{
\includegraphics[width=0.3\textwidth]{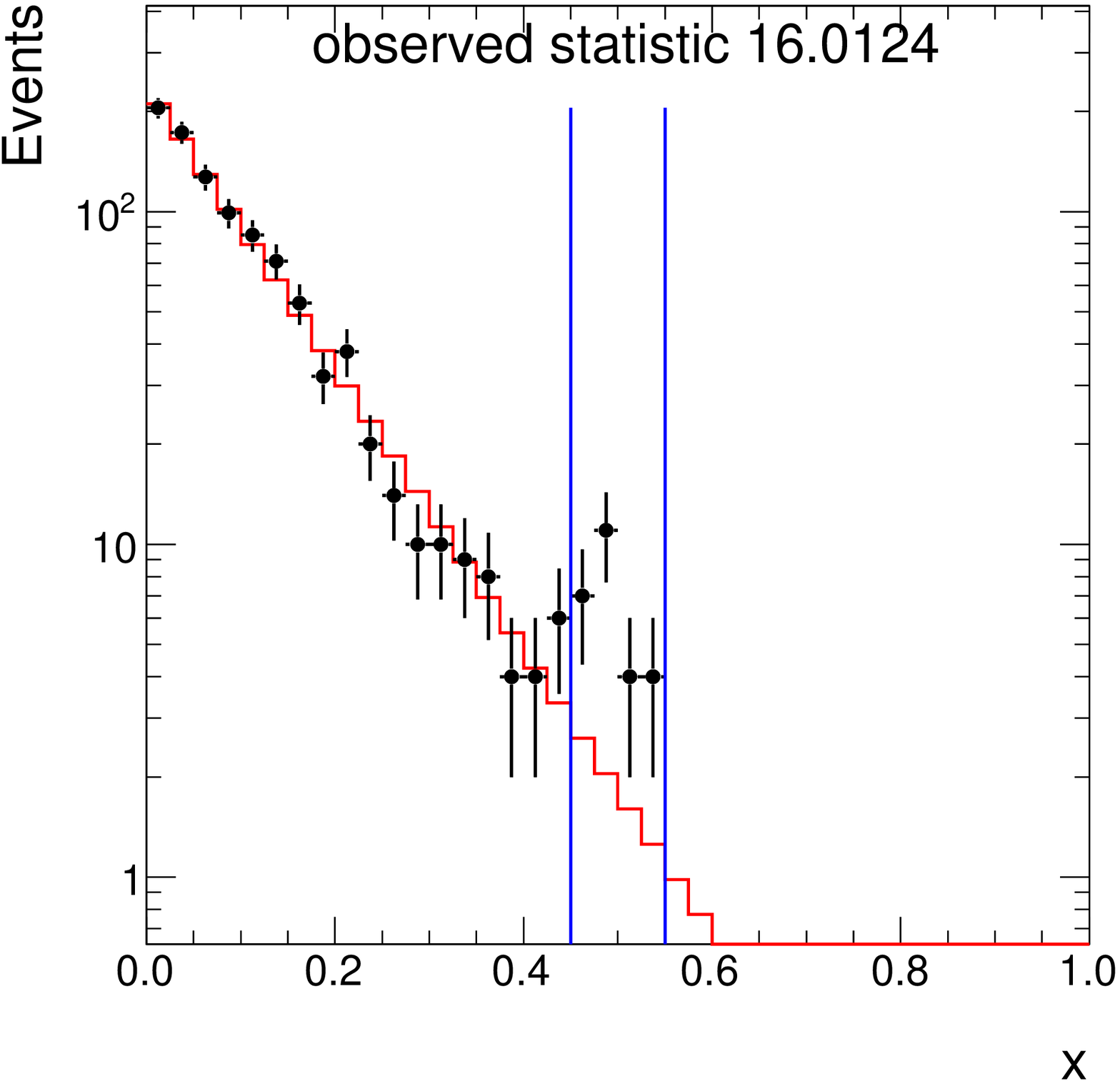}
\label{fig:example2}
}
\subfigure[]{
\includegraphics[width=0.3\textwidth]{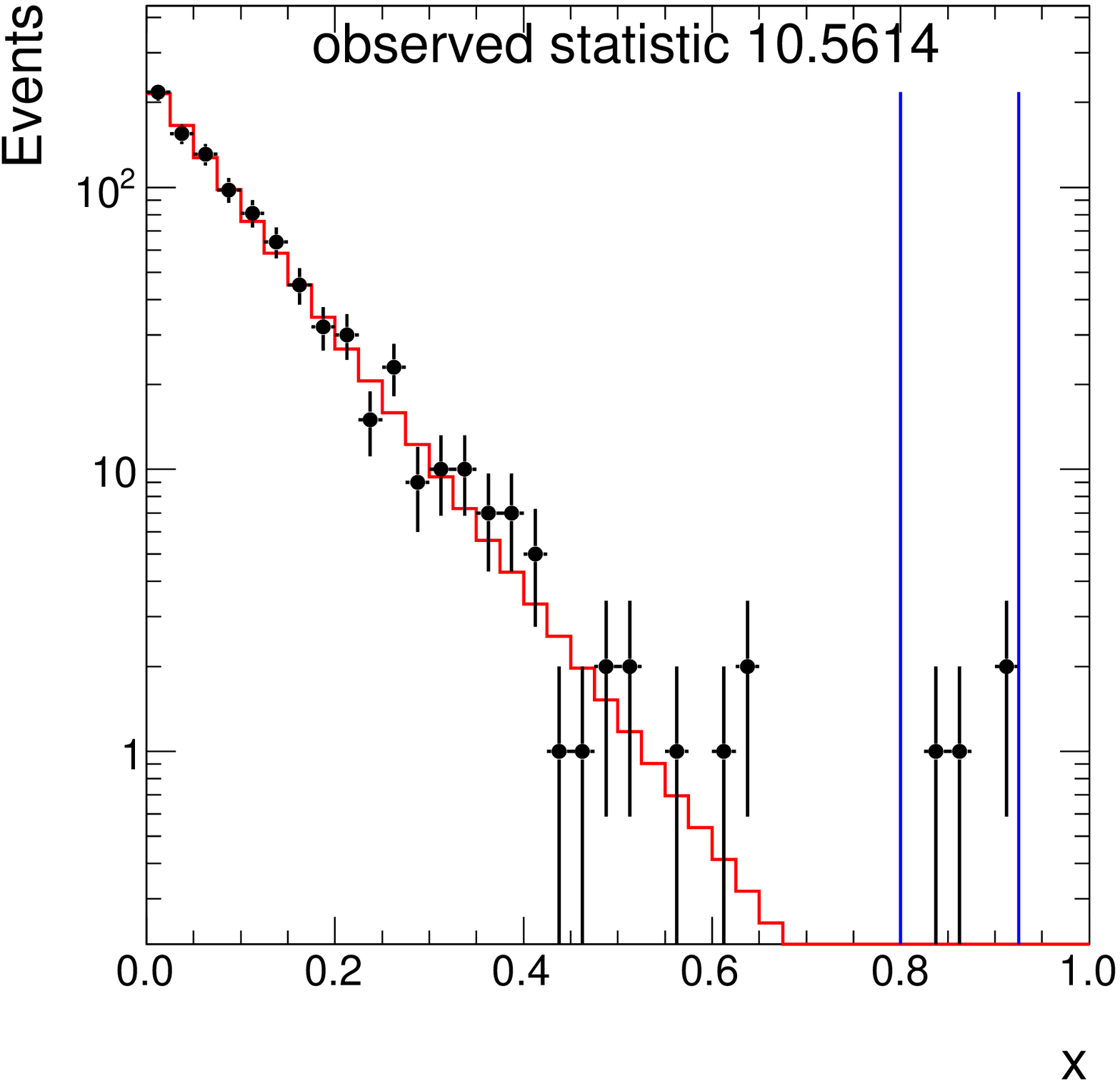} 
\label{fig:example3}
}\\
\caption{\label{fig:test}  The expected spectrum in each of the three sensitivity tests listed in paragraph \ref{sec:tests}.   In each case an example of pseudo-data is shown, with the blue vertical lines bracketing the central window of the most discrepant bump found in each pseudo-spectrum.}
\end{figure}

In the first test, we have $\{D,E\} = \{1010,0.1\}$.  Integrating the signal function in $x\in[0,1]$, we have a total of 75.9 events.  Out of 300 pseudo-experiments, generated from the distribution in Fig.~\ref{fig:test}\subref{fig:distr1}, the \bh \pval was less than 0.01 in 64 pseudo-experiments.  That implies discovery probability of about 21.3\%.

The results of the second and third test are summarized in Table~\ref{table:sensitivity1}.

Fig.~\ref{fig:test} summarizes the expected distributions in the three sensitivity tests, and shows an example of pseudo-data from each expected spectrum.  

\subsection{Comparison to the case of known signal shape and position}
\label{sec:compareToTargetted}

For the sake of comparison, what would our sensitivity be if we knew the location of the signal and its exact shape, and we only ignored its amount (which is proportional to $D$)?
In that case, obviously, the \bh would be unnecessary; why look at many places, and pay the penalty of the trials factor, when knowing exactly where the signal is?  

In that ideal case, we could compare the null hypothesis to the hypothesis which includes the specific signal and best fits the data.  We could define as test statistic the ``log likelihood ratio'':
\begin{equation}
t = \log \frac{L(\text{Data}|\hat{D})}{L(\text{Data}|D=0)},
\end{equation}
where  $L(\text{Data}|D)$ is the probability of observing the data, bin by bin, assuming the given signal shape with parameter $D$, and $\hat{D}$ is the value of $D$ which maximizes this likelihood.

Running this hypothesis test, we found that in the first test we found a $\pval \le 0.01$ in $\frac{175}{300}$ pseudo-experiments (probability about 58\%).  In the second test the same success rate was $\frac{173}{300} \simeq 58$\%.  In the third test, the result was $\frac{112}{300} \simeq 37$\%.  These numbers are added to Table~\ref{table:sensitivity1} as an extra column.

Comparing these success rates to the ones mentioned in paragraph~\ref{sec:tests}, one confirms that the \bh is less sensitive than a test to which the location and shape of the signal have been disclosed.  This lower sensitivity is a consequence of the greater trials factor in the \bh, as expected from the discussion in paragraph~\ref{sec:howToAccount}.  Nevertheless, in research one doesn't know in advance what he is going to discover, unless some confirmation is sought instead of discovery.   Between the less sensitive \bh, which covers a large range of possibilities, and an arbitrary hypothesis test that is sensitive to just one arbitrary signal and insensitive to almost everything else, the \bh seems to be a better choice.

\begin{table}
\centering
  \begin{tabular}{ccc|c|c}
    $E$ & $D$ & Total signal & \bh $P(\pval < 0.01)$ & likelihood ratio test \\ \hline
    0.1  & 1010 & 75.9 & 64/300 $\simeq$ 21.3\% & 175/300 $\simeq$ 58\% \\
    0.5  & 137 & 10.3 & 87/300 $\simeq$ 29.0\% & 173/300 $\simeq$ 58\% \\
    0.9  & 18 & 1.35 & 32/300 $\simeq$ 10.7\% & 112/300 $\simeq$ 37\% \\ 
  \end{tabular}
\caption{\label{table:sensitivity1} Summary of \bh sensitivity to the three tests posed by the Banff Challenge.  The last column shows, for comparison, the results of the targeted likelihood ratio test of paragraph~\ref{sec:compareToTargetted}.}
\end{table}

\subsection{Sensitivity of different tunings, without re-fitting}
\label{sec:powers}

In this section we will compare the sensitivity of the \bh when it is tuned in the following ways:  
\begin{enumerate}
  \item \label{tuning1} Not using sidebands criteria, and trying all window sizes, as described in paragraph~\ref{sec:bhAlgo}.
  \item \label{tuning2} Not using sidebands criteria, and constraining the window size between 3 and 5 bins.  This is the tuning used to address the Banff Challenge, as described in paragraph~\ref{sec:bhAlgo}.
  \item \label{tuning3} Using sidebands criteria, and trying all window sizes, as described in paragraph~\ref{sec:bhAlgo}.  This tuning was used in \cite{dijetResonanceSearch}.
  \item \label{tuning4} Using sidebands criteria, and constraining the window size between 3 and 5 bins. 
\end{enumerate}

In this paragraph, \Ho is not obtained by re-fitting eq.~\ref{eq:BanffBkg} to the data (or pseudo-data), but is always the same spectrum, which corresponds to $10^4 e^{-10x}$.

The sensitivity of the various \bh tunings are compared to that of the targeted test of paragraph~\ref{sec:compareToTargetted}.  The sensitivity of Pearson's traditional $\chi^2$ is also shown, where the test statistic is that of eq.~\ref{eq:chi2stat}.

\begin{figure}[p]
\centering
\subfigure[]{
\includegraphics[width=0.4\textwidth]{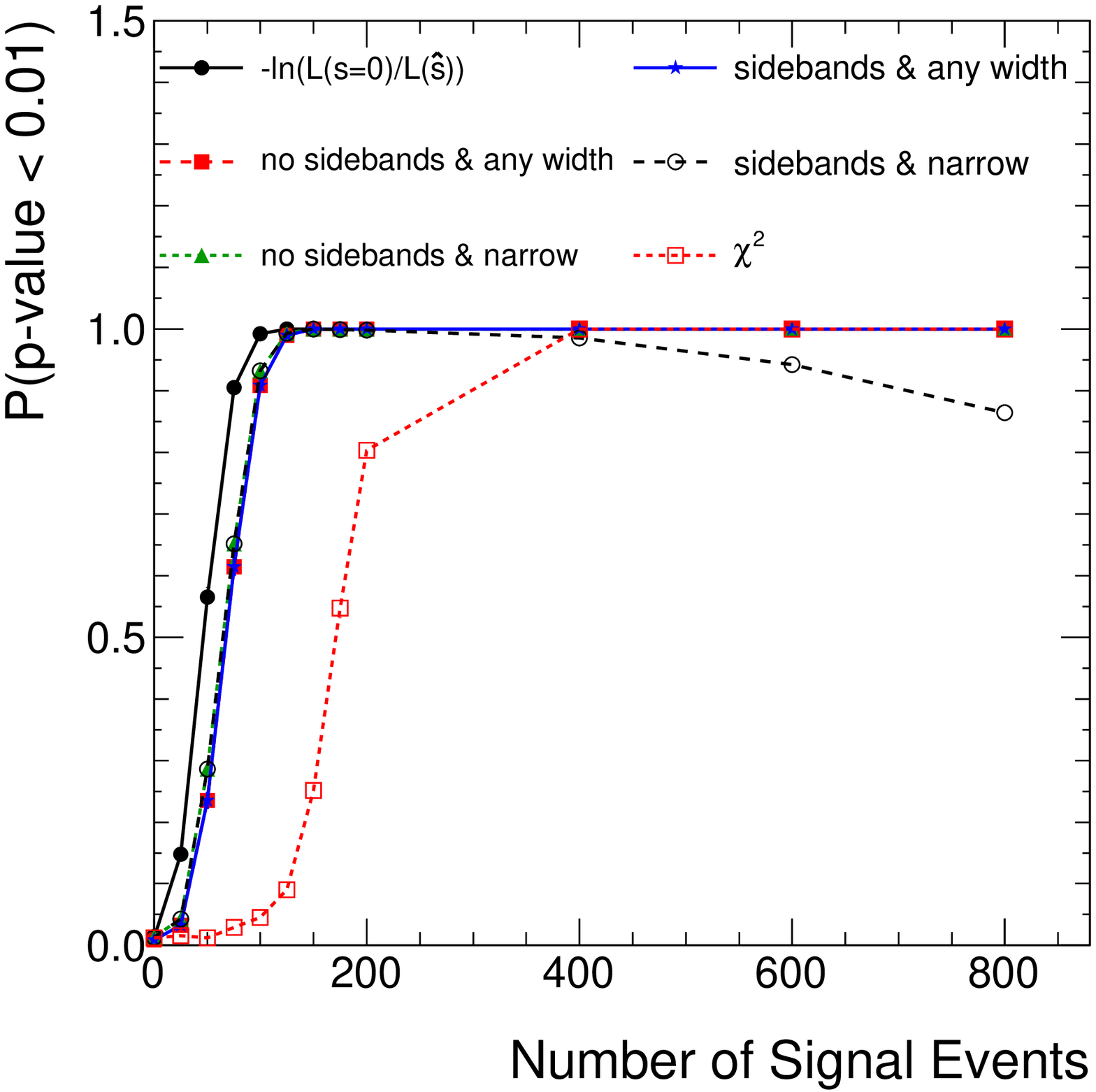}
\label{fig:power01}
}
\subfigure[]{
\includegraphics[width=0.4\textwidth]{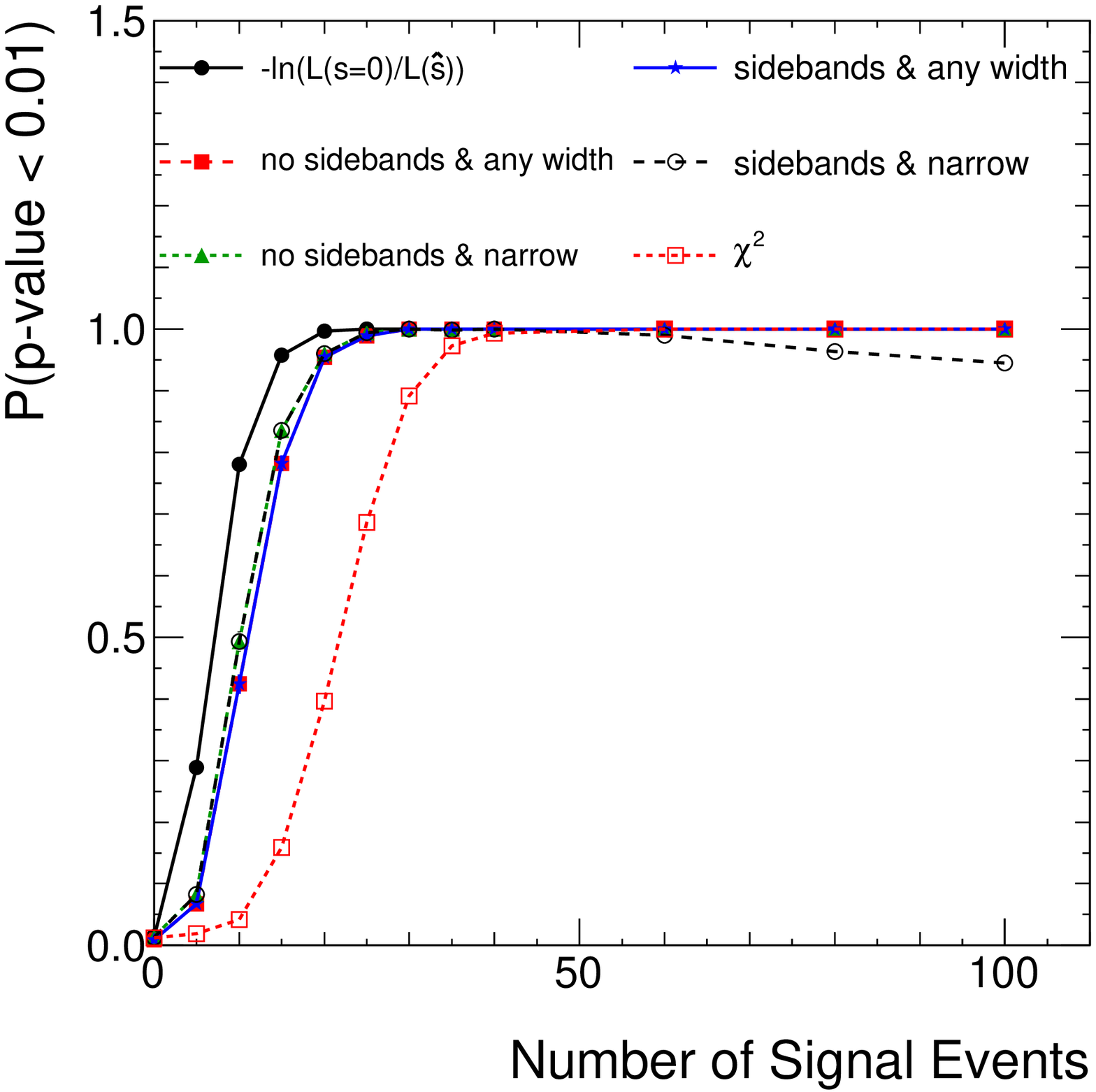}
\label{fig:power05}
}
\subfigure[]{
\includegraphics[width=0.4\textwidth]{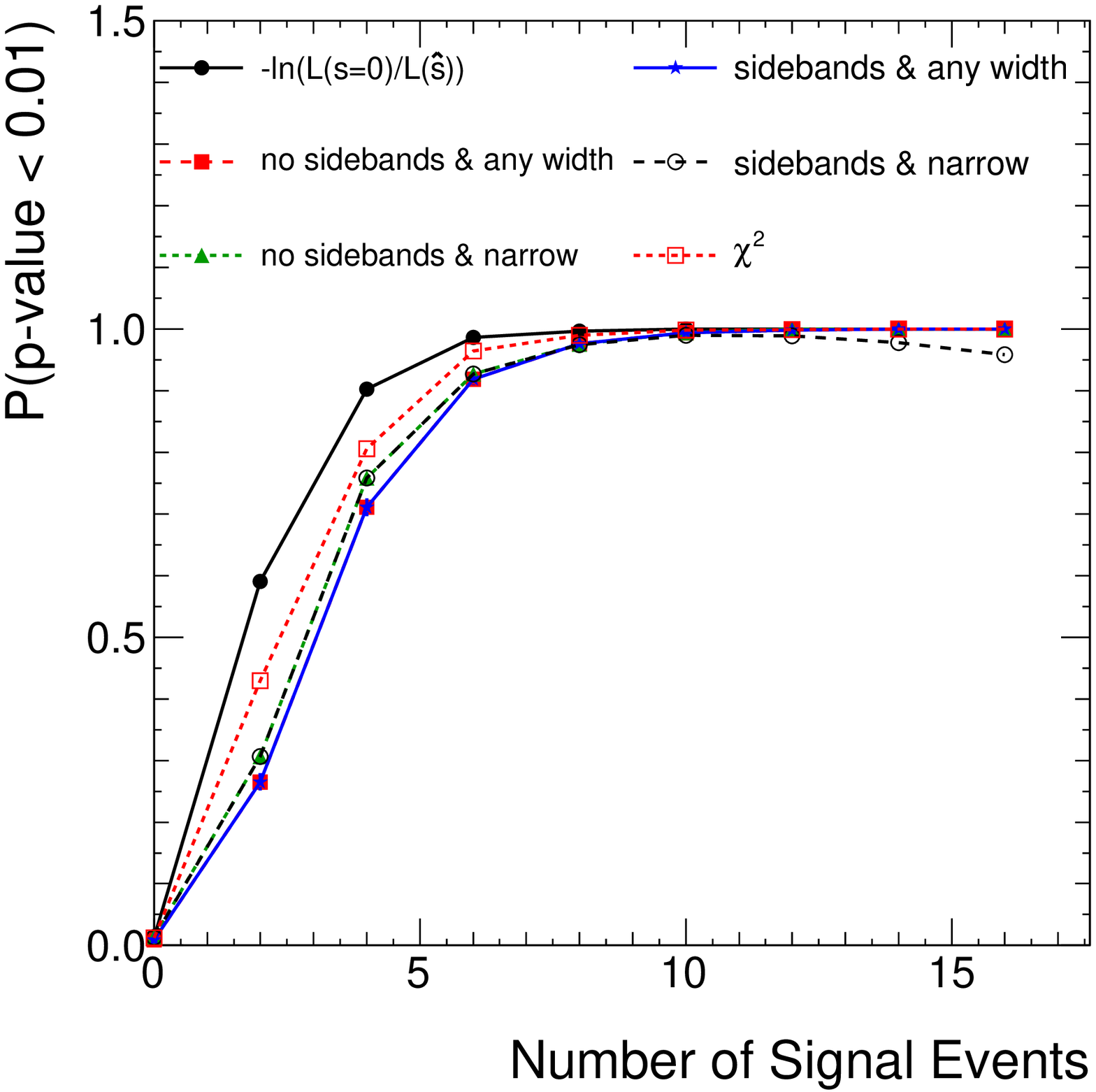}
\label{fig:power09}
}
\caption{\label{fig:powersTunings} The sensitivity of various \bh tunings and of the targeted test of paragraph~\ref{sec:compareToTargetted} as a function of the expected value of signal events.  In the legend, ``narrow'' implies bump width ($W_C$) constrained between 3 and 5 bins (see paragraph~\ref{sec:bhAlgo}).  Figures \subref{fig:power01}, \subref{fig:power05} and \subref{fig:power09} correspond to Gaussian signal injected with $\sigma=0.03$ and mean value 0.1, 0.5 and 0.9 respectively.  The targeted likelihood ratio test of paragraph~\ref{sec:compareToTargetted} is shown in addition to four \bh tunings, and the $\chi^2$ test. }
\end{figure}

Fig.~\ref{fig:powersTunings} shows the probability of observing $\pval < 0.01$ in three cases of signal, as a function of the expected number of signal events.  The three signal cases used correspond to Gaussians of $\sigma = 0.03$ and means $\{0.1,0.5,0.9\}$, according to the Banff sensitivity tests discussed in \ref{sec:tests} and \ref{sec:compareToTargetted}.  

In Fig.~\ref{fig:powersTunings} we see, as expected, that the \bh is always less sensitive than the targeted test.  It is much more sensitive, though, than a simple $\chi^2$ test, except when the signal is at 0.9.

In Fig.~\ref{fig:powersTunings} it may be surprising is that the \bh sensitivity does not reach asymptotically 100\% when the sidebands criteria are taken into account and the width of the central window is constrained.  This is the risk talked about in paragraph~\ref{sec:bhAlgo}, step~\ref{item:sb}.  The explanation is simple.  When the signal increases a lot, and the central window is not allowed to become wider, the sidebands start accumulating so many signal events that they become discrepant, so the bump candidate often disqualifies.  We see that this doesn't happen when the sidebands are ignored, or when the size of the central window can vary freely.

One may compare the sensitivity of the \bh without sidebands and constrained width in Fig.~\ref{fig:powersTunings} to Table~\ref{table:sensitivity1}.  In Fig.~\ref{fig:powersTunings}, for the same amount of injected signal shown in the table (i.e. 75.9, 10.3 and 1.35), the sensitivity appears higher.  The difference is that in Fig.~\ref{fig:powersTunings} the background is known and fixed, rather than obtained by fitting as in Table~\ref{table:sensitivity1}.

It is worth reminding here that, for any hypothesis test, sensitivity depends on the kind of signal.  The conclusions of this paragraph may not apply to different signal shapes.

\subsection{Locating the right interval}
\label{sec:interval}

Here will be demonstrated how the \bh locates the position of injected signal.  We will refer to two of the \bh tunings of paragraph~\ref{sec:powers}; tuning \ref{tuning1} (no sidebands and unconstrained width) and tuning \ref{tuning2} (no sidebands and width constrained between 3 and 5 bins).  The signal injected will be Gaussian of $\sigma=0.03$ and mean 0.5; the results are similar at mean 0.1 and 0.9.  Various amounts of signal will be tried to show how the ability to locate the right $x$ interval progresses.

\begin{figure}[p]
\centering
\subfigure[]{
\includegraphics[width=0.3\textwidth]{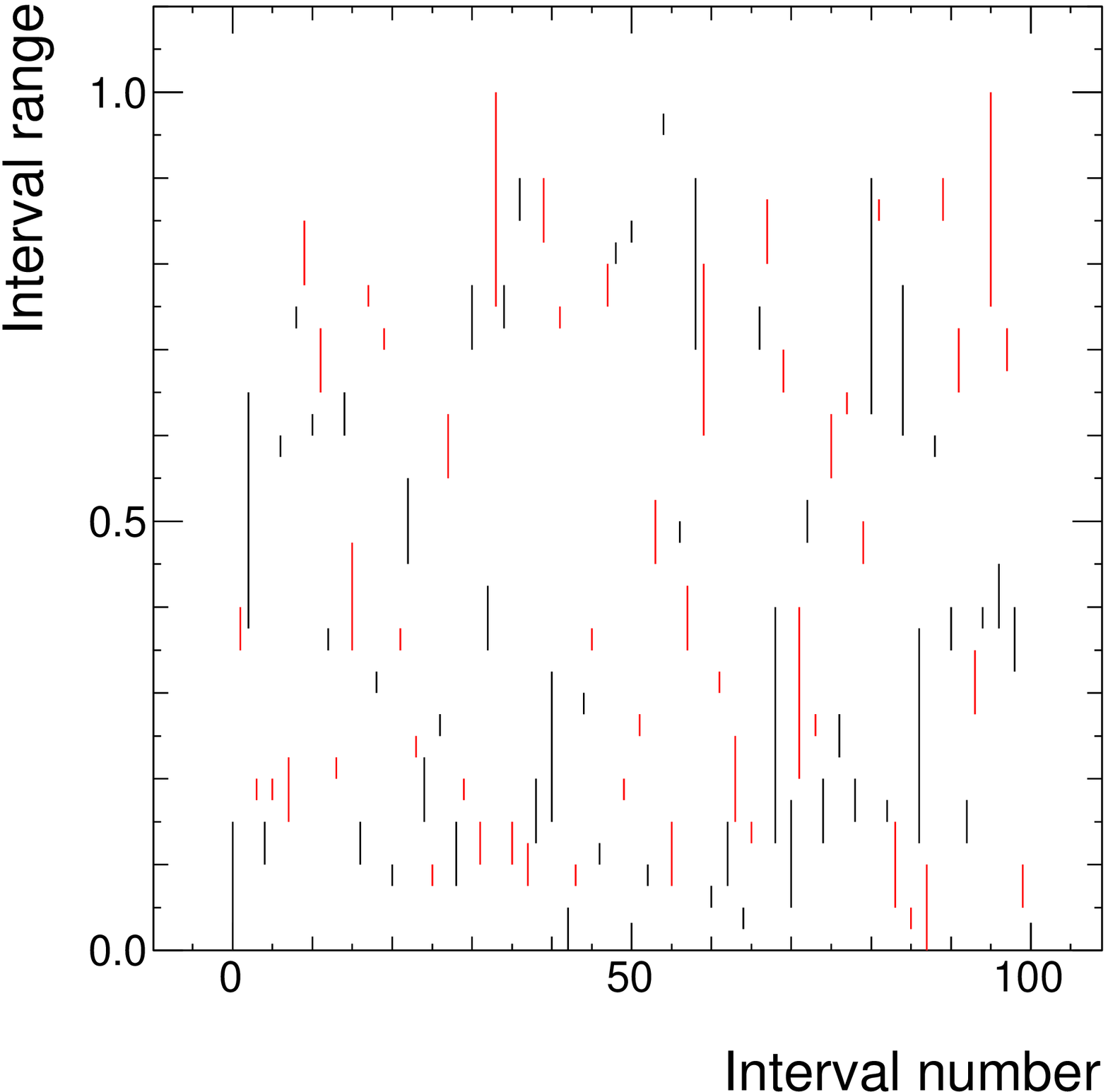}
\label{fig:intervals0a}
}
\subfigure[]{
\includegraphics[width=0.3\textwidth]{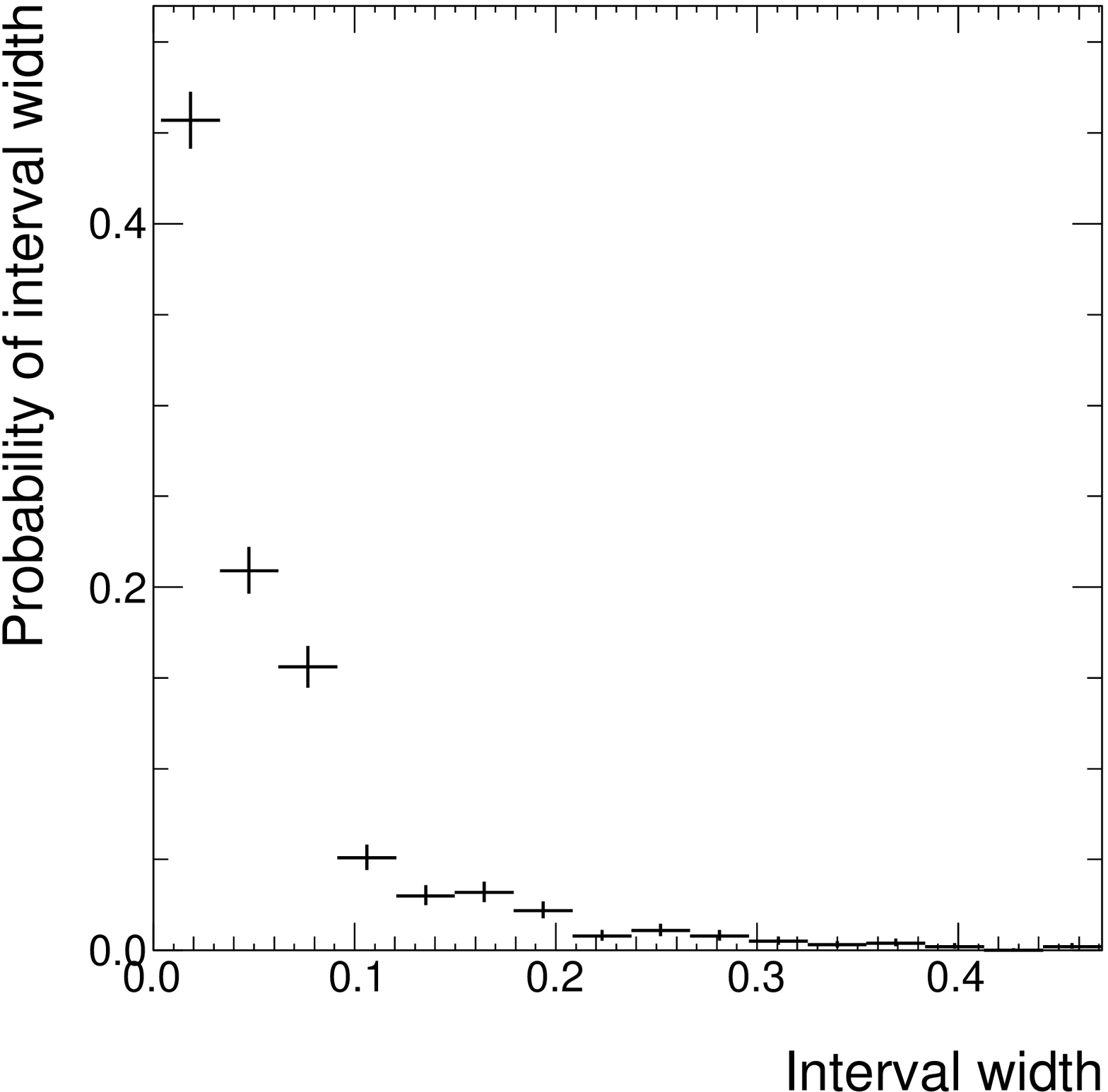}
\label{fig:intervals0b}
}
\subfigure[]{
\includegraphics[width=0.3\textwidth]{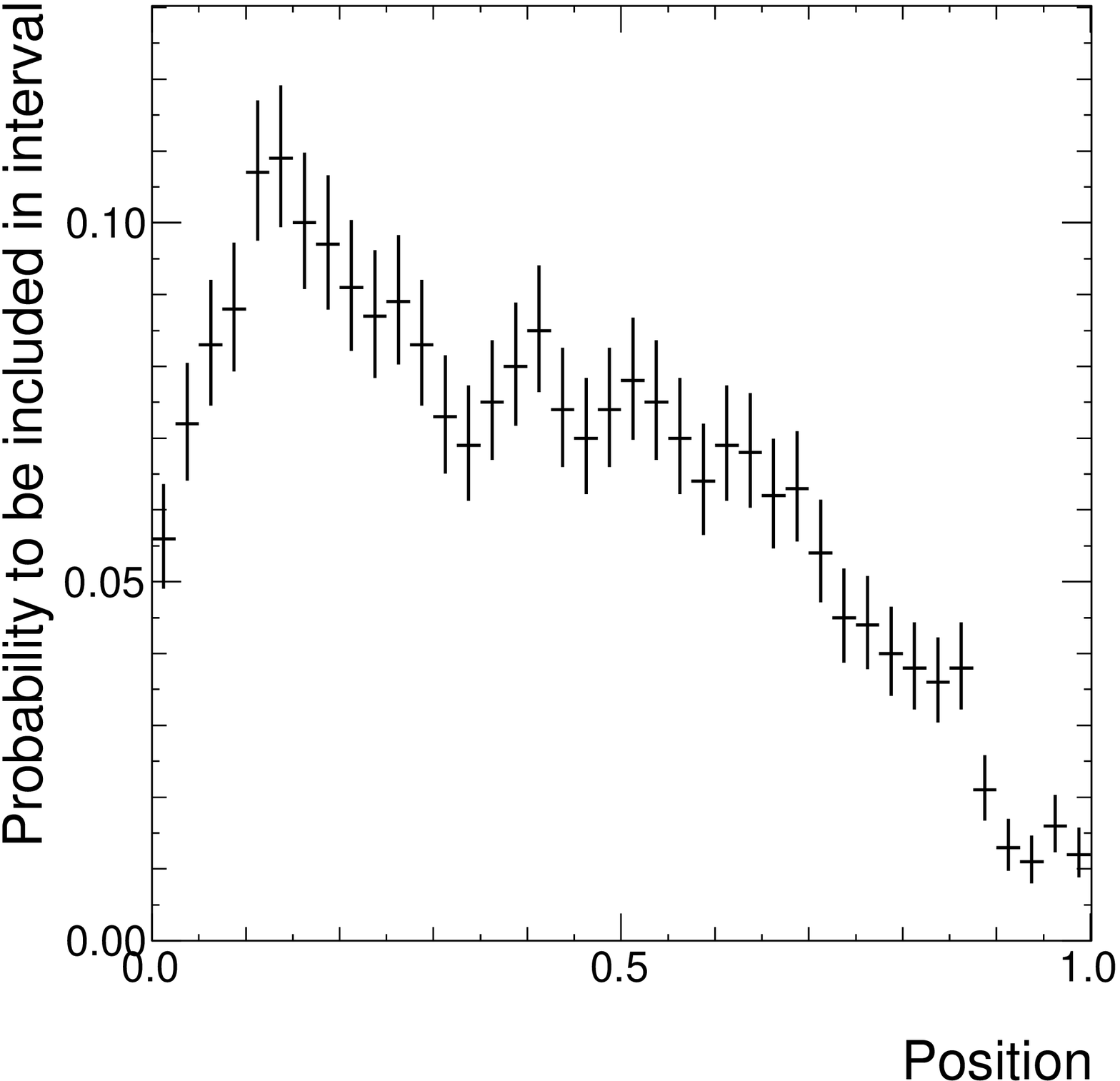}
\label{fig:intervals0c}
}
\subfigure[]{
\includegraphics[width=0.3\textwidth]{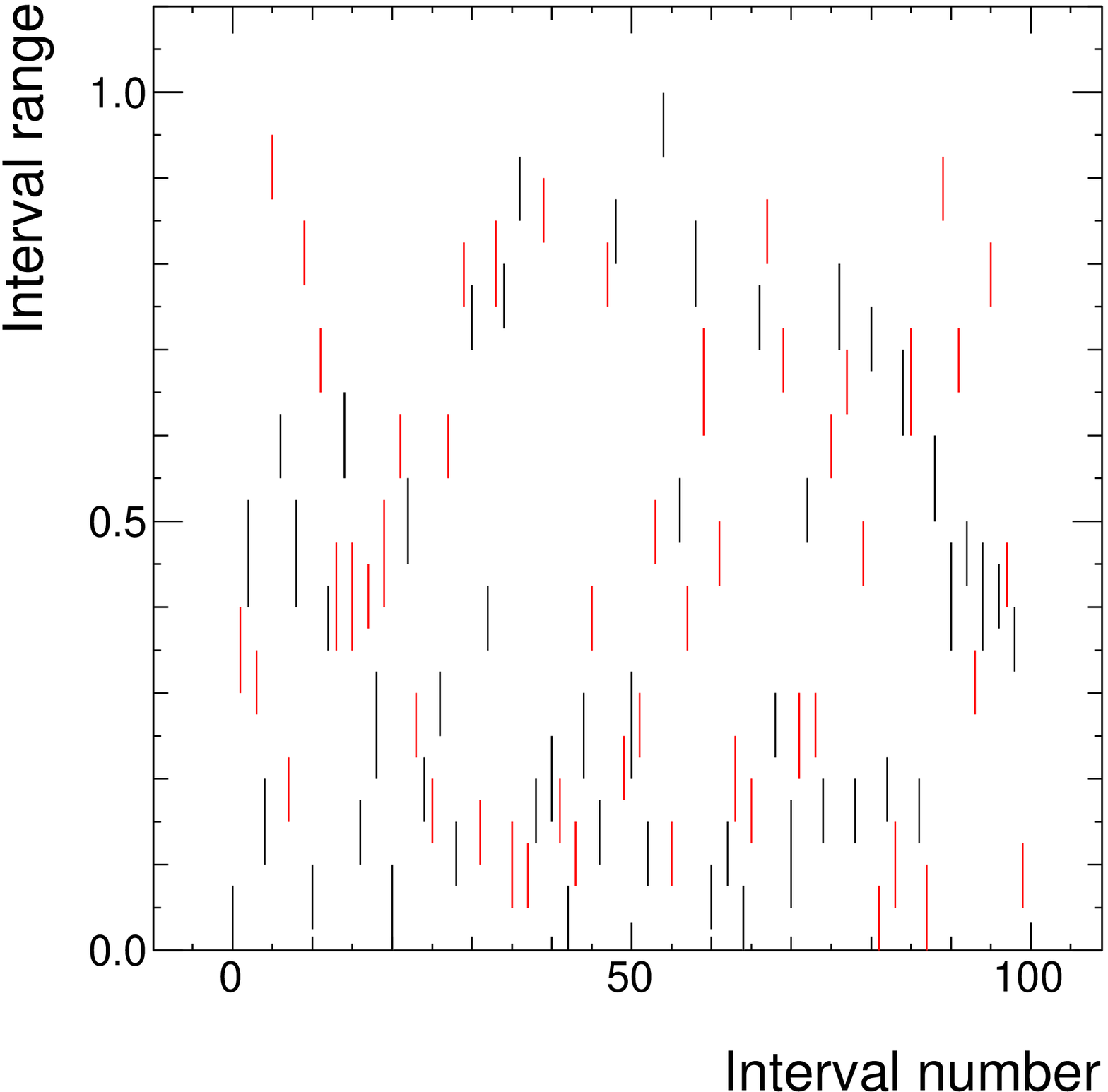}
\label{fig:intervals0d}
}
\subfigure[]{
\includegraphics[width=0.3\textwidth]{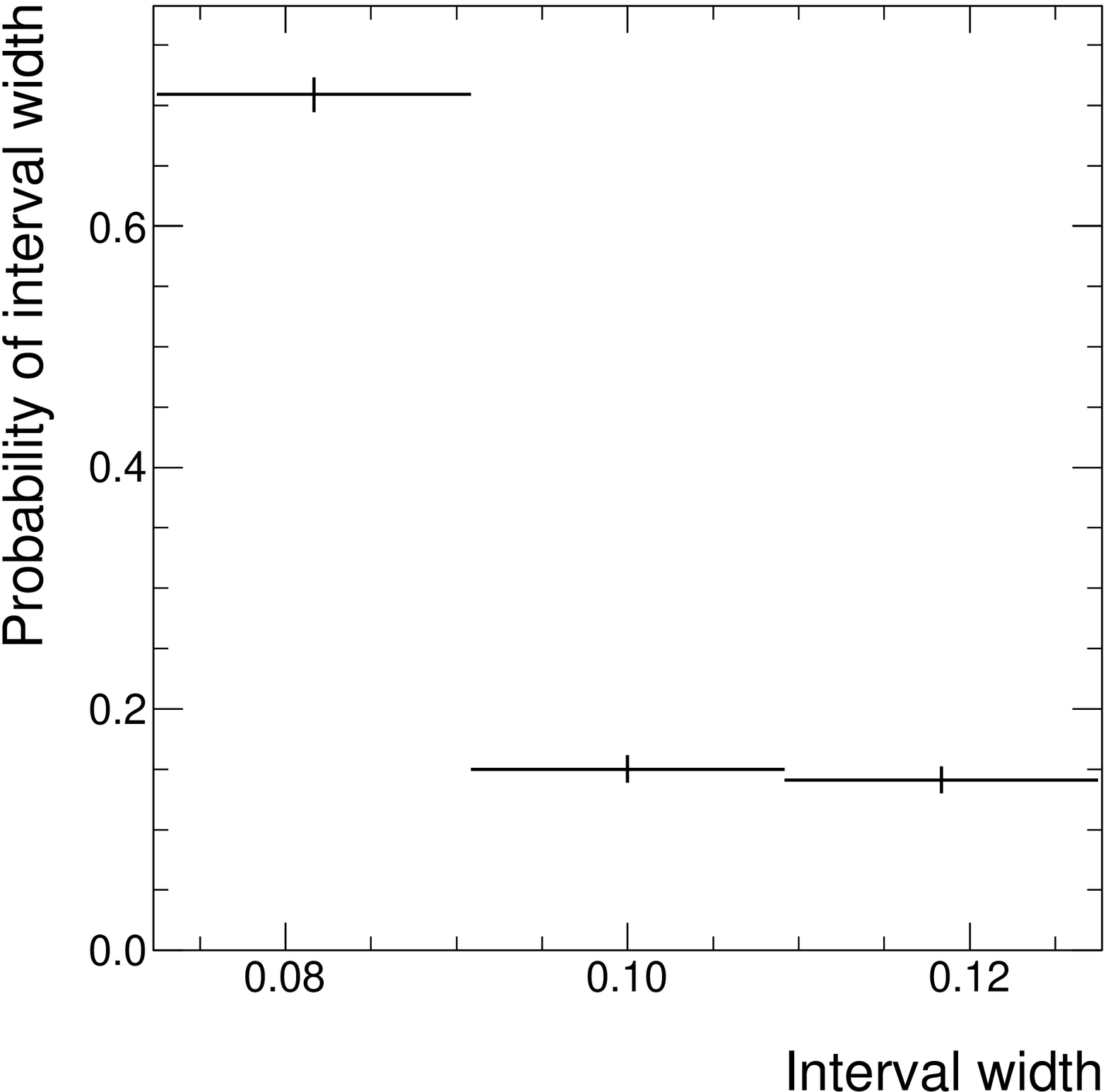}
\label{fig:intervals0e}
}
\subfigure[]{
\includegraphics[width=0.3\textwidth]{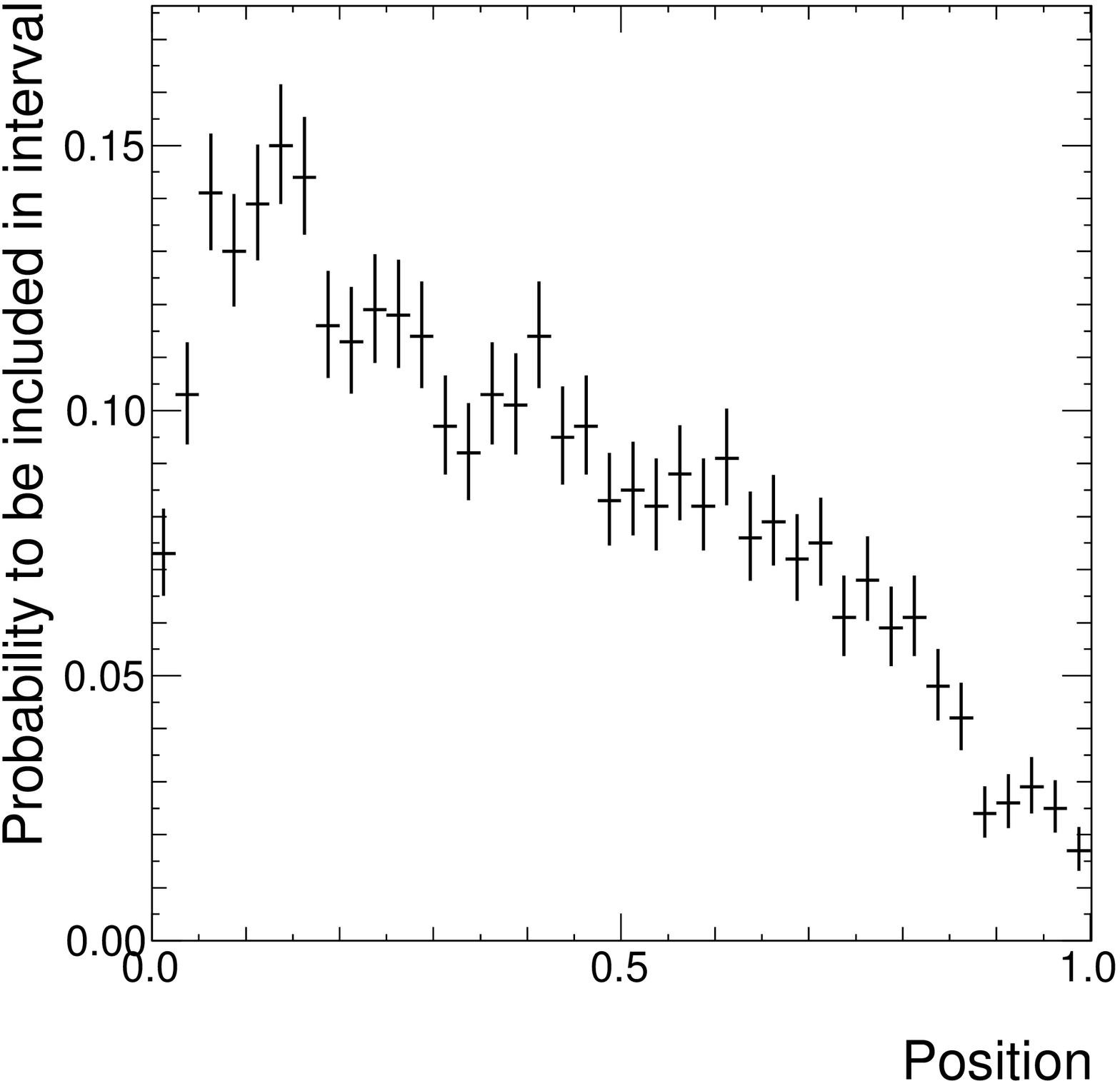}
\label{fig:intervals0f}
}
\caption{\label{fig:intervals0signal} This figure is made using pseudo-experiments where 0 signal is injected on top of the background given by $10^4 e^{-10x}$.  The first row (Fig.~\subref{fig:intervals0a}, \subref{fig:intervals0b}, \subref{fig:intervals0c}) show the results of \bh tuning \ref{tuning1} (unconstrained window size), and the second row (Fig.~\ref{fig:intervals0d}, \ref{fig:intervals0e}, \ref{fig:intervals0f}) show the results of tuning \ref{tuning2} (window between 3 and 5 bins, which corresponds to intervals of length 0.075 to 0.125 in $x$).  Fig.~\ref{fig:intervals0a} shows the most discrepant intervals found in 100 pseudo-experiments.  The x-coordinate is an integer that enumerates each interval, and the y-axis shows the $x$-range of each interval with a black line for odd pseudo-experiments and a red line for even pseudo-experiments.  Fig.~\ref{fig:intervals0b} shows the probability distribution of the width of the most discrepant interval located in a pseudo-experiment.  The probability distribution is estimated from a sample of 1000 pseudo-experiments, and the uncertainties are binomial.  Fig.~\ref{fig:intervals0c} shows the probability each bin of $x$ has to be included in the most discrepant interval located in a pseudo-experiment.  This is {\em not} a probability distribution; the sum of its bins is not equal to 1.  It may be understood looked at bin-by-bin; for example the 1st bin, $x\in[0,0.025]$, has probability $\simeq$4.5\% to be included in the most discrepant interval the \bh locates in a pseudo-experiment.  These probabilities are estimated using 1000 pseudo-experiments, and the uncertainties are binomial.}
\end{figure}

\begin{figure}[p]
\centering
\includegraphics[width=0.3\textwidth]{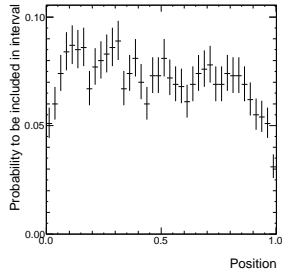}
\caption{\label{fig:intervals0_bigBkg} Same as Fig.~\ref{fig:intervals0signal}\subref{fig:intervals0c}, except that the pseudo-experiments are generated from the background $10^8 e^{-10x}$, so as to have large event counts in all $x$ bins.}
\end{figure}

Let's first examine what intervals are located as most discrepant when there is no signal injected on top of the background of the Banff Challenge, $10^4 e^{-10x}$.  Fig.~\ref{fig:intervals0signal} shows two examples; one with \bh tuning \ref{tuning1} and one with \ref{tuning2}.   Fig.~\ref{fig:intervals0signal}\subref{fig:intervals0c} and Fig.~\ref{fig:intervals0signal}\subref{fig:intervals0f} show that higher $x$ values are less likely to be included in the most discrepant interval.  The reason has to do with expecting too few events beyond $x\simeq 0.6$ (see Fig.~\ref{fig:test}).  To demonstrate that, Fig.~\ref{fig:intervals0_bigBkg} shows the same as Fig.~\ref{fig:intervals0signal}\subref{fig:intervals0c}, but for a background function $10^8e^{-10x}$ instead, so as to expect over 100 events even in the highest $x$ bin.  Consequently, Fig.~\ref{fig:intervals0_bigBkg} shows more constant probabilities, indicating that the most interesting window is uniformly distributed in the $[0,1]$ range.  In Fig.~\ref{fig:intervals0_bigBkg} one can still see a reduction of probability close to $x$=0 and 1.  These edge effects are there because the $x$ bins that are not so close to the edges have more possibilities to be included in the most discrepant interval; they may be in its middle, or near its end.  Marginal bins, however, have fewer possibilities to be included; for the very last $x$ bin, only one way exists: the most discrepant interval has to reach to the edge of the $[0,1]$ range.

\begin{figure}[p]
\centering
\subfigure[]{
\includegraphics[width=0.3\textwidth]{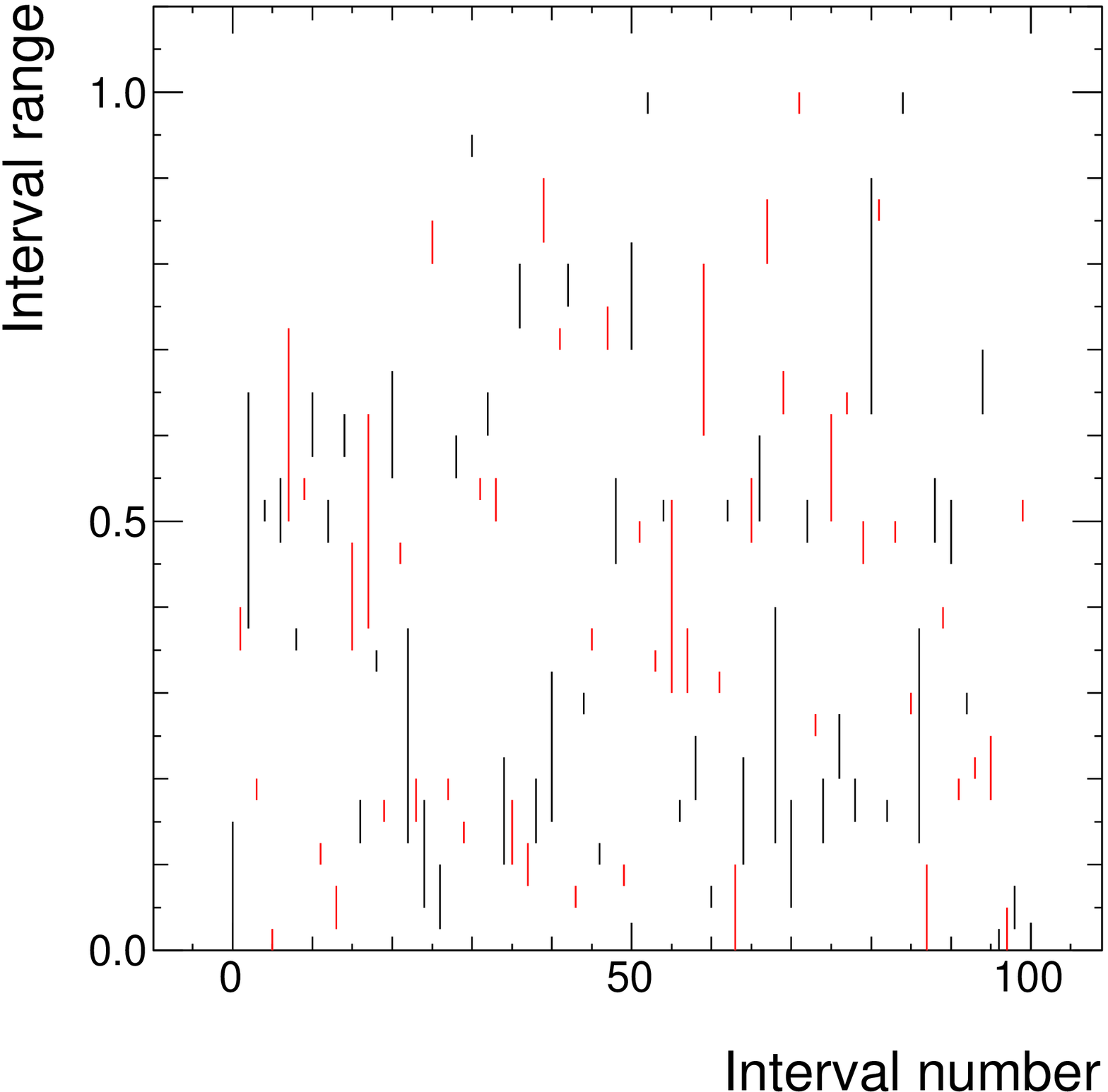}
\label{fig:intervals1a}
}
\subfigure[]{
\includegraphics[width=0.3\textwidth]{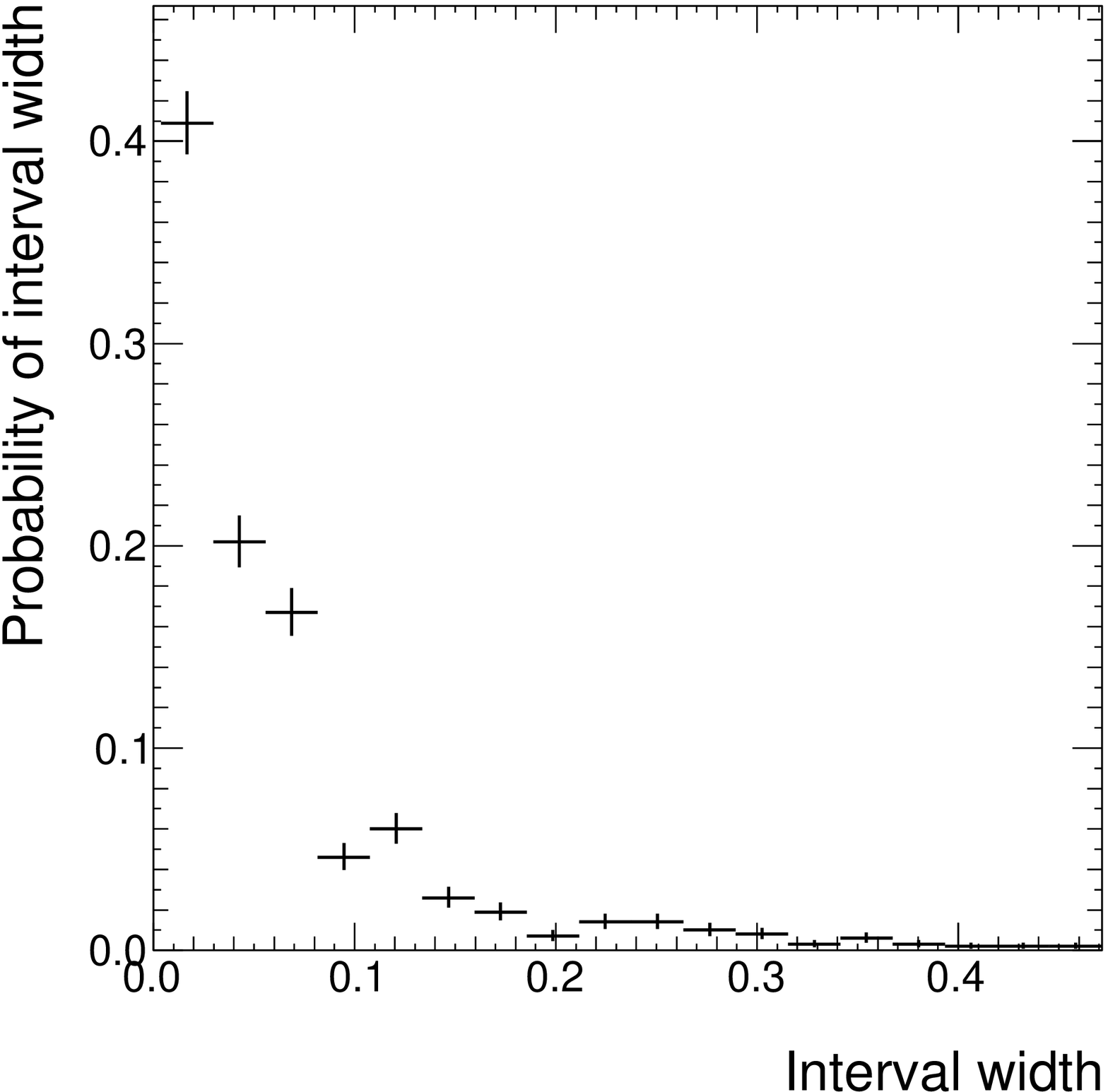}
\label{fig:intervals1b}
}
\subfigure[]{
\includegraphics[width=0.3\textwidth]{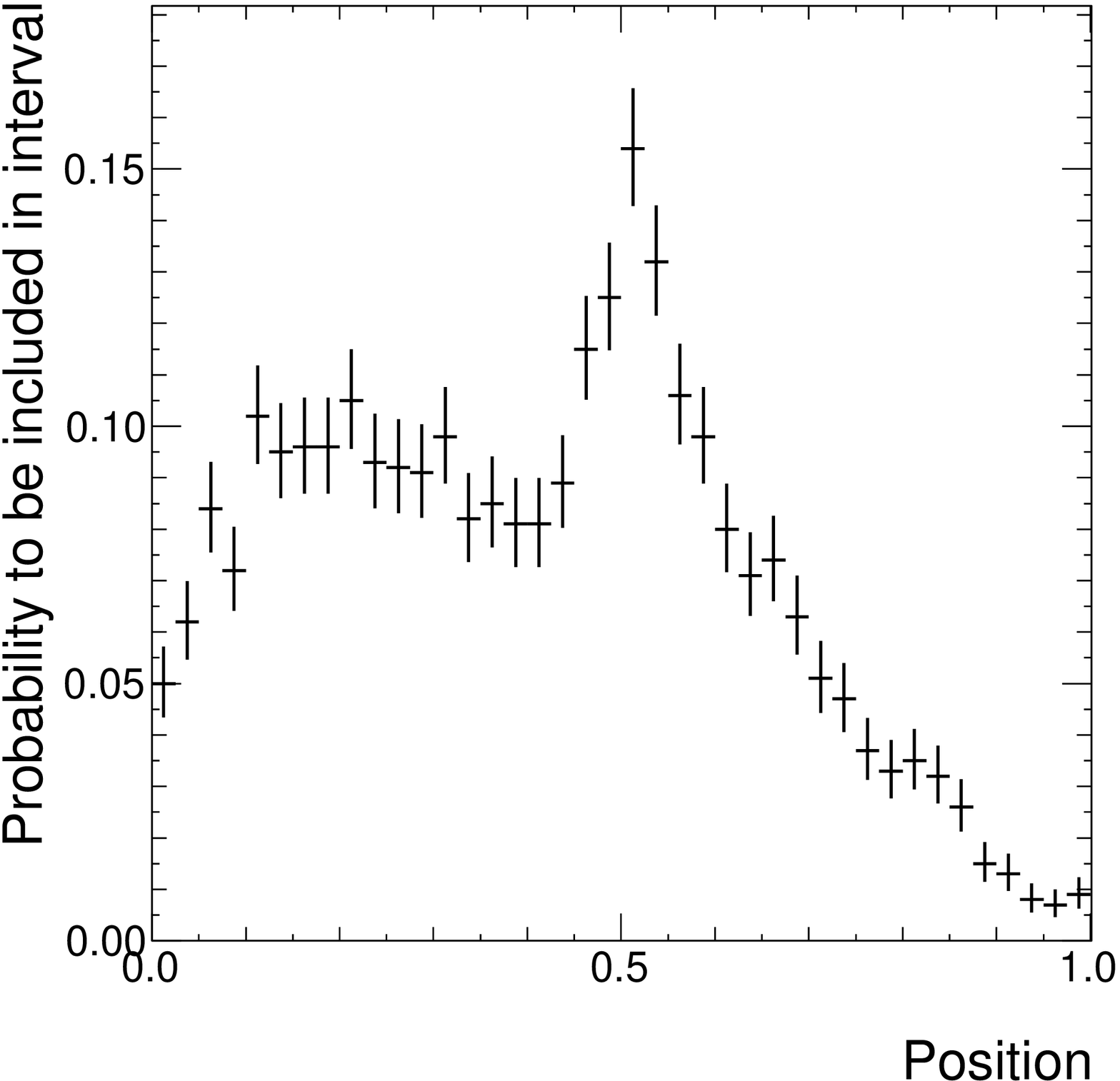}
\label{fig:intervals1c}
}
\subfigure[]{
\includegraphics[width=0.3\textwidth]{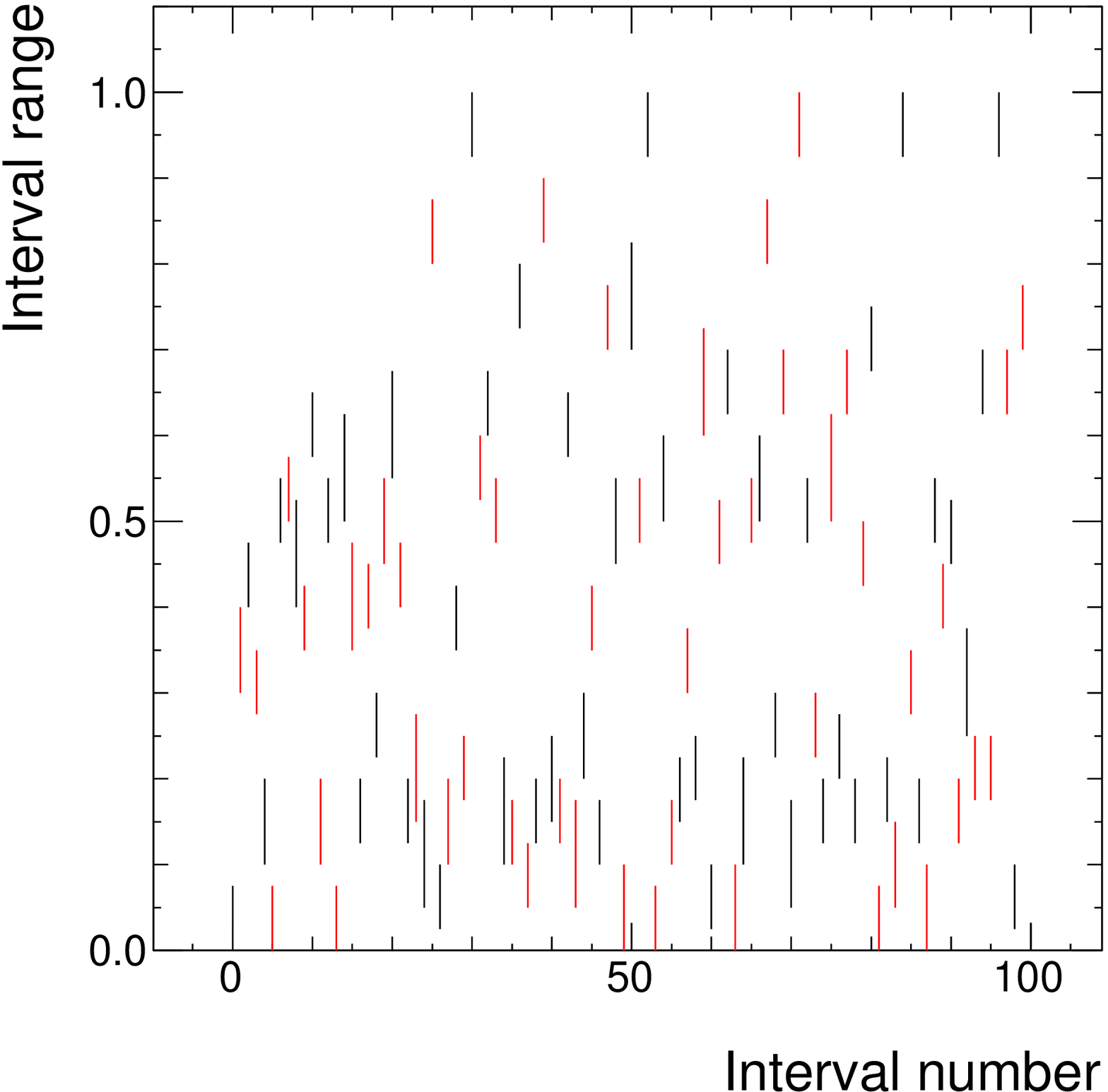}
\label{fig:intervals1d}
}
\subfigure[]{
\includegraphics[width=0.3\textwidth]{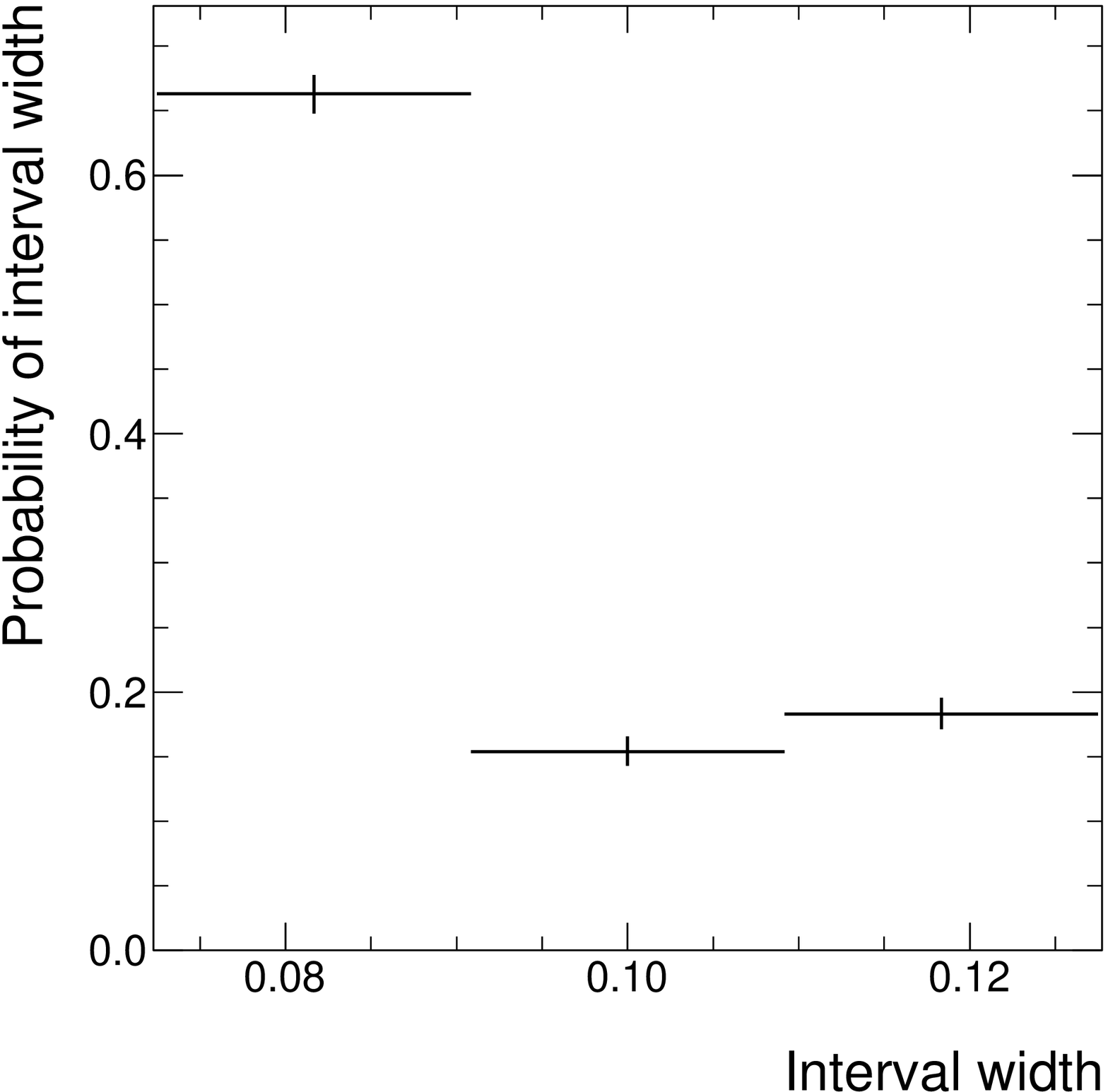}
\label{fig:intervals1e}
}
\subfigure[]{
\includegraphics[width=0.3\textwidth]{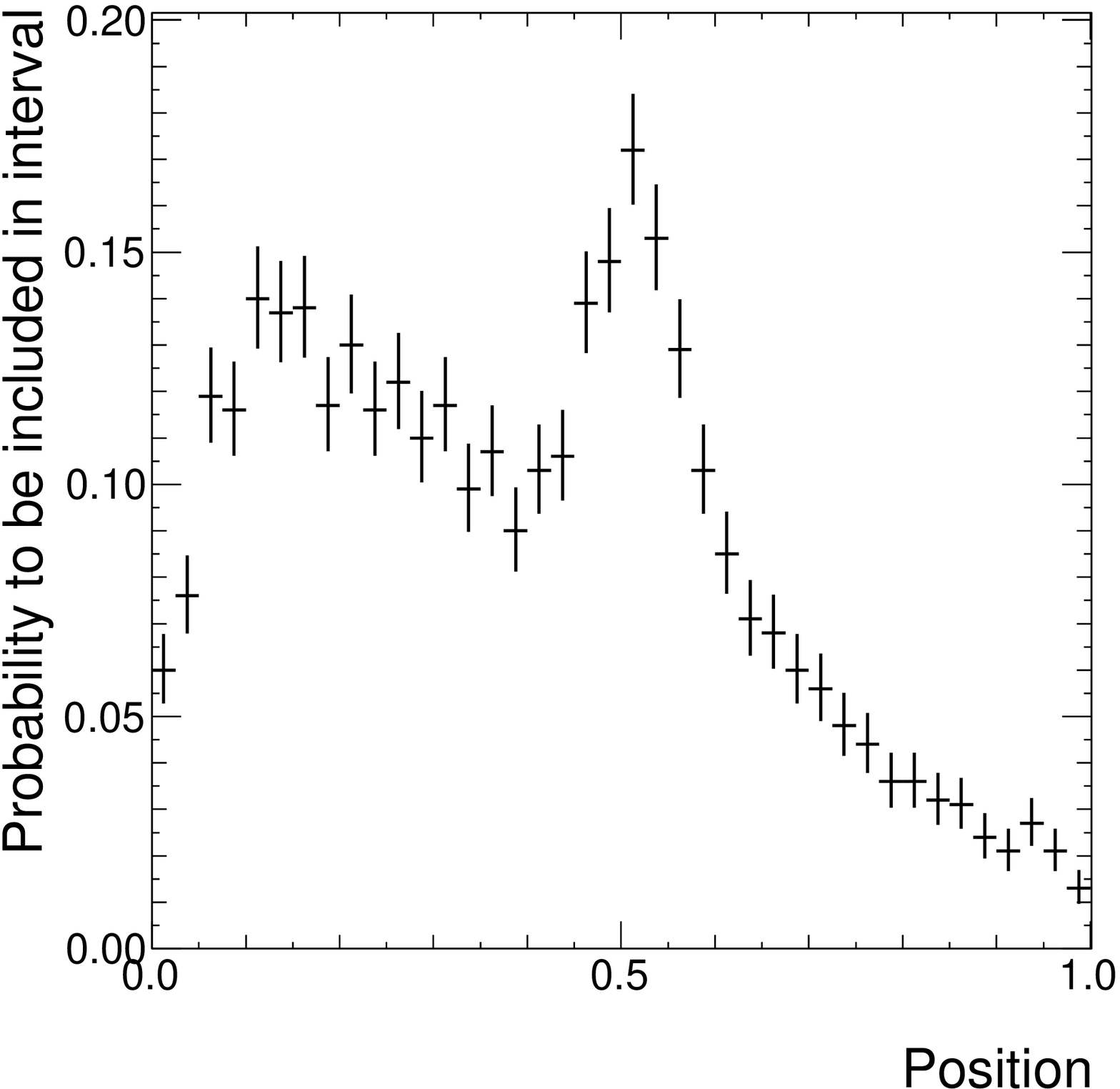}
\label{fig:intervals1f}
}
\caption{\label{fig:intervals1signal} Same as Fig.~\ref{fig:intervals0signal}, except that just 1 signal event is expected, distributed as a Gaussian with $\sigma=0.03$ and mean 0.5.}
\end{figure}

\begin{figure}[p]
\centering
\subfigure[]{
\includegraphics[width=0.3\textwidth]{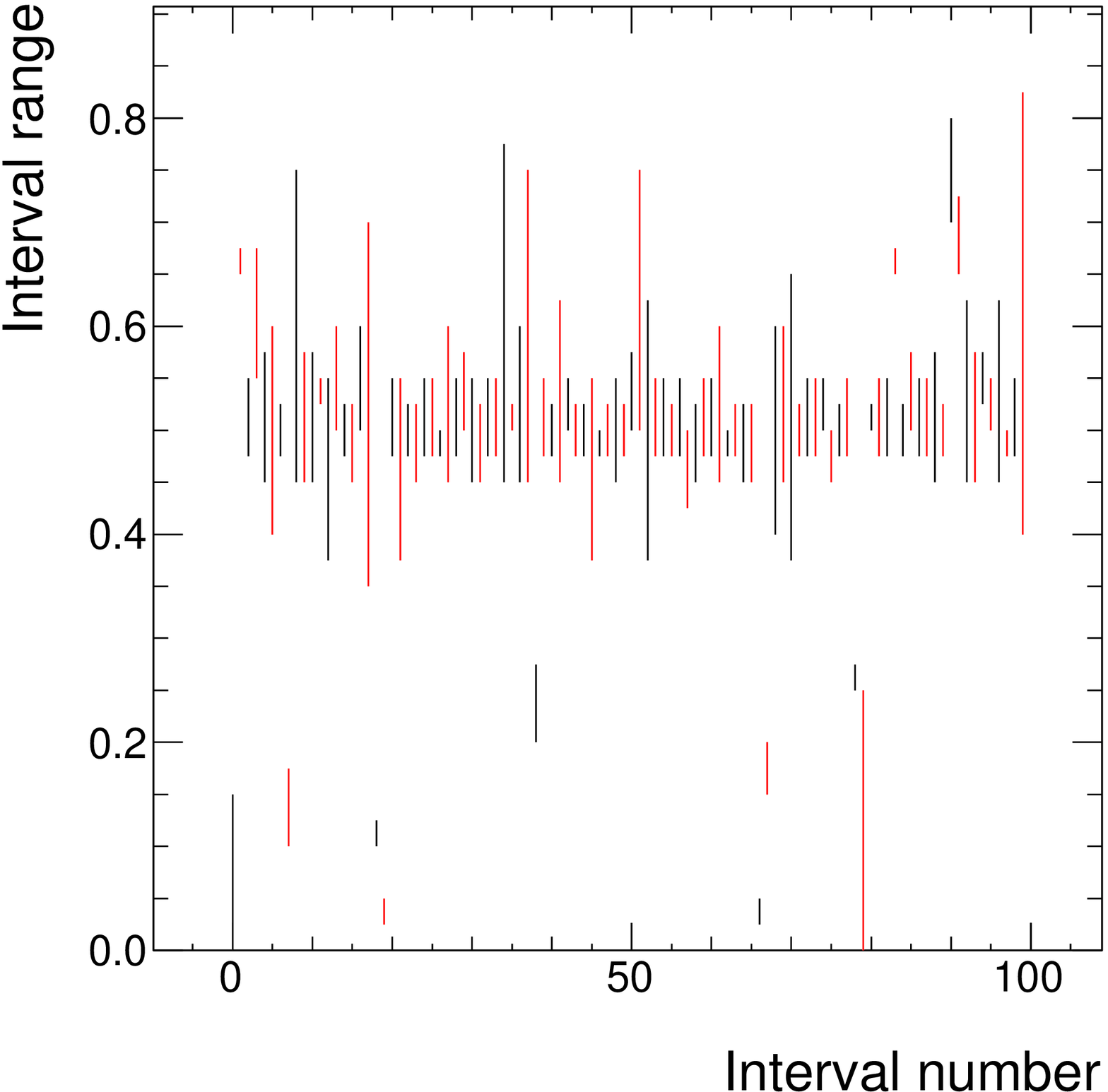}
\label{fig:intervals10a}
}
\subfigure[]{
\includegraphics[width=0.3\textwidth]{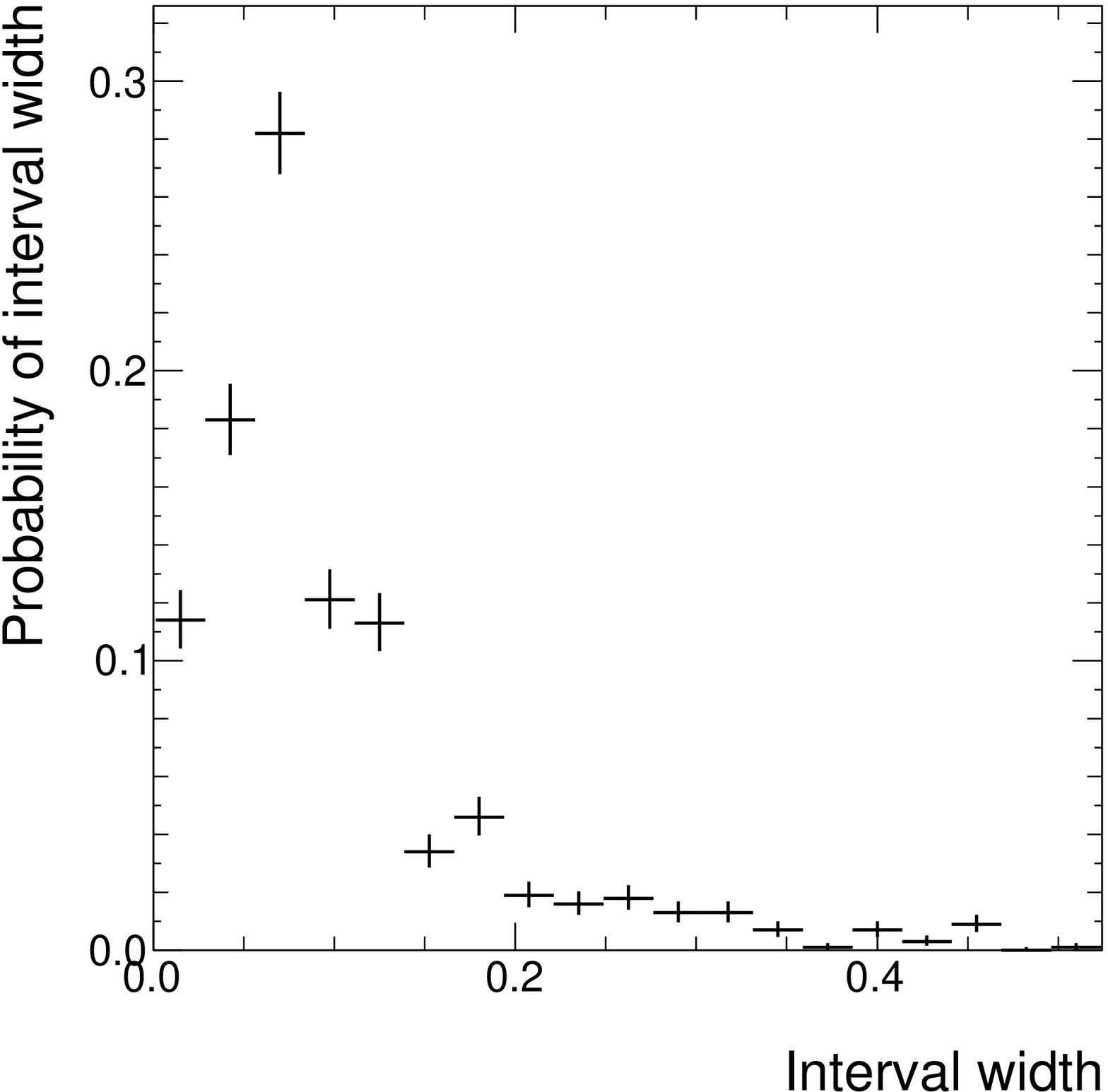}
\label{fig:intervals10b}
}
\subfigure[]{
\includegraphics[width=0.3\textwidth]{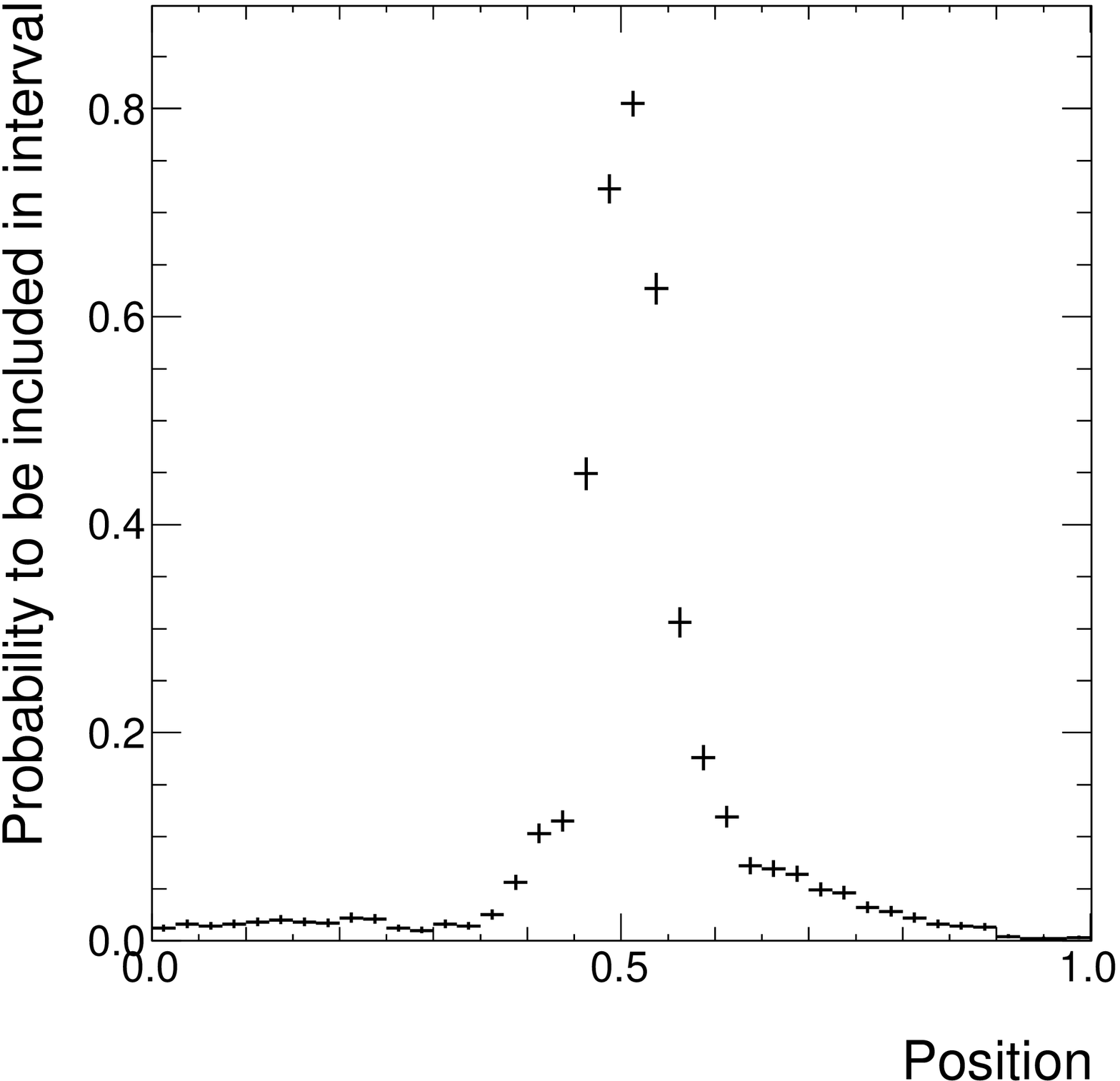}
\label{fig:intervals10c}
}
\subfigure[]{
\includegraphics[width=0.3\textwidth]{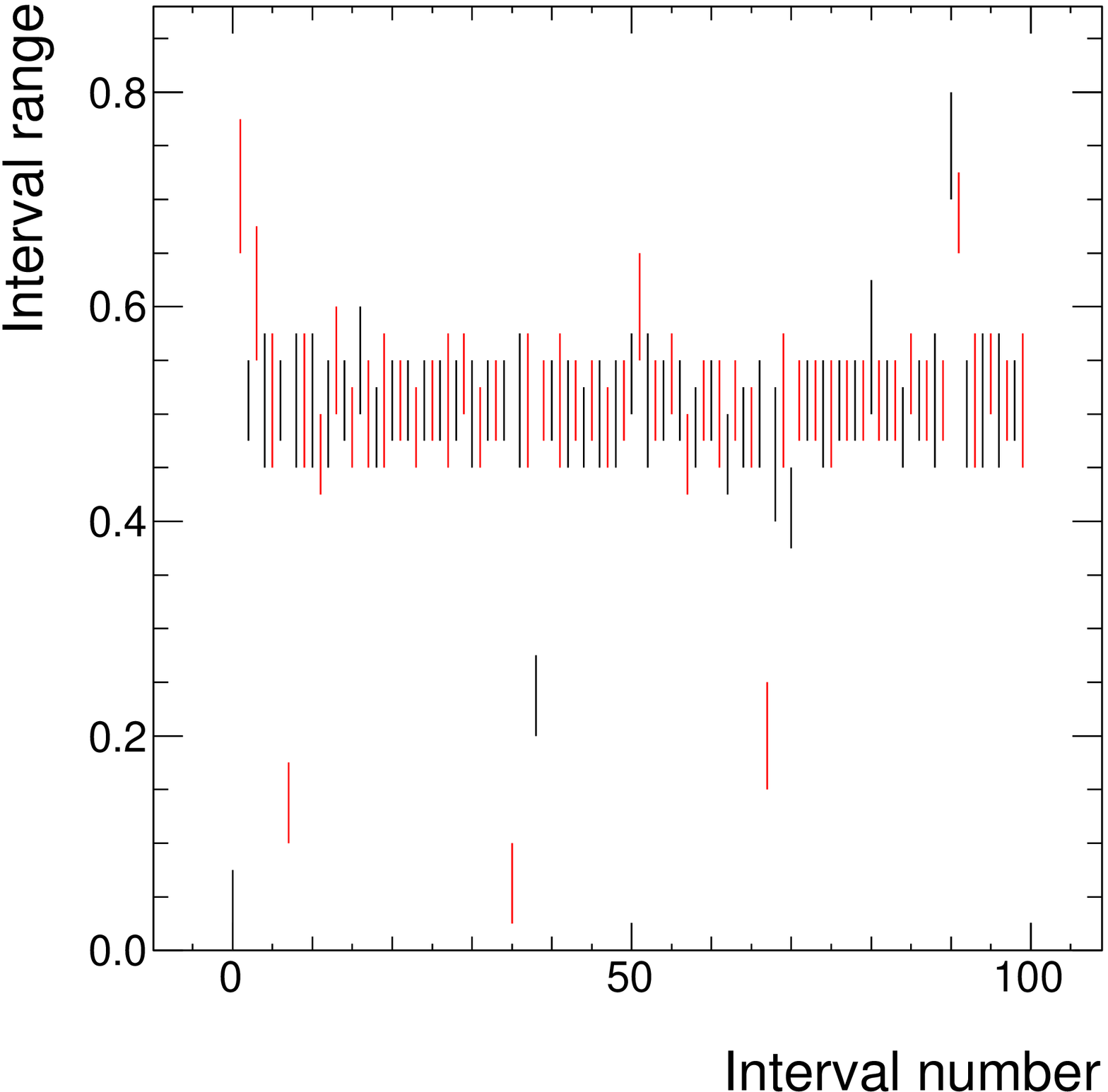}
\label{fig:intervals10d}
}
\subfigure[]{
\includegraphics[width=0.3\textwidth]{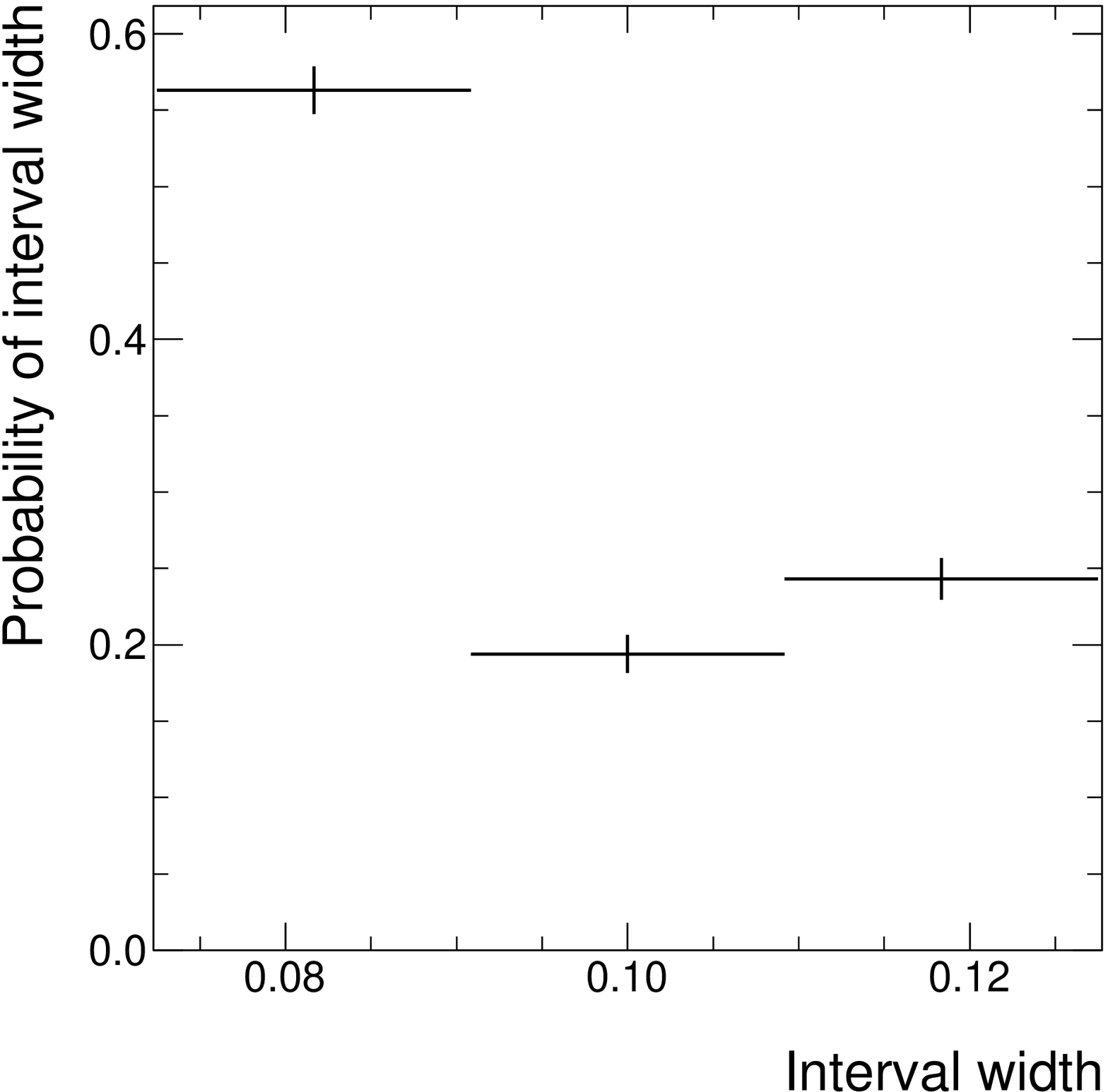}
\label{fig:intervals10e}
}
\subfigure[]{
\includegraphics[width=0.3\textwidth]{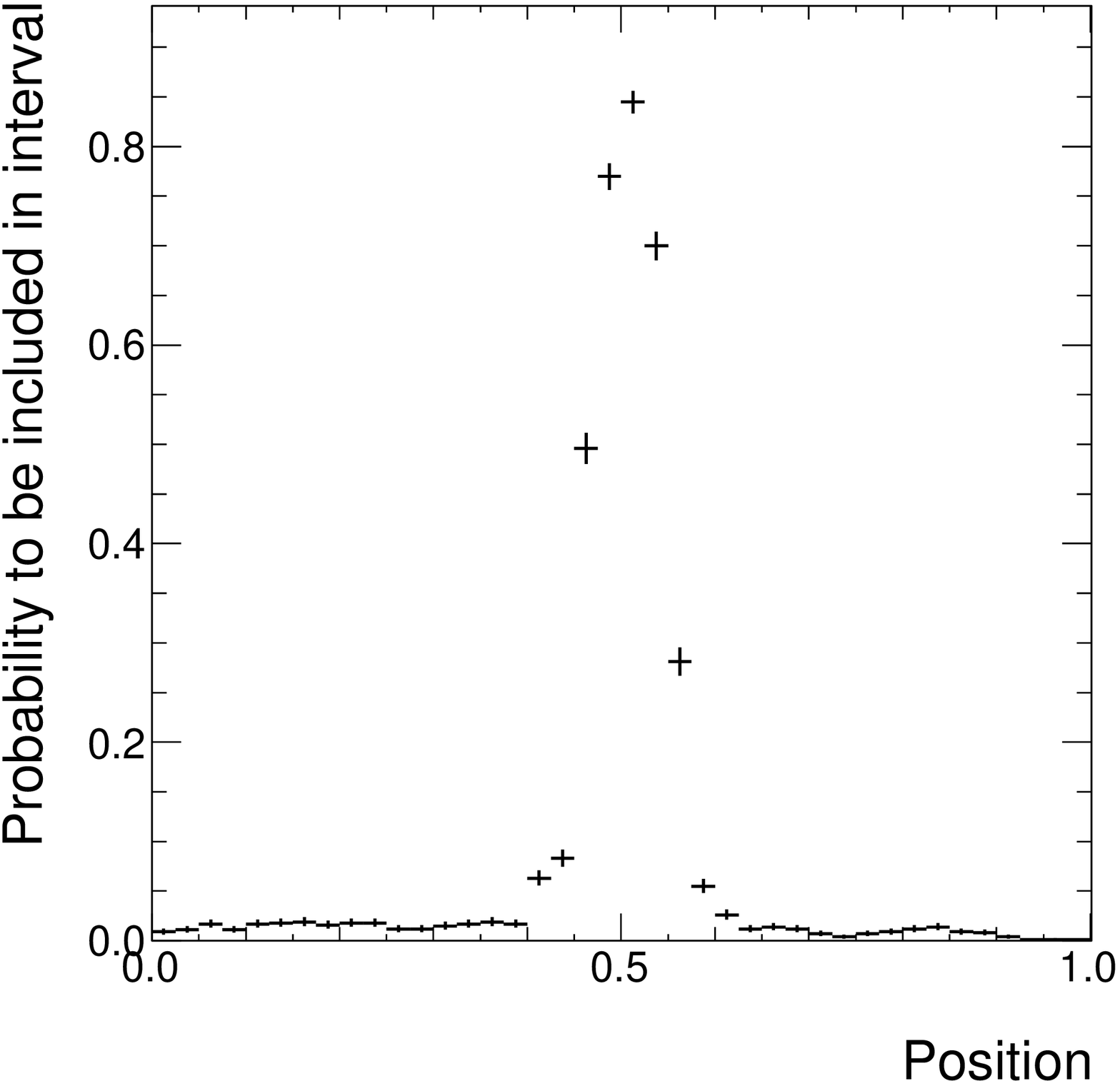}
\label{fig:intervals10f}
}
\caption{\label{fig:intervals10signal} Same as Fig.~\ref{fig:intervals1signal}, except that 10 signal events are expected, distributed as a Gaussian with $\sigma=0.03$ and mean 0.5.}
\end{figure}

\begin{figure}[p]
\centering
\subfigure[]{
\includegraphics[width=0.3\textwidth]{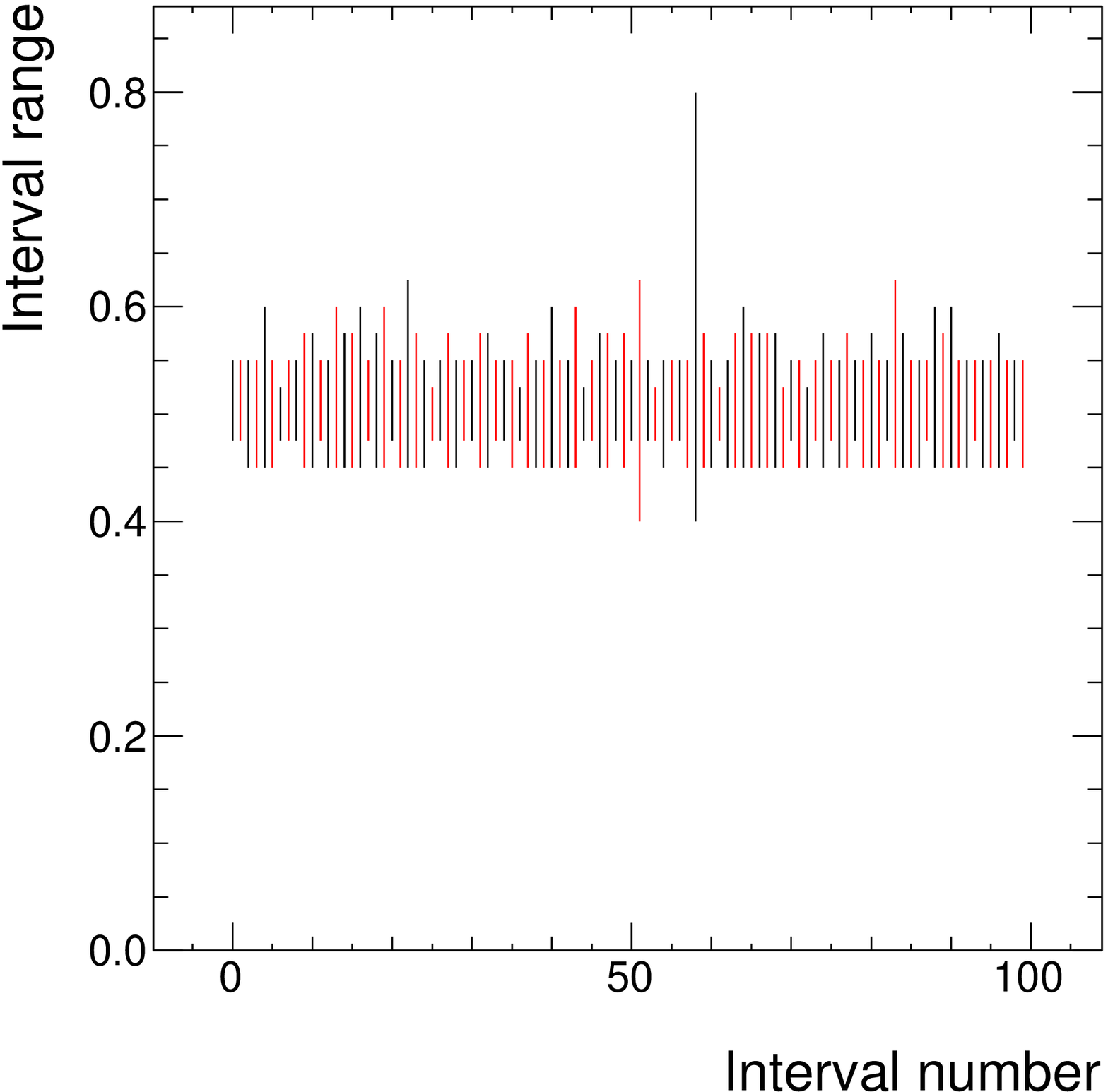}
\label{fig:intervals40a}
}
\subfigure[]{
\includegraphics[width=0.3\textwidth]{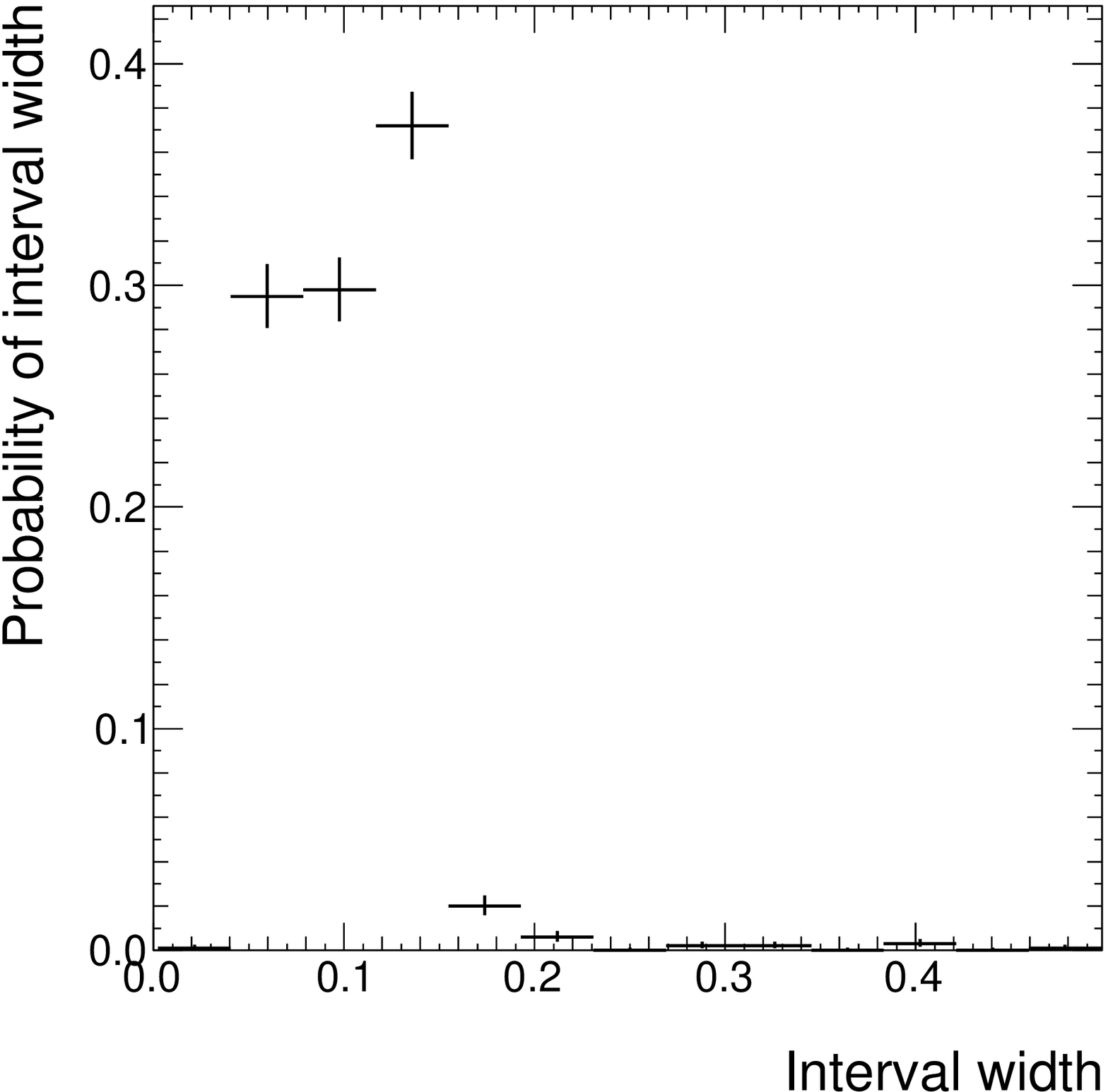}
\label{fig:intervals40b}
}
\subfigure[]{
\includegraphics[width=0.3\textwidth]{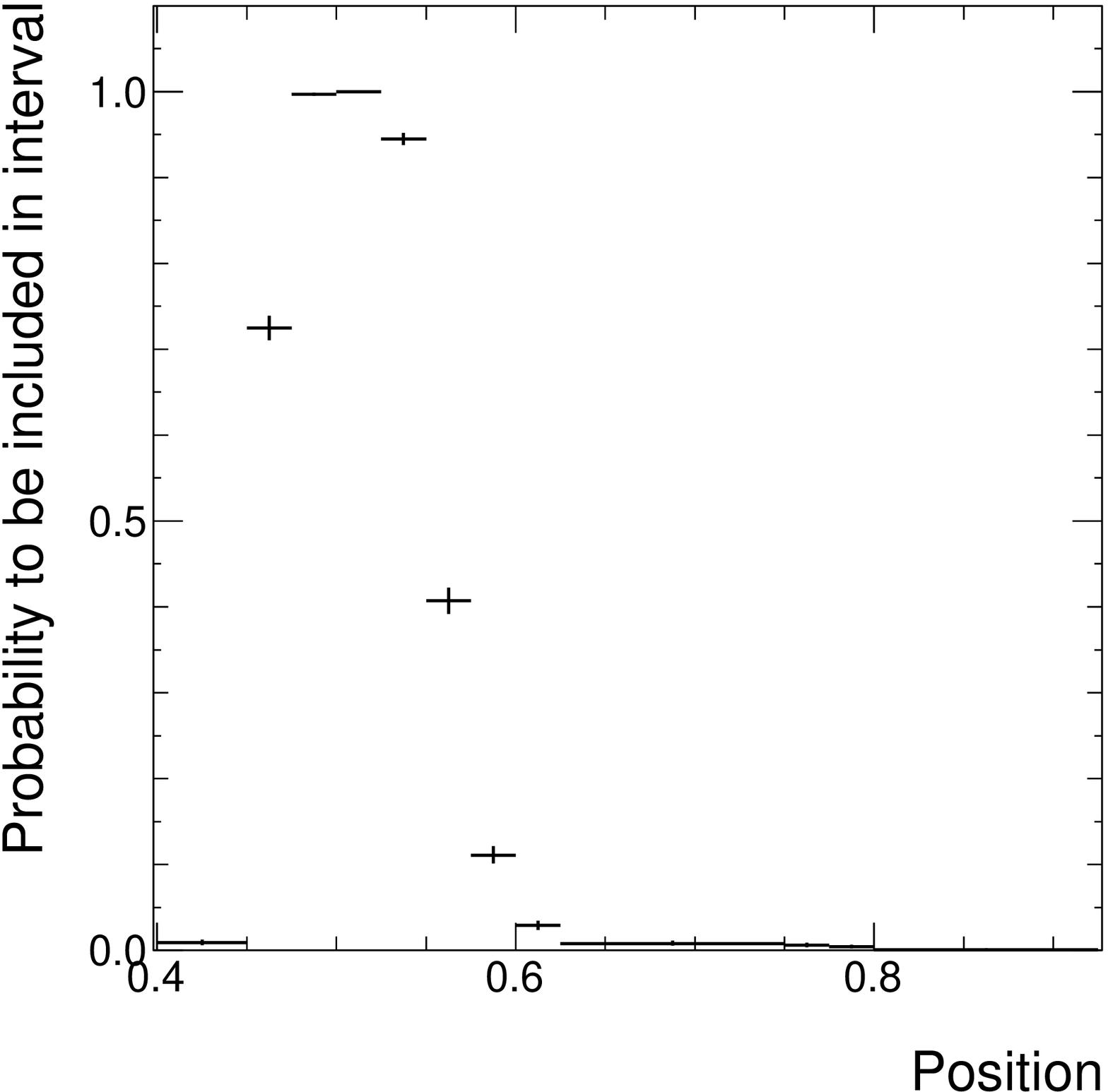}
\label{fig:intervals40c}
}
\subfigure[]{
\includegraphics[width=0.3\textwidth]{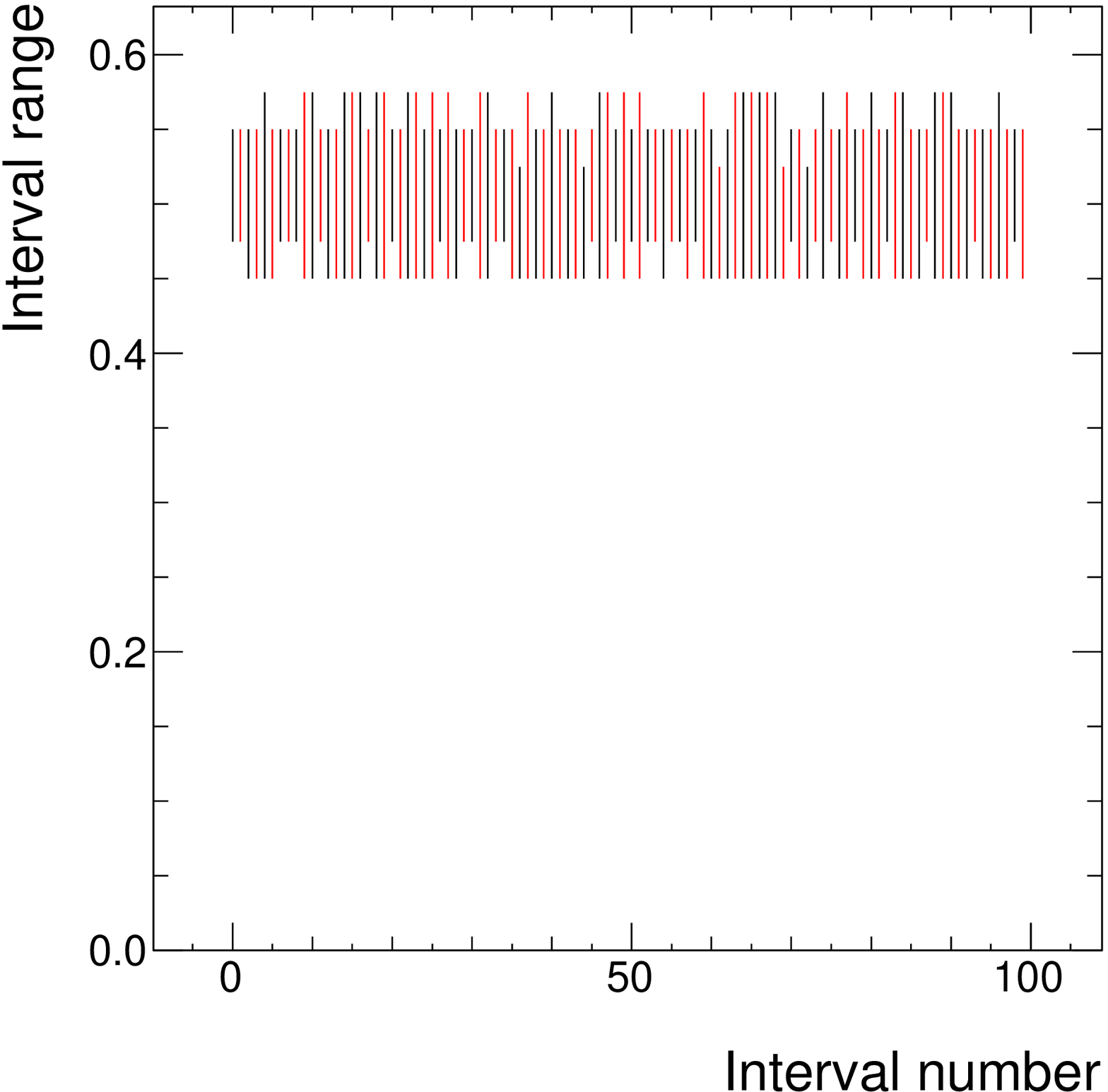}
\label{fig:intervals40d}
}
\subfigure[]{
\includegraphics[width=0.3\textwidth]{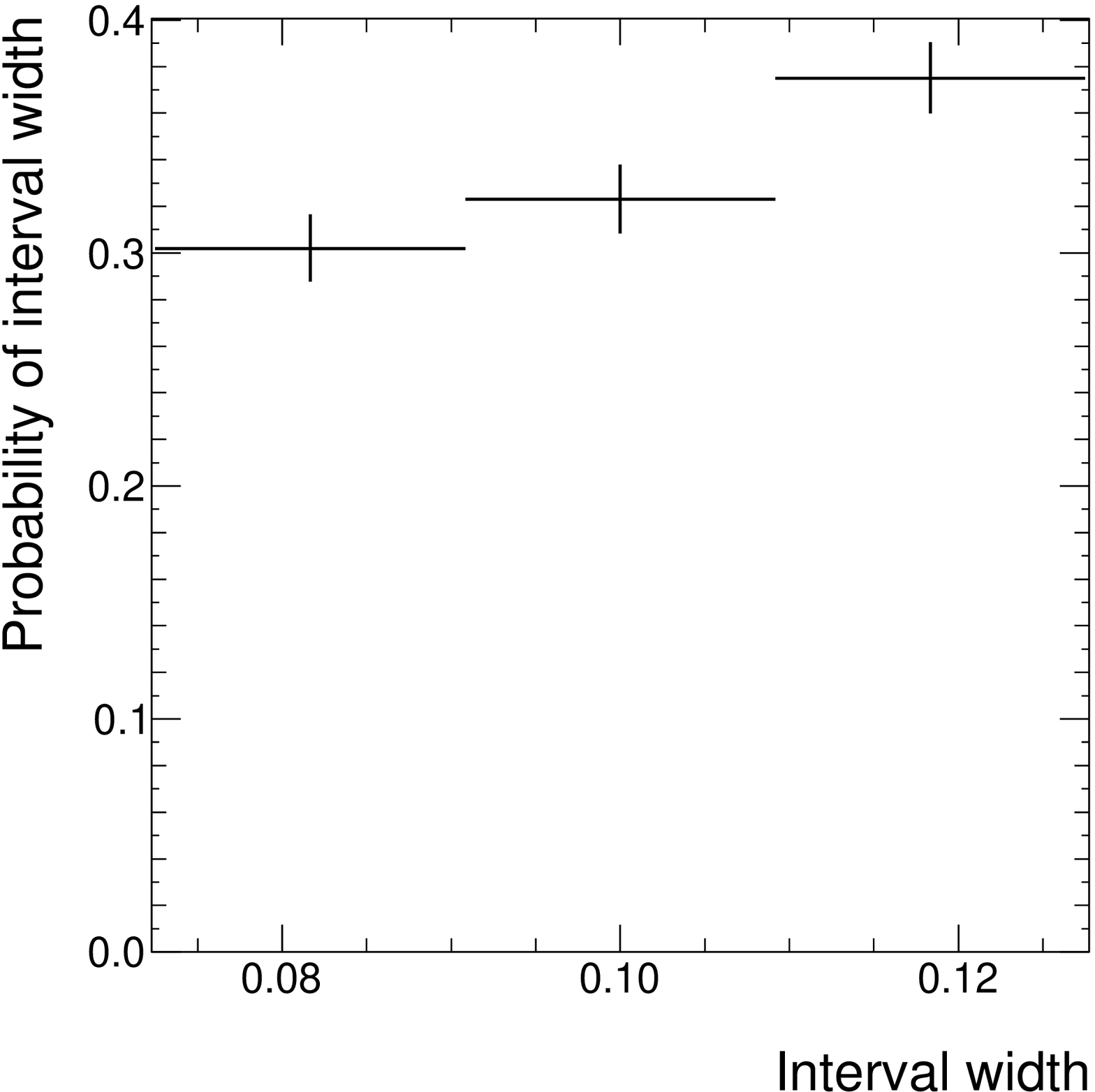}
\label{fig:intervals40e}
}
\subfigure[]{
\includegraphics[width=0.3\textwidth]{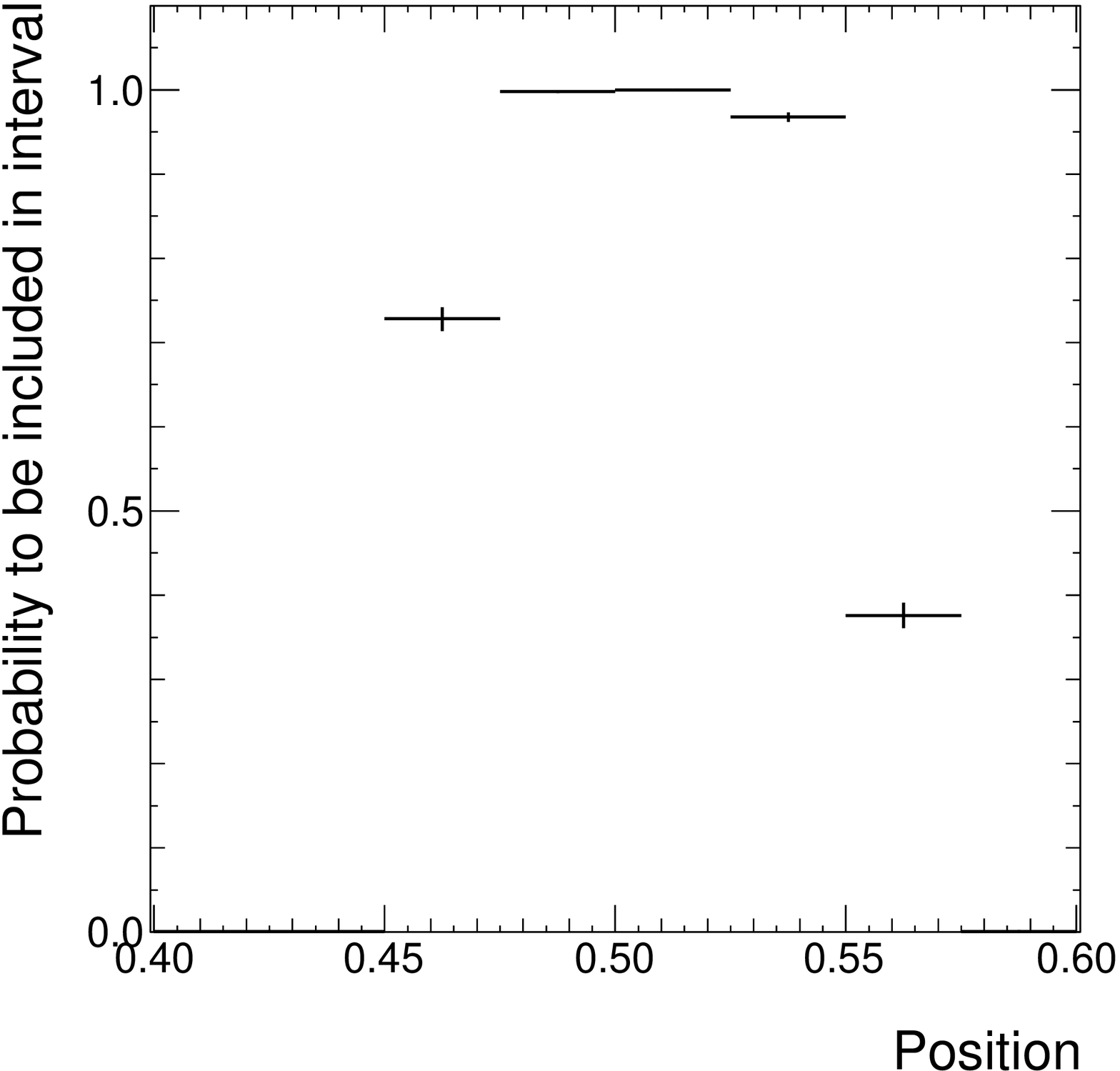}
\label{fig:intervals40f}
}
\caption{\label{fig:intervals40signal} Same as Fig.~\ref{fig:intervals1signal}, except that 40 signal events are expected, distributed as a Gaussian with $\sigma=0.03$ and mean 0.5.}
\end{figure}

Fig.~\ref{fig:intervals1signal} shows the same as Fig.~\ref{fig:intervals0signal}, except that just one signal event is injected (on average) on top of the $10^4 e^{-10x}$ background.  According to Fig.~\ref{fig:powersTunings}\subref{fig:power05}, the sensitivity to 1 signal event is very low.  However, in Fig.~\ref{fig:intervals1signal}\subref{fig:intervals1c} and \ref{fig:intervals1signal}\subref{fig:intervals1f} one sees that this signal is enough to give the right $x$ bins a much greater probability to be included in the most discrepant interval.
Fig.~\ref{fig:intervals10signal} shows the same, but with 10 signal events injected on average, which makes the effects more prominent.  In Fig.~\ref{fig:intervals10signal}\subref{fig:intervals10b} one sees that the intervals tend to have approximately the width of the injected signal.
Fig.~\ref{fig:intervals40signal} shows the same, but with 40 signal events injected on average, which means that the \bh has $~\sim$100\% probability to return $\pval \le 0.01$, according to Fig.~\ref{fig:powersTunings}\subref{fig:power05}.  In this case all intervals are located at the right position, and have the right width, given the finite size of $x$ bins which discretizes the width of the intervals returned by the \bh.

\section{Generalizing the \bh concept}

The \bh is not the only hypertest one could use, as explained in paragraph \ref{sec:theSet}.  Understanding the logic behind the \bh allows one to think of generalizations of this idea.  One such generalization is the \tailHunter (paragraph \ref{sec:th}).  Another is a hypertest that combines multiple distributions (paragraph \ref{sec:multi}).  Another hypertest, very similar to the \bh, was developped previously in the H1 experiment \cite{Aktas:2004pz}, where data deficits were also considered as potential signs of new physics, and no sideband criteria were used.  The H1 hypertest\footnote{This is not the terminology used by H1, but looking at it from the perspective of this work, it was indeed a hypertest, taking the trials factor into account correctly.}, which obviously predates this work, can be viewed a-posteriori as a particular tuning of the \bh.

\subsection{\tailHunter}
\label{sec:th}

A simple hypertest, analogous to the \bh, is the \tailHunter, which is used in \cite{dijetResonanceSearch}, and is also similar to the {\sc Sleuth} algorithm \cite{Knuteson:2001dq,Choudalakis:2007nb} used in \cite{vistaPRD,vistaPRDRC}.\footnote{Besides small technical differences, the biggest difference is that {\sc Sleuth} combined many final states, and didn't use fixed bins.  Regarding the combination of many final states, see paragraph \ref{sec:multi}.}

One can think of the \tailHunter as a \bh without sidebands, where the right edge of every window is always at the last bin that contains data.  The only requirement that remains in the definition of $t$ (eq.~\ref{eq:correctT}) is to have an excess of data with respect to the background.  All tails are examined by local hypothesis tests, the smallest \pval is used to define the statistic of the \tailHunter hypertest, and the \pval of the \tailHunter is found as explained in paragraph \ref{sec:allTests}.

\begin{figure}[p]
\centering
\subfigure[]{
\includegraphics[width=0.4\textwidth]{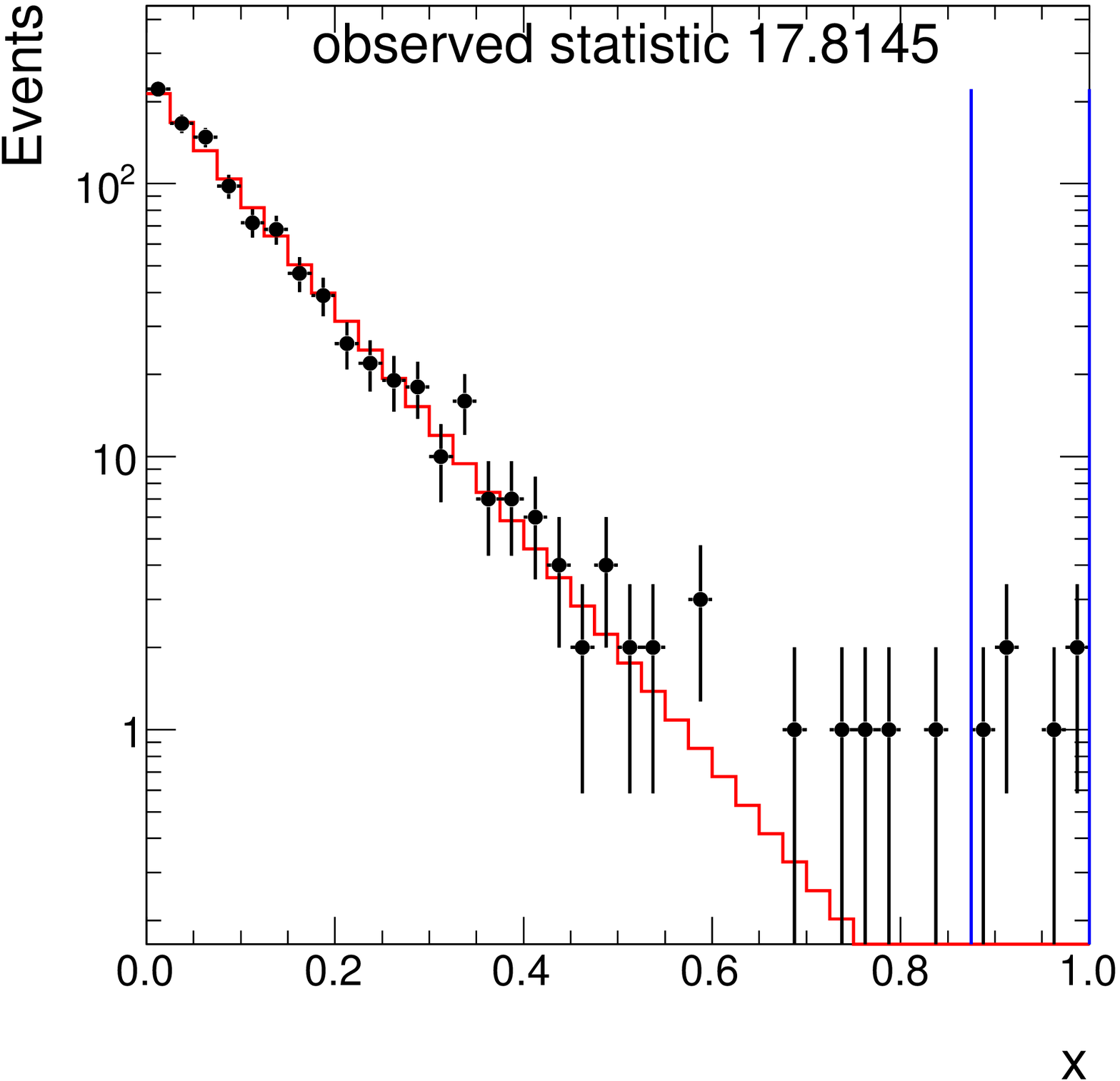}
\label{fig:TailHunterSpectrum}
}
\subfigure[]{
\includegraphics[width=0.4\textwidth]{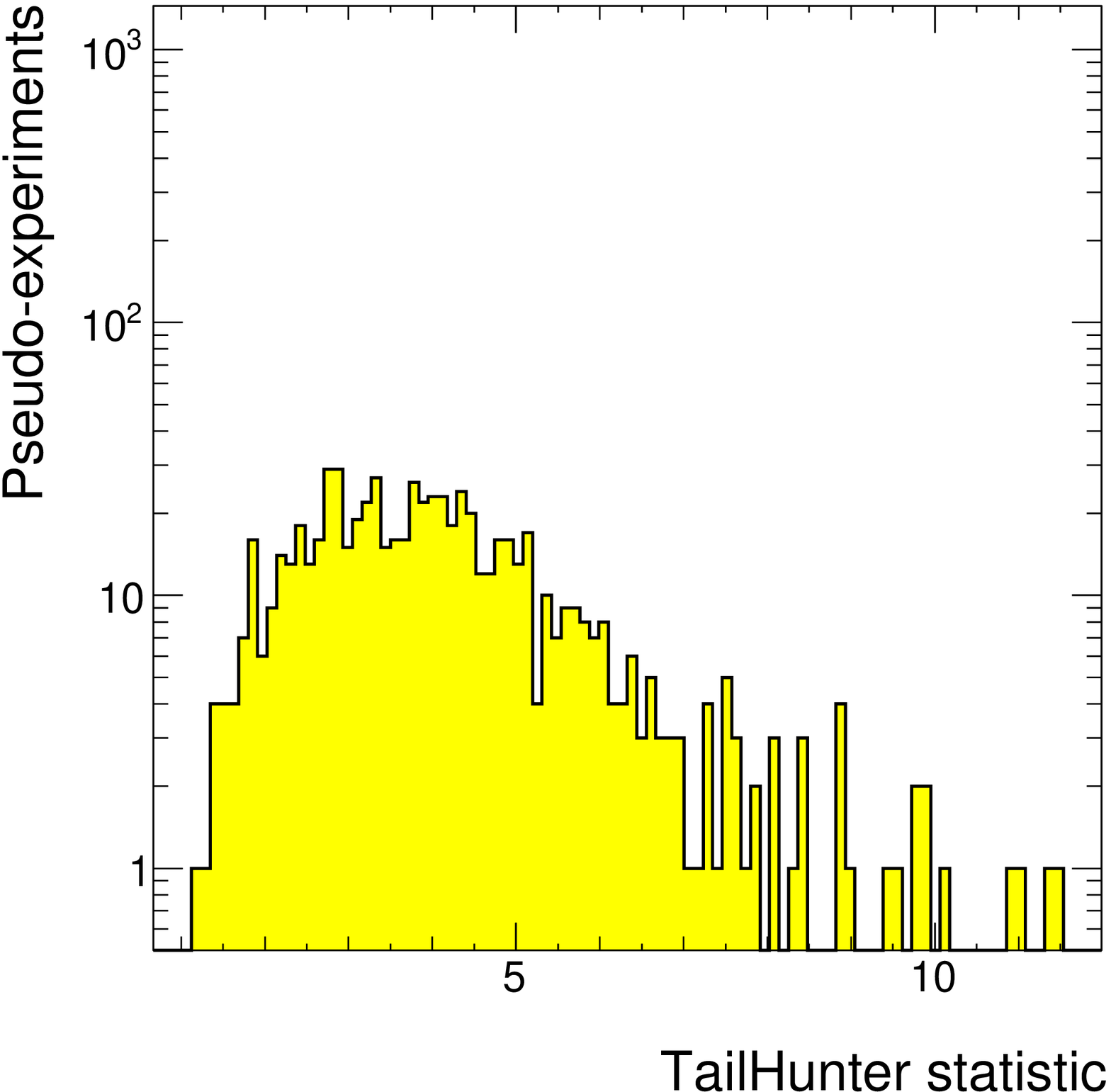}
\label{fig:TailHunterDistr}
}
\caption{\label{fig:TailHunter} Fig.~\subref{fig:TailHunterSpectrum} shows an example of spectrum where the \tailHunter locates a significant high-$x$ tail.  The spectrum has been constructed to contain indeed signal uniformly distributed between 0 and 1 (see paragraph~\ref{sec:th}).  Fig.~\subref{fig:TailHunterDistr} shows the distribution of the \tailHunter statistic under \Ho.  The observed \tailHunter statistic is 17.8, unmatched by any of the 690 pseudo-experiments generated, implying a \pval less than 0.01 with probability that exceeds 0.999.}
\end{figure}

Fig.~\ref{fig:TailHunter}\subref{fig:TailHunterSpectrum} presents an example of a spectrum where the \tailHunter finds \pval less than 0.01 with credibility greater than 0.999.  The spectrum is created by adding to dataset 0 of the Banff Challenge some signal events that follow a uniform distribution between 0 and 1, with 40 signal events expected in the whole interval.  The observed \tailHunter statistic in this example is 17.8, far beyond the values obtained in pseudo-data, shown in Fig.~\ref{fig:TailHunter}\subref{fig:TailHunterDistr}.

\subsection{Combining spectra}
\label{sec:multi}

Another hypertest (let's refer to it as mBH for ``multi-\bh''), allows the combination of two or more spectra to be scanned simultaneously.   
In some particle physics analyses this is useful, because an exotic particle may decay in many ways (e.g.\ $Z'\to e^+e^-$ and $Z'\to \mu^+\mu^-$), so the signal may populate two or more statistically independent distributions.  When we search for bumps in the mass spectrum of decay products, e.g.\ in $m_{ee}$ and $m_{\mu\mu}$, all spectra should indicate an excess at roughly the same mass, namely the mass of the new particle.  The width of the signal, though, is not expected to be the same in all distributions, since different decay products may be measured with different experimental resolution.

One way to extend the \bh into mBH is the following: The \bh statistic is first computed independently in each spectrum, and then the mBH statistic is defined as the sum of all \bh statistics\footnote{Remember that the \bh statistic is the negative logarithm of a \pval, so the sum of many \bh statistics is the negative logarithm of a product of \pvals.  Adding \bh statistics is equivalent to multiplying \pvals.}, with the extra requirement that all spectra must have their most interesting intervals within some distance from each other.  The exact distance criterion can be adjusted.  If bumps are found at different masses, then we can characterize the mBH's finding maximally uninteresting, by setting the mBH statistic to 0.  At last, the mBH \pval is estimated as explained in paragraph \ref{sec:allTests}.

The mBH is highly sensitive to signals that appear simultaneously in two (or more) spectra, because all signal significances are combined at the step where the \bh statistics are summed.  
Obviously, the mBH described so far makes a strong assumption; that the signal has to appear simultaneously in all examined spectra.  If this is indeed a characteristic of the signal, then mBH is more sensitive to it; otherwise it is not a well-motivated test.  As explained in paragraph~\ref{sec:theSet}, there is not a universally best hypertest.

If one relaxes the extra requirement which compares the interval locations in different spectra, and uses as mBH statistic the biggest \bh statistic instead of their sum, then mBH naturally reduces to the approach taken in \cite{vistaPRD, vistaPRDRC} and \cite{Aktas:2004pz} to search in multiple spectra without making strong assumptions.  There, each spectrum is examined independently, without checking for patterns across spectra, and without making any attempt to combine the significance of the findings in different spectra.  The smallest \pval from all spectra is noted (this corresponds to defining the mBH statistic as the maximum \bh statistic found across the examined spectra), and the probability is estimated of seeing a \pval as small as that, or smaller, in pseudo-data that follow \Ho in all distributions (and this corresponds to finding the \pval of the mBH).

\section{Conclusion}

After an introduction to hypothesis testing and the meaning of \pvals, the issue of the trials factor was illustrated, and a method to deal with it was proposed, by the introduction of hypothesis {\em hypertests}.  
One such hypertest is the \bh, inspired by searches for exotic phenomena in high energy physics.

The \bh algorithm is presented, and its performance is demonstrated with the opportunity of the Banff Challenge, Problem 1 \cite{BanffChallenge}.

\vspace{0.4cm}
Besides documenting the \bh (and \tailHunter) algorithm in detail, the author is open to collaborating with people who need his code.  Hopefully, it will soon be incorporated in a standard library, like {\tt ROOStats} \cite{roostats}.

I wish to thank Pekka Sinervo, Pierre Savard, Tom Junk, and Bruce Knuteson, for our fruitful discussions.


\bibliographystyle{utphys}
\bibliography{bumpHunterDescription}{}

%
%

\appendix

\section{First 100 lines from the Banff Challenge, problem 1}
\label{sec:log}

\tiny
\begin{verbatim}
0	0	0.9 = 9/10	 P(pval>0.01)= 1	0	0	0	0	0	0	0	0	0
1	0	0.9 = 9/10	 P(pval>0.01)= 1	0	0	0	0	0	0	0	0	0
2	0	0.1 = 3/30	 P(pval>0.01)= 0.999746	0	0	0	0	0	0	0	0	0
3	0	1 = 10/10	 P(pval>0.01)= 1	0	0	0	0	0	0	0	0	0
4	0	0.8 = 8/10	 P(pval>0.01)= 1	0	0	0	0	0	0	0	0	0
5	0	0.1 = 3/30	 P(pval>0.01)= 0.999746	0	0	0	0	0	0	0	0	0
6	0	0.0666667 = 4/60	 P(pval>0.01)= 0.999626	0	0	0	0	0	0	0	0	0
7	0	0.4 = 4/10	 P(pval>0.01)= 1	0	0	0	0	0	0	0	0	0
8	0	0.0666667 = 6/90	 P(pval>0.01)= 0.999961	0	0	0	0	0	0	0	0	0
9	0	0.4 = 4/10	 P(pval>0.01)= 1	0	0	0	0	0	0	0	0	0
10	1	0 = 0/690	P(pval<0.01)= 0.99904	0.663528	0.645274	0.681782	0.128468	0.061079	0.195858	242.076	230.556	253.595
11	0	0.166667 = 5/30	 P(pval>0.01)= 0.999999	0	0	0	0	0	0	0	0	0
12	0	0.4 = 4/10	 P(pval>0.01)= 1	0	0	0	0	0	0	0	0	0
13	0	0.6 = 6/10	 P(pval>0.01)= 1	0	0	0	0	0	0	0	0	0
14	0	0.5 = 5/10	 P(pval>0.01)= 1	0	0	0	0	0	0	0	0	0
15	0	0.25 = 5/20	 P(pval>0.01)= 1	0	0	0	0	0	0	0	0	0
16	0	0.2 = 4/20	 P(pval>0.01)= 0.999998	0	0	0	0	0	0	0	0	0
17	0	0.5 = 5/10	 P(pval>0.01)= 1	0	0	0	0	0	0	0	0	0
18	0	1 = 10/10	 P(pval>0.01)= 1	0	0	0	0	0	0	0	0	0
19	0	0.4 = 4/10	 P(pval>0.01)= 1	0	0	0	0	0	0	0	0	0
20	0	0.2 = 2/10	 P(pval>0.01)= 0.999845	0	0	0	0	0	0	0	0	0
21	0	0.8 = 8/10	 P(pval>0.01)= 1	0	0	0	0	0	0	0	0	0
22	1	0.00431373 = 11/2550	P(pval<0.01)= 0.999027	0.0907464	0.0800601	0.101433	2.5333	1.78409	3.28251	236.094	221.044	251.144
23	0	0.3 = 3/10	 P(pval>0.01)= 0.999997	0	0	0	0	0	0	0	0	0
24	0	0.1 = 3/30	 P(pval>0.01)= 0.999746	0	0	0	0	0	0	0	0	0
25	1	0.00357143 = 7/1960	P(pval<0.01)= 0.99901	0.497455	0.488462	0.506448	0.392582	0.266279	0.518885	267.207	255.062	279.351
26	0	0.0136605 = 103/7540	 P(pval>0.01)= 0.999016	0	0	0	0	0	0	0	0	0
27	0	0.0165385 = 43/2600	 P(pval>0.01)= 0.999245	0	0	0	0	0	0	0	0	0
28	0	0.6 = 6/10	 P(pval>0.01)= 1	0	0	0	0	0	0	0	0	0
29	0	1 = 10/10	 P(pval>0.01)= 1	0	0	0	0	0	0	0	0	0
30	0	0.9 = 9/10	 P(pval>0.01)= 1	0	0	0	0	0	0	0	0	0
31	0	0.4 = 4/10	 P(pval>0.01)= 1	0	0	0	0	0	0	0	0	0
32	0	0.0428571 = 6/140	 P(pval>0.01)= 0.999411	0	0	0	0	0	0	0	0	0
33	0	0.6 = 6/10	 P(pval>0.01)= 1	0	0	0	0	0	0	0	0	0
34	0	0.5 = 5/10	 P(pval>0.01)= 1	0	0	0	0	0	0	0	0	0
35	1	0 = 0/690	P(pval<0.01)= 0.99904	0.391657	0.385773	0.397541	1.0584	0.834799	1.282	297.514	284.591	310.436
36	0	0.3 = 3/10	 P(pval>0.01)= 0.999997	0	0	0	0	0	0	0	0	0
37	0	0.2 = 2/10	 P(pval>0.01)= 0.999845	0	0	0	0	0	0	0	0	0
38	0	0.2 = 2/10	 P(pval>0.01)= 0.999845	0	0	0	0	0	0	0	0	0
39	0	0.2 = 2/10	 P(pval>0.01)= 0.999845	0	0	0	0	0	0	0	0	0
40	0	0.9 = 9/10	 P(pval>0.01)= 1	0	0	0	0	0	0	0	0	0
41	1	0.00855856 = 266/31080	P(pval<0.01)= 0.999893	0.543414	0.52964	0.557188	0.194656	0.105582	0.283731	220.339	209.713	230.964
42	1	0 = 0/690	P(pval<0.01)= 0.99904	0.143887	0.134607	0.153166	2.21724	1.70884	2.72564	287.405	273.936	300.875
43	0	0.6 = 6/10	 P(pval>0.01)= 1	0	0	0	0	0	0	0	0	0
44	0	0.2 = 2/10	 P(pval>0.01)= 0.999845	0	0	0	0	0	0	0	0	0
45	0	0.1 = 3/30	 P(pval>0.01)= 0.999746	0	0	0	0	0	0	0	0	0
46	0	0.0177778 = 32/1800	 P(pval>0.01)= 0.999091	0	0	0	0	0	0	0	0	0
47	0	0.7 = 7/10	 P(pval>0.01)= 1	0	0	0	0	0	0	0	0	0
48	0	0.3 = 3/10	 P(pval>0.01)= 0.999997	0	0	0	0	0	0	0	0	0
49	0	0.5 = 5/10	 P(pval>0.01)= 1	0	0	0	0	0	0	0	0	0
50	0	0.3 = 3/10	 P(pval>0.01)= 0.999997	0	0	0	0	0	0	0	0	0
51	0	0.7 = 7/10	 P(pval>0.01)= 1	0	0	0	0	0	0	0	0	0
52	0	0.2 = 6/30	 P(pval>0.01)= 1	0	0	0	0	0	0	0	0	0
53	0	0.3 = 3/10	 P(pval>0.01)= 0.999997	0	0	0	0	0	0	0	0	0
54	0	1 = 10/10	 P(pval>0.01)= 1	0	0	0	0	0	0	0	0	0
55	0	0.133333 = 4/30	 P(pval>0.01)= 0.999986	0	0	0	0	0	0	0	0	0
56	0	0.2 = 2/10	 P(pval>0.01)= 0.999845	0	0	0	0	0	0	0	0	0
57	0	0.05 = 5/100	 P(pval>0.01)= 0.999437	0	0	0	0	0	0	0	0	0
58	0	0.4 = 4/10	 P(pval>0.01)= 1	0	0	0	0	0	0	0	0	0
59	0	0.2 = 2/10	 P(pval>0.01)= 0.999845	0	0	0	0	0	0	0	0	0
60	0	0.4 = 4/10	 P(pval>0.01)= 1	0	0	0	0	0	0	0	0	0
61	0	0.7 = 7/10	 P(pval>0.01)= 1	0	0	0	0	0	0	0	0	0
62	0	0.4 = 4/10	 P(pval>0.01)= 1	0	0	0	0	0	0	0	0	0
63	0	0.15 = 3/20	 P(pval>0.01)= 0.999948	0	0	0	0	0	0	0	0	0
64	0	0.6 = 6/10	 P(pval>0.01)= 1	0	0	0	0	0	0	0	0	0
65	0	0.3 = 3/10	 P(pval>0.01)= 0.999997	0	0	0	0	0	0	0	0	0
66	0	0.0368421 = 7/190	 P(pval>0.01)= 0.999247	0	0	0	0	0	0	0	0	0
67	0	0.25 = 5/20	 P(pval>0.01)= 1	0	0	0	0	0	0	0	0	0
68	0	1 = 10/10	 P(pval>0.01)= 1	0	0	0	0	0	0	0	0	0
69	0	0.4 = 4/10	 P(pval>0.01)= 1	0	0	0	0	0	0	0	0	0
70	0	0.3 = 3/10	 P(pval>0.01)= 0.999997	0	0	0	0	0	0	0	0	0
71	0	0.7 = 7/10	 P(pval>0.01)= 1	0	0	0	0	0	0	0	0	0
72	0	0.0571429 = 4/70	 P(pval>0.01)= 0.999247	0	0	0	0	0	0	0	0	0
73	1	0.00431373 = 11/2550	P(pval<0.01)= 0.999027	0.507966	0.496324	0.519609	0.391159	0.262968	0.51935	275.858	263.193	288.522
74	0	0.4 = 4/10	 P(pval>0.01)= 1	0	0	0	0	0	0	0	0	0
75	0	0.1 = 4/40	 P(pval>0.01)= 0.999944	0	0	0	0	0	0	0	0	0
76	0	0.2 = 2/10	 P(pval>0.01)= 0.999845	0	0	0	0	0	0	0	0	0
77	0	0.2 = 2/10	 P(pval>0.01)= 0.999845	0	0	0	0	0	0	0	0	0
78	0	0.5 = 5/10	 P(pval>0.01)= 1	0	0	0	0	0	0	0	0	0
79	0	0.4 = 4/10	 P(pval>0.01)= 1	0	0	0	0	0	0	0	0	0
80	0	0.4 = 4/10	 P(pval>0.01)= 1	0	0	0	0	0	0	0	0	0
81	1	0.00230769 = 3/1300	P(pval<0.01)= 0.99902	0.508847	0.495754	0.521941	0.24085	0.135517	0.346183	251.84	240.059	263.621
82	0	0.5 = 5/10	 P(pval>0.01)= 1	0	0	0	0	0	0	0	0	0
83	0	0.8 = 8/10	 P(pval>0.01)= 1	0	0	0	0	0	0	0	0	0
84	0	0.7 = 7/10	 P(pval>0.01)= 1	0	0	0	0	0	0	0	0	0
85	0	0.2 = 4/20	 P(pval>0.01)= 0.999998	0	0	0	0	0	0	0	0	0
86	0	0.3 = 3/10	 P(pval>0.01)= 0.999997	0	0	0	0	0	0	0	0	0
87	0	0.2 = 2/10	 P(pval>0.01)= 0.999845	0	0	0	0	0	0	0	0	0
88	0	0.5 = 5/10	 P(pval>0.01)= 1	0	0	0	0	0	0	0	0	0
89	1	0 = 0/690	P(pval<0.01)= 0.99904	0.961137	0.885444	1.03683	6.06533	-670.853	682.983	245.924	234.977	256.871
90	0	0.166667 = 5/30	 P(pval>0.01)= 0.999999	0	0	0	0	0	0	0	0	0
91	0	0.8 = 8/10	 P(pval>0.01)= 1	0	0	0	0	0	0	0	0	0
92	0	0.075 = 3/40	 P(pval>0.01)= 0.999246	0	0	0	0	0	0	0	0	0
93	0	0.0473684 = 9/190	 P(pval>0.01)= 0.999973	0	0	0	0	0	0	0	0	0
94	0	0.7 = 7/10	 P(pval>0.01)= 1	0	0	0	0	0	0	0	0	0
95	0	0.5 = 5/10	 P(pval>0.01)= 1	0	0	0	0	0	0	0	0	0
96	0	0.6 = 6/10	 P(pval>0.01)= 1	0	0	0	0	0	0	0	0	0
97	0	0.0333333 = 9/270	 P(pval>0.01)= 0.999529	0	0	0	0	0	0	0	0	0
98	0	0.0571429 = 4/70	 P(pval>0.01)= 0.999247	0	0	0	0	0	0	0	0	0
99	0	0.9 = 9/10	 P(pval>0.01)= 1	0	0	0	0	0	0	0	0	0
\end{verbatim}

\end{document}